\documentclass[preprint,aps,nofootinbib]{revtex4-1}

\usepackage{graphicx}
\usepackage{amsmath,slashed}
\usepackage{amsfonts}
\usepackage{hyperref}
\usepackage[capitalise]{cleveref}
\usepackage{epstopdf}
\usepackage{subcaption}
\usepackage{multirow}
\usepackage{fancyhdr}
\usepackage[utf8]{inputenc}
\usepackage{xcolor}

\def\be{\begin{equation}}
\def\ee{\end{equation}}

\def\vep{\varepsilon}

\def\tr{\mathrm{tr}}
\def\ordr{\mathcal{O}}

\def\sib{\begin{align}r{\sigma}}
\def\<{\langle}
\def\>{\rangle}

\def\pole{ {\mathrm{pole}}}
\def\gut{ {\mathrm{gut}}}

\def\mt{{\mathcal{T}}}

\def\lag{{\mathcal{L}}}

\def\non{\nonumber}
\def\bea{\begin{eqnarray}}
\def\eea{\end{eqnarray}}
\def\bean{\begin{eqnarray*}}
\def\eean{\end{eqnarray*}}

\def\nlsm{\text{NLSM}}
\def\dbi{\text{DBI}}

\def\ym{\text{YM}}

\def\ee{\eta}

\def\ta{{\tilde{a}}}
\def\tb{{\tilde{b}}}
\def\tc{{\tilde{c}}}

\def\wzw{{\text{wzw}}}
\def\SO{{\text{SO}}}
\def\SU{{\text{SU}}}
\def\U{{\text{U}}}

\def\alpd{\dot{\alpha}}
\def\Ar{{\text{Ar}}}
\def\pf{\mathrm{Pf}}
\def\A{\textsf{A}}
\def\sgn{{\text{sgn}}}
\def\J{\mathcal{J}}
\def\ta{{\tilde{a}}}
\def\tb{{\tilde{b}}}
\def\tc{{\tilde{c}}}
\def\tdi{{\tilde{i}}}
\def\tdj{{\tilde{j}}}

\def\tT{{\tilde{T}}}

\def\ff{\mathcal{F}}

\def\M{{\cal M}}

\def\sff{\text{f}}
\def\gT{\textsf{T}}
\def\gX{\textsf{X}}

\def\R{{\cal R}}
\def\V{{\cal V}}
\def\S{{\cal S}}
\def\gd{\mathsf{d}}
\def\eom{\text{EoM}}
\def\bN{\mathbf{N}}

\newcommand{\Rmnum}[1]{\expandafter\@slowromancap\romannumeral #1@}

\newcommand*\DAl{\mathop{}\!\mathbin\Box}

\begin{document}

\vspace*{.5cm}

\title{Exploring the Landscape for Soft Theorems of Nonlinear Sigma Models}

\author{\vspace{0.5cm} Laurentiu Rodina$^{\, a}$, Zhewei Yin$^{\, b,c}$}
\affiliation{\vspace{0.5cm}
\mbox{$^a$ Department of Physics, National Taiwan University,  Taipei 10617, Taiwan}\\
\mbox{$^b$ Department of Physics and Astronomy, Uppsala University, 75108 Uppsala, Sweden}\\ 
\mbox{$^c$ Department of Physics and Astronomy, Northwestern University, Evanston, IL 60208, USA}
 \vspace{0.5cm}
}

\begin{abstract}
We generalize soft theorems of the nonlinear sigma model beyond the $\ordr (p^2)$ amplitudes and the coset of $\SU (N) \times \SU (N) / \SU (N) $. We first discuss the universal flavor ordering of the amplitudes for the Nambu-Goldstone bosons,  so that we can reinterpret the known $\ordr (p^2)$ single soft theorem for $\SU (N) \times \SU (N) / \SU (N) $ in the context of a general symmetry group representation. We then investigate the special case of the fundamental representation of $\SO (N)$, where a special flavor ordering of the ``pair basis'' is available. We provide novel amplitude relations and a Cachazo-He-Yuan formula for such a basis, and derive the corresponding single soft theorem. Next, we extend the single soft theorem for a general group representation to $\ordr (p^4)$, where for at least two specific choices of the $\ordr (p^4)$ operators, the leading non-vanishing pieces can be interpreted as new extended theory amplitudes involving bi-adjoint scalars, and the corresponding soft factors are the same as at $\ordr (p^2)$. Finally, we compute the general formula for the double soft theorem, valid to all derivative orders, where the leading part in the soft momenta is fixed by the $\ordr(p^2)$ Lagrangian, while any possible corrections to the subleading part are determined by the $\ordr(p^4)$ Lagrangian alone. Higher order terms in the derivative expansion do not contribute any new corrections to the double soft theorem. 

\end{abstract}

\preprint{UUITP-10/21}

\maketitle

\tableofcontents

\section{Introduction}

The non-linear sigma model (NLSM) \cite{GellMann:1960np,Coleman:1969sm,Callan:1969sn}, originally designed to describe light mesons in the chiral perturbation theory \cite{Weinberg:1978kz,Gasser:1983yg,Gasser:1984gg}, serves as an effective field theory (EFT) for Nambu-Goldstone bosons (NGB's), which are consequences of spontaneous symmetry breaking. It was gradually realized \cite{Susskind:1970gf,Ellis:1970nn} and made concrete in the recent years \cite{Low:2014nga,Low:2014oga} that a universal Lagrangian can be constructed for NLSM using entirely infrared (IR) information, without input from the symmetry breaking pattern in the ultraviolet (UV). The IR universality has already been shown to have profound phenomenological implications \cite{Liu:2018vel,Liu:2018qtb,Liu:2019rce} for composite Higgs models \cite{Kaplan:1983fs,Agashe:2004rs,Panico:2015jxa}.

On the other hand, the past decade has seen much activity in studying on-shell properties of the NLSM \cite{ArkaniHamed:2008gz,Kampf:2012fn,Kampf:2013vha,Chen:2013fya,Chen:2014dfa,He:2016vfi,Du:2016tbc,Chen:2016zwe,Cheung:2017ems,Cheung:2017yef,Mizera:2018jbh,Carrillo-Gonzalez:2018pjk,Bjerrum-Bohr:2018jqe,Gomez:2019cik,Carrasco:2019qwr}. The bulk of these endeavors focus on the NLSM amplitudes at the leading $\ordr (p^2)$ in the derivative expansion of the EFT, as well as of the coset $\SU (N) \times \SU (N)/\SU (N)$ seen in the chiral perturbation theory. This appears as a disconnection to the IR universality of the EFT Lagrangian. There are notable exceptions \cite{Carrillo-Gonzalez:2019aao,Bijnens:2019eze}, such as  interesting higher derivative terms connected to the Z-theory \cite{Carrasco:2016ldy,Carrasco:2016ygv}, which involves a very specific set of operators starting at $\ordr (p^6)$. However, the general amplitudes at the subleading $\ordr(p^4)$, as well as other kinds of flavor symmetries, are less understood. 

A particularly important aspect of the NLSM amplitudes is the soft theorems, which dictate the on-shell behaviors when there exists a hierarchy among the energy of the external states. Interesting soft theorems are seen in a variety of quantum field theories, and have received much attention in the recent years for their connections to other subjects such as asymptotic symmetries and memory effects \cite{Strominger:2017zoo}. For NLSM, in the single soft limit, i.e. when one of the external momenta is taken to zero, the amplitudes vanish\footnote{This applies to the common case when the NGB's transform under a group representation that can be embedded into a symmetric coset, which will be the focus of this work. For a recent discussion on the soft theorems when this is not the case, see Ref. \cite{Kampf:2019mcd}.}, a behavior known as the Adler zero \cite{Adler:1964um}. Acting as a defining property of the EFT by enforcing a nonlinear shift symmetry, the Adler zero is a key ingredient in the IR construction of the universal Lagrangian. An on-shell equivalent of such a construction is the soft bootstrap \cite{Cheung:2014dqa,Cheung:2015ota,Cheung:2016drk,Elvang:2018dco,Low:2019ynd}, which utilizes recursion relations that are only valid because of the Adler zero, and has been used to explore higher derivative corrections \cite{Dai:2020cpk}. Other constructions in similar spirit have been realized as well \cite{Kampf:2013vha,Arkani-Hamed:2016rak,Rodina:2016jyz,Rodina:2018pcb}.

The leading non-vanishing term in the single soft theorem of the tree level $\ordr (p^2)$ NLSM amplitudes involves an extended theory with additional field content of bi-adjoint scalars, which was first discovered \cite{Cachazo:2016njl} using the Cachazo-He-Yuan (CHY) formalism \cite{Cachazo:2013gna,Cachazo:2013hca,Cachazo:2013iea,Cachazo:2014xea}. The appearance of the extended theory was later understood as a consequence of the Ward identity corresponding to the shift symmetry, which enables a direct calculation of the relevant Feynman rules \cite{Low:2017mlh,Low:2018acv,Yin:2018hht}. These previous results have all been presented in the context of a symmetry breaking pattern of $\SU (N) \times \SU (N)/\SU (N)$. We argue here that such a restriction is unnecessary, because the flavor ordering at this derivative order is universal. The extended theory emerging from the soft theorem thus also can be interpreted in more general group representations.

A specific application of such a generalization is for NGB's in the fundamental representation $\bN$ of $\SO (N)$, as seen in composite Higgs models, where the Higgs doublet is treated as pseudo NGB's furnishing $\mathbf{4}$ of the custodial $\SO (4)$. Apart from the single trace ordering for a general group representation, a special ``pair basis'' is available in this case. An extended theory amplitude in such a basis has been shown to exist in the corresponding single soft theorem \cite{Low:2019wuv}, for which we provide detailed derivations, uncovering interesting properties of the pair basis amplitudes along the way.

By using the Ward identity, there is no obstruction to calculate the leading non-vanishing piece in the single soft theorem to higher derivative orders. What is less clear is whether there still exist the interpretations of extended theories. Recent work by the authors \cite{Low:2020ubn} asked a similar question in the context of the double copy \cite{Bern:2019prr}. It was demonstrated that at least one $\ordr (p^4)$ operator, which results in a theory dubbed $\nlsm^{d_2}$, admits a double copy construction, through novel color-kinematic numerators  \cite{Carrasco:2019yyn,Low:2019wuv} which do not necessarily imply Bern-Carrasco-Johansson (BCJ) relations \cite{Bern:2008qj}. 

Naturally, one would expect an extended theory emerging from the single soft theorem of $\nlsm^{d_2}$, as previously its appearance has been known to be intricately related to the existence of double copy structures and CHY representations of the amplitudes. We find that this is indeed the case. However, we are also able to discover that at least another $\ordr (p^4)$ operator  gives rise to an extended theory as well. This is surprising, as the corresponding  amplitudes have no known double copy structures or CHY representations, marking the first instance when an extended theory emerging in the single soft theorem is uncovered using the Ward identity alone.

The double soft theorem for the NLSM has also been well established, when two of the external states take much less energy than the rest. Unlike the single soft case, the double soft limits of NLSM do not generate extended theory amplitudes with new field content, but just the usual lower point amplitudes as in the well-known single and double soft limits of gauge theory and gravity \cite{Strominger:2017zoo}. The double soft theorem has been known to the leading and subleading orders in the soft expansion \cite{Cachazo:2015ksa,Du:2015esa}, and has been studied in a completely general group representation as well \cite{Low:2015ogb}, but still only for $\ordr (p^2)$ in the EFT expansion. It is natural then to extend it to higher derivative orders. We find that at tree level, the double soft theorems can actually be computed to all orders in the derivative expansion, by judiciously applying various single soft limits in the spirit of Ref. \cite{Low:2015ogb}. Interestingly, they are fully determined by the $\ordr(p^2)$ and $\ordr(p^4)$ order Lagrangians, implying that higher derivative corrections to NLSM all satisfy the same soft theorems.

The paper is organized as follows. In Section \ref{sec2} we review the usual trace decomposition of the NLSM amplitudes governed by the linear flavor symmetry, as well as the single soft theorem as a consequence of the nonlinear shift symmetry and the corresponding Ward identity. The universality of the flavor ordering for a general symmetry group representation enables us to reinterpret the known extended theory in the single soft limit. In Section \ref{sec:son} we inspect the specific case of NGB's furnishing $\bN$ of $\SO (N)$, where we study the alternative flavor decomposition in the pair basis. We present new amplitude relations and a CHY formula for the amplitudes in such a basis, then derive the corresponding single soft theorem at $\ordr (p^2)$ and the associated extended theory amplitudes. In Section \ref{sec4} we compute the single soft theorems for $\ordr(p^4)$ amplitudes, and study the possible existence of the extended theories, first for the $\nlsm^{d_2}$ and then for the general case. In Section \ref{sec5} we compute the double-soft theorems valid to arbitrary derivative order. We conclude in Section \ref{sec6}. We also provide supporting materials in the appendices: useful derivations in Appendix \ref{app:der}, as well as examples related to the single soft theorems in Appendix \ref{app:lpv}.

\section{Symmetries and the amplitudes of the NLSM}\label{sec2}

As an EFT, the NLSM is valid below a high energy scale $\Lambda$, and its Lagrangian admits a derivative expansion of $\partial/\Lambda$. Up to the 4-derivative level, we can write
\bea
\lag^\nlsm = \lag^{(2)} + \lag^{(4)} +\ordr \left( \frac{1}{\Lambda^4} \right),
\eea
with $\lag^{(2)} = \ordr (1/\Lambda^0)$ and $\lag^{(4)} = \ordr (1/ \Lambda^2)$.

Let us consider a general NLSM of NGB's furnishing the representation $R$ of a linearly realized flavor group $H$, with  associated generators $(T^i)_{ab}$\footnote{We choose a totally imaginary and anti-symmetric basis for any group generators throughout this work, and the normalization is given by $\tr (T^i T^j) = \delta^{ij}$, $\tr (\gX^a \gX^b) = \delta^{ab}$, etc.}, and the NGB $\pi^a$ carries the flavor index $a$. The basic building blocks for the Lagrangian are
\bea
d_\mu \equiv d_\mu^a \gX^a,\qquad   E_\mu \equiv E_\mu^i \gT^i,
\eea
where $\gX^a$ and $\gT^i$ are generators of some group $G$ containing the subgroup $H$. In the traditional coset construction of Callan, Coleman, Wess and Zumino \cite{Coleman:1969sm,Callan:1969sn}, the NGB's are known to be coming from the spontaneous symmetry breaking of group $G$ to the unbroken group $H$, $\gT^i$ are the ``unbroken generators'' of the subgroup $H$, while $\gX^a$ are the ``broken generators'' associated with the coset $G/H$. On the other hand, the same Lagrangian can also be constructed using entirely IR information of the linearly realized group $H$ and its representation $R$ \cite{Low:2014nga,Low:2014oga}, where $\gT^i$ and $\gX^a$ are constructed using the generators $(T^i)_{ab}$ of $R$ and the structure constants $f^{ijk}$ of $H$ \cite{Low:2019ynd}. We are also assuming that the generators $T^i$ satisfy the ``closure condition'':
\bea
T^i_{ab} T^i_{cd} + T^i_{ac} T^i_{db}+T^i_{ad} T^i_{bc} = 0,\label{eq:clocon}
\eea
which guarantees that $R$ of $H$ can be embedded into a symmetric coset $G/H$, so that one can identify $T^i_{ab} = -if^{iab}$ as the structure constants for $G$, whose Lie algebra is given by
\bea
[\gX^a, \gX^b] = if^{iab} \gT^i,\qquad [\gT^i, \gX^a] = if^{iab} \gX^{b},\qquad  [\gT^i, \gT^j] = if^{ijk} \gT^k.\label{eq:crg}
\eea

In general, $d_\mu^a$ and $E_\mu^i$ can be expressed as
\bea
d_\mu^a = \frac{1}{f} [ F_1 (\mt) ]_{ab}  \partial_\mu \pi^b,  \qquad E_\mu^i = \frac{1}{f^2} \partial_\mu \pi^{a}  [ F_2(\mt) ]_{a b} T^i_{b c} \pi^c , \label{eq:dexpo}
\eea
where
\bea
F_1 (\mt) &=& \frac{\sin \sqrt{\mt}}{\sqrt{ \mt} } = \sum_{n=0}^\infty \frac{(-1)^n}{(2n+1)! } \mt^n,\label{eq:ccf1}\\
F_2 (\mt ) &=& -\frac{2i}{\mt} \sin^2 \frac{\sqrt{\mt}}{2} =-i \sum_{n=0}^\infty \frac{(-1)^n}{(2n+2)! } \mt^n, \label{eq:ccf2}\\
(\mt)_{a b} &=& \frac{1}{f^2} T^i_{ac} T^i_{db} \pi^c \pi^{d} .
\eea
Notice that both $d_\mu$ and $E_\mu$ are linear in $\partial$, while being a series expansion of $\pi/f$, where $f$ is the coupling constant with the same mass dimension as $\pi$. One can construct a ``covariant derivative'' $\nabla_\mu$ so that
\bea
\nabla_\mu d_\nu \equiv \partial_\mu d_\nu + i[E_\mu, d_\nu].
\eea
Then the Lagrangian can be expressed as:
\bea
\lag^\nlsm = f^2 \Lambda^2 \tilde{\lag} \left(\frac{d}{\Lambda},\frac{\nabla}{\Lambda} \right),\label{eq:nlsmp}
\eea
where $\tilde{\lag}$ is a dimensionless function. Up to $\ordr (p^4)$, the Lagrangian is given by
\bea
 \lag^{(2)}= \frac{f^2}{2} \tr \left( d_\mu d^\mu \right),\qquad
 \lag^{(4)}= \frac{f^2}{\Lambda^2} \left( \sum_{i=1}^4 C_i O_i + C_- O_\wzw \right)  ,\label{eq:nlsml4}
\eea
where $C_i$ and $C_-$ are dimensionless Wilson coefficients,
\bea
O_1  =  [ \tr ( d_\mu d^\mu ) ]^2, \quad O_2  =  [\tr ( d_\mu d_\nu ) ]^2, \quad O_3  =  \tr ( [ d_\mu, d_\nu]^2 ) ,\quad O_4  =  \tr ( \{ d_\mu, d_\nu \}^2 ) \label{eq:o14}
\eea
are the parity (P) even generators at $\ordr (p^4)$, and
\bea
  O_\wzw =  \vep^{\mu\nu\rho\sigma}\  \tr \left( \Pi  \partial_\mu\Pi \partial_\nu\Pi \partial_\rho\Pi \partial_\sigma\Pi \right)+ \ordr (\Pi^7), \qquad \Pi \equiv \frac{\pi^a \gX^a}{f}
\eea
is the P-odd Wess-Zumino-Witten (WZW) term \cite{Wess:1971yu,Witten:1983tw} that can exist when the spacetime dimension $d=4$, and can be expressed using $d_\mu$ by compactifying a 5-dimensional spacetime:
\bea
S_{\wzw}  \propto \int d^5y \  \vep^{\alpha \beta \gamma \delta \epsilon} \tr  ( d_\alpha d_\beta d_\gamma d_\delta d_\epsilon ) = \int d^4 x\  O_\wzw .\label{eq:lagwzwdef}
\eea

The form of $\lag^\nlsm$ is dictated by the linearly realized symmetry of $H$, as well as a nonlinearly realized shift symmetry, which we will discuss in detail in Section \ref{sec:st}. Notice that up to $\ordr (p^4)$, the Lagrangian can be expressed entirely using $d_\mu$ without involving $\nabla_\mu$. One may write down other operators for the Lagrangian, e.g. $\tr ( d_\mu d^\nu \nabla^\mu d^\nu)$, but they will not be independent of the operators in Eq. (\ref{eq:nlsml4}): they are related by total derivatives, symmetry transformations or the equation of motion (EoM) \cite{Low:2019ynd}. If we specify the $R$ of $H$, the basis of independent operators may further be reduced. For example, in the chiral perturbation theory, the coset is   $G/H = \SU (N_\sff) \times \SU (N_\sff)/ \SU (N_\sff)$, so that $R$ is the adjoint representation of $\SU (N_\sff)$,  where $N_\sff = 2$ or $3$ is the number of the flavors for the light quarks. For $N_\sff = 2$, $O_\wzw$ vanishes while $O_3 $ and $O_4$ can be expressed in terms of linear combinations of $O_1$ and $O_2$, so that there are only 2 independent $\ordr (p^4)$ operators. For $N_\sff = 3$, $O_\wzw$ is non-vanishing, though $O_4$ can be expressed as a linear combination of $O_1$, $O_2$ and $O_3$, thus there are 3 independent P-even operators at $\ordr (p^4)$.

The amplitudes for the NLSM also exhibit a derivative expansion, and at tree level one can write
\bea
{\cal M}_n^\nlsm &=& {\cal M}_n^{(2)} + {\cal M}_n^{(4)} + \ordr  \left( \frac{1}{\Lambda^4} \right),\label{eq:dena}
\eea
where $n$ is the multiplicity, with
\bea
 {\cal M}_{n}^{(m)} = \ordr (f^{2-n} \Lambda^{2-m} ),
\eea
which is  controlled by the Lagrangian up to $\ordr (p^m)$. Below we discuss the consequences of the symmetries in NLSM at the amplitude level. In Section \ref{sec:fo} we review the flavor decomposition of the amplitudes, while in Section \ref{sec:st} we review the single soft theorem resulting from the nonlinearly realized shift symmetry.

\subsection{Flavor symmetry and flavor ordering}
\label{sec:fo}

The existence of the linearly realized flavor symmetry of $H$ leads to a convenient separation of flavor and kinematics for the tree level amplitudes. We will see that up to $\ordr (p^4)$, the general NLSM can be expressed in a single or double trace basis. An additional ``pair basis'' is available when we specify $R$ of $H$ to be $\bN$ of $\SO (N)$, which will be discussed later in Section \ref{sec:son}.

\label{sec:rb}
Let us consider the leading order Lagrangian
\bea
\lag^{(2)}= \frac{f^2}{2} \tr \left( d_\mu d^\mu \right) = \frac{f^2}{2} d_\mu^a d^{a\mu} ,\label{eq:lolg}
\eea
and denote the corresponding theory $\nlsm^{(2)}$. Using the relations in Eq. (\ref{eq:crg}) we can rewrite $\lag^{(2)}$ as \cite{Kampf:2013vha}
\bea
\lag^{(2)} &=& \frac{f^2}{8} \tr \left( \partial_\mu U^\dagger \partial^\mu U \right),\label{eq:lagst}
\eea
where
\bea
U = \exp \left(2i \Pi \right) = \exp \left(2i \pi^a \gX^a /f \right).
\eea
The interactions given by Eq. (\ref{eq:lagst}) are even powers of $\pi^a$ contracted with a single trace of generators $\gX^a$.

Therefore, it is convenient to consider flavor-ordered partial amplitudes, as used in the soft bootstrap as the on-shell construction of the NLSM  \cite{Cheung:2015ota,Cheung:2016drk,Elvang:2018dco,Low:2019ynd}. These partial amplitudes are similar to the color-ordered amplitudes of the Yang-Mills (YM) theory \cite{Dixon:1996wi}, where the interactions involve the structure constant $f^{ijk}$, which corresponds to $T^i_{ab} $ in Eq. (\ref{eq:lolg}). From the perspective of just the group $H$, $T^i_{ab}$ is a group generator in some general representation; however, from the perspective of the broken group $G$ and coset $G/H$, $T^i_{ab} = -if^{iab}$ is the structure constant of $G$, i.e. the generator of $G$ in the adjoint representation. Similarly, the gauge bosons in YM theories furnish the adjoint representation as well.

The color-decomposition of the YM theory can thus be directly applied to general $\nlsm^{(2)}$. The flavor structure of the full amplitude can be expanded in the trace basis as
\bea
{\cal M}_n^{(2),a_1 \cdots a_n} (p_1, \cdots, p_n) = \sum_{\alpha \in S_{n-1}} \tr \left( \gX^{a_1} \gX^{a_{\alpha (1)}} \cdots \gX^{a_{\alpha (n-1)}} \right) M_n^{(2)} (1,\alpha),\label{eq:fdtb}
\eea
where $\alpha$ is a permutation of $\{2, 3, \cdots, n\}$ and $M_n^{(2)} (1,\alpha)$ is the single-trace flavor-ordered partial amplitude. The right-hand side (RHS) of Eq. (\ref{eq:fdtb}) is a sum of $(n-1)!$ terms. The lesson we learn from YM theories is that the flavor expansion in Eq. (\ref{eq:fdtb}) is over-complete, and can be further reduced to the Del Duca-Dixon-Maltoni (DDM) basis \cite{DelDuca:1999rs} as a sum of $(n-2)!$ terms:
\bea
&&{\cal M}_n^{(2),a_1 \cdots a_n} (p_1, \cdots, p_n)\non\\
 &=& \sum_{\alpha \in S_{n-2}} (-1)^{n/2-1} f^{a_1 a_{\alpha (1)} i_1 }\left( \prod_{j=1}^{n/2-2} f^{i_{j} a_{ \alpha (2j)} b_j} f^{b_j a_{\alpha (2j+1)} i_{j+1}} \right)   f^{i_{n/2-1} a_{\alpha (n-2)} a_n} M_n^{(2)} (1,\alpha, n)\non\\
&=& \sum_{\alpha \in S_{n-2}}  T^{i_1}_{a_1 a_{\alpha (1)}  }\left( \prod_{j=1}^{n/2-2} T^{i_{j}}_{ a_{ \alpha (2j)} b_j} T^{i_{j+1}}_{b_j a_{\alpha (2j+1)} } \right)  T^{i_{n/2-1}}_{ a_{\alpha (n-2)} a_n} M_n^{(2)} (1,\alpha, n),\label{eq:ddm}
\eea
where $\alpha $ is a permutation of $\{2,3,\cdots, n-1\}$. The ordered amplitudes thus need to satisfy the Kleiss-Kuijf (KK) relations \cite{Kleiss:1988ne}. If we further take flavor-kinematics duality into account, using the BCJ relations \cite{Bern:2008qj} we can reduce the number of independent ordered amplitudes to $(n-3)!$.

\begin{figure}[t]
\centering
\includegraphics[width=0.5\textwidth]{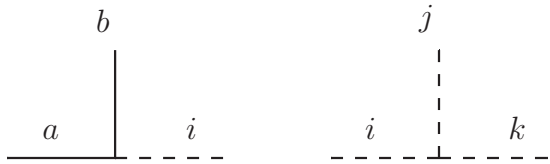}
\caption{\label{fig:tf} The graphic presentation of $T^i_{ab}$ and $f^{ijk}$ as vertices.}
\end{figure}

\begin{figure}[t]
\centering
\includegraphics[width=0.7\textwidth]{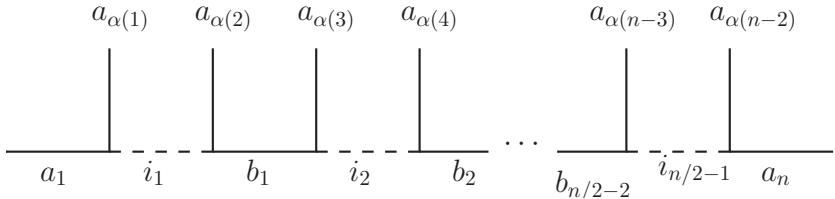}
\caption{\label{fig:ddm} The flavor factors in the DDM basis. The internal lines represent the indices that are contracted and summed over.}
\end{figure}

It is helpful to express the flavor factors diagrammatically, where solid lines represent the broken indices $a$ in the representation $R$, and dashed lines denote the unbroken indices $i$  in the adjoint of $H$. The generator $T^i_{ab}$ and the structure constant $f^{ijk}$ are then vertices given by Fig. (\ref{fig:tf}). Under this notation, the flavor factor of each term in Eq. (\ref{eq:ddm}) is given by a half-ladder graph shown in Fig. \ref{fig:ddm}.

A common practice of calculating the ordered amplitude is working out the ordered Feynman vertices, and then summing up all distinct planar Feynman diagrams \cite{Dixon:1996wi}. It is important to note that in principle, such operation does not work for a general group representation of the NLSM, as it relies on the correct factorization of traces, which is only valid in some cases such as the adjoint of $\SU (N)$ \cite{Kampf:2013vha}. This applies to recursion relations as in the soft bootstrap as well \cite{Low:2019ynd}. However, in practice, we can still always calculate the on-shell amplitudes correctly using these methods, as the flavor ordering is universal, thus the partial amplitudes for the adjoint of $\SU (N)$ are not different from the partial amplitudes of any other group representations\footnote{Again, with the assumption that the closure condition Eq. (\ref{eq:clocon}) is satisfied.}. This is similar to the case of gauge theory: the ordered amplitudes are universal whatever the gauge group is.

The general trace ordering can be extended to $\ordr (p^4)$. At tree level, when we consider the Feynman diagrams, the $\ordr (p^4)$ vertices generated by $\lag^{(4)}$ enter once and only once in each diagram for ${\cal M}_{n}^{(4)}$. Consequently, from Eqs. (\ref{eq:nlsml4}), (\ref{eq:o14}) and (\ref{eq:lagwzwdef}) we see that the contributions of $O_1$ and $O_2$ admit a double-trace ordering, while $O_3$, $O_4$ and $O_\wzw$ still have the same single-trace structure as the $\ordr (p^2)$ piece. In general, one can define the multi-trace flavor decomposition as
\bea
&&\M^{a_1 \cdots a_{n} }_n (p_1, \cdots, p_{n})\non\\
& \equiv&
\sum_{t=1}^{\lfloor n/2 \rfloor } \sum_{ l}  \sum_{\sigma \in S_{n} / S_{n;l} } \left[ \prod_{i=1}^{t} \tr \left( \gX^{a_{\sigma  (l_{i-1} +1)}} \cdots  \gX^{a_{\sigma (l_i)}} \right)  \right]  \non\\
&&\times M_n ( \alpha (1), \cdots, \alpha (l_1) | \alpha(l_1 + 1), \cdots , \alpha (l_2 ) | \cdots | \alpha (l_{t-1} + 1), \cdots \alpha (n) ),\label{eq:mttfo}
\eea
where $l = \{ l_0, \cdots ,l_t \}$ labels partitions of ordered indices $\{1, 2, \cdots , n\}$ into $t$ subsets, so that $l_0 = 0$, $l_t = n$ and $l_{i+1} - l_i \le l_{i+2} - l_{i+1}$, $i=0,1, \cdots, t-2$; $S_{n;l}$ are the permutations of $\{1, 2, \cdots , n\}$ that leave the flavor factor invariant. The partial amplitude
\bea
M_n (1,2, \cdots l_1|l_1 + 1, \cdots , l_2 | \cdots | l_{t-1} + 1, \cdots, n)
\eea
is invariant not only under the cyclic permutations separately for the sets of indices $\{1,2, \cdots, l_1\}$, $\{ l_1+1, \cdots, l_2 \}$ and so on, but also when we exchange the sets $\{ l_i+1, \cdots ,l_{i+1} \}$ and $\{ l_{i+1} + 1, \cdots , l_{i+2} \}$ if they are of the same size, i.e.  $l_{i+1} - l_i = l_{i+2} - l_{i+1}$.

\subsection{Shift symmetry and the single soft theorem}
\label{sec:st}

The NLSM effective Lagrangian is determined by a non-linear shift symmetry. From the UV perspective of spontaneous symmetry breaking, this shift symmetry is the non-linear realization of the broken symmetry associated with the coset $G/H$. However, the shift symmetry can also be  fixed without knowing the UV information of the broken group $G$. This is directly related to the fact that the amplitudes of NLSM satisfy the Adler zero condition: for an on-shell amplitude $\M_{n}^{a_1\cdots a_{n-1} a} (p_1, \cdots, p_{n-1},  q)$\footnote{In this work we assume all momenta are ingoing, so that the momentum conservation here is given by $\sum_{i=1}^{n-1} p_i = -q$.}, if we take the soft limit of $q$, i.e. replace $q$ with $\tau q$ and take the limit of $\tau \to 0$, the amplitude vanishes linearly in $\tau$:
\bea
\M_{n}^{a_1\cdots a_{n-1} a} (p_1, \cdots, p_{n-1},  \tau q) = \ordr (\tau).\label{eq:azc}
\eea
Such a condition can be treated as the defining property of the NLSM, and is the most basic of the soft theorems of NLSM amplitudes. Upon recognizing the Adler zero condition, the non-linear shift symmetry can be derived without the UV information of the broken group $G$. The associated transformation of the shift symmetry for the NGB's is \cite{Low:2014nga,Low:2014oga,Low:2017mlh,Low:2018acv}
\bea
\pi^a \to \pi^a + \left[ F_3 (\mt) \right]^{ab}  \vep^b,
\eea
where
\bea
F_3 (\mt) &=& \sqrt{\mt} \cot \sqrt{\mt},
\eea
and $\vep^a$ are constants that parameterize the shift. Starting at $\ordr (p^4)$, there are both P-even and P-odd parts in the Lagrangian, the latter of which are the WZW terms that capture the effects of the anomalies. Under the shift transformation, the P-even parts of the Lagrangian are invariant, while the WZW terms change by a total derivative.

\begin{figure}[t]
\centering
\includegraphics[width=0.4\textwidth]{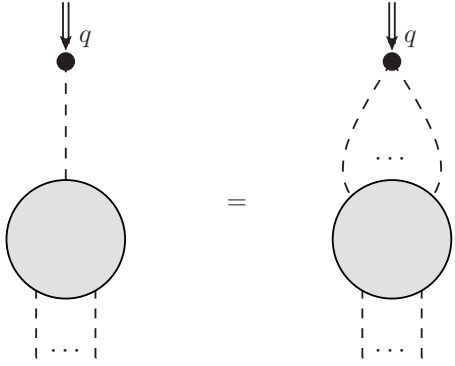}
\caption{A graphic representation of Eq. (\ref{eq:wia}), where the grey blobs denote collections of NLSM Feynman diagrams, while the black dot indicates the insertion of the current $\J$ carrying the momentum $q$, with the LHS given by the one-particle pole in $\J$, and the RHS given by $\tilde{\J}$. \label{fig:cur}}
\end{figure}

\begin{figure}[t]
\centering
\includegraphics[width=0.16\textwidth]{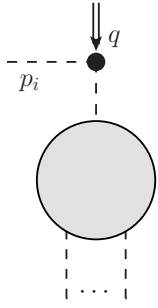}
\caption{The only possible kind of diagrams that develops a pole for soft $q$ in $\R$. As $p_i$ is on-shell, this kind of diagrams gives the propagator $i/(2 p_i \cdot q)$. This apparently requires terms in $\tilde{\J}$ that are quadratic in $\pi$. \label{fig:sinp}}
\end{figure}

We can then calculate the current associated with the shift symmetry: we promote the shift parameter to a local one: $\vep^a \to \vep^a (x)$, and find out that up to total derivatives, the variation of the Lagrangian is given by
\bea
\delta \lag =\vep^a (x) \partial^\mu \mathcal{J}_\mu^a,
\eea
where
\bea
\mathcal{J}_\mu^a = \partial_\mu \pi^a + \ordr \left (\frac{1}{f^2} \right). 
\eea
Classically, the current is conserved because the action should not change:
\bea
\partial^\mu \mathcal{J}_\mu^a = 0.
\eea
As we have taken care of the quantum anomalies with the WZW term in the Lagrangian, the  current remains conserved at the quantum level, leading to the Ward identity for the correlation functions:
\bea
&&i \partial^\mu \< \Omega| \J^a_\mu (x) \prod_{i=1}^n \pi^{a_i} (x_i) |\Omega \>\nonumber \\
&=&  \sum_{r=1}^n \<\Omega| \pi^{a_1} (x_1) \cdots \left[F_3 (\mt)\right]_{a_r a}(x_r) \delta^{(4)}(x-x_r) \cdots \pi^{a_n} (x_n) |\Omega \> .
\eea
Performing the LSZ reduction and taking the on-shell limit, the RHS of the above vanishes, and we arrive at a single soft theorem of the  on-shell amplitude:
\bea
\M_n^{a_1\cdots a_{n} a} (p_1, \cdots, p_n,  q) =  q \cdot {\cal R}^{a_1 \cdots a_{n} a} (p_1, \cdots, p_n;  q),\label{eq:wia}
\eea
where the momentum $q$ is carried by the current. As shown in Fig. \ref{fig:cur}, the left-hand side (LHS) of the above comes from the one particle pole in $\J$, while the rest of the current enter the remainder function $\R^{a_1 \cdots a_{n} a}_\mu$:
\bea
 \R_\mu^{a_1 \cdots a_{n} a} (p_1, \cdots, p_n;  q) = \frac{i}{\sqrt{Z}}   \<0| \int d^4x\, e^{-i q \cdot x} \tilde{\J}_\mu^a | \pi^{a_1} \cdots \pi^{a_n}\>,\label{eq:genrf}
\eea
where $Z$ is the field strength renormalization factor, with $Z=1$ at tree level, which is what we will assume in the rest of the work, and
\bea
\tilde{\J}_\mu^a = - (\mathcal{J}_\mu^a -  \partial_\mu \pi^a) .
\eea
As there are no cubic interactions in the Lagrangian, which is the case when the closure condition of Eq. (\ref{eq:clocon}) is satisfied, we have $\tilde{\J}  = \ordr (\pi^3)$. Therefore, the only divergence possible when we take $q \to 0$, given by the ``pole diagrams'' shown in Fig. (\ref{fig:sinp}), cannot exist, so that the remainder function ${\cal R}$ is finite. This leads to the Adler zero condition given by Eq. (\ref{eq:azc}).


Notice that in the flavor decomposition, the flavor factors are linearly independent, thus each of the ordered amplitudes also need to satisfy the Adler zero condition, similar to the YM theory where the ordered amplitudes are also gauge invariant.

\begin{figure}[t]
\centering
\includegraphics[width=0.5\textwidth]{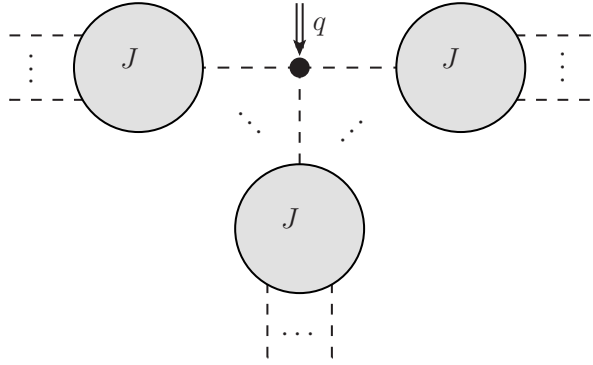}
\caption{The remainder function at tree level. The black dot now labels the vertex $\V$ created by the current $\J$, while the grey blobs are now semi-on-shell amplitudes $J$ where the only off-shell external leg is connected to $\V$. \label{fig:curt}}
\end{figure}

Now we can go beyond the Adler zero and compute the $\ordr (\tau)$ piece in the soft limit, which gives us the subleading single soft theorem. This amounts to calculating the vertices in $\R$ given by the current $\J$. At  $\ordr (p^2)$, this leads to the single-trace version of the subleading single soft theorem, which is already known. The current from $\lag^{(2)}$ is:
\bea
 \mathcal{J}_\mu^{(2),a} = \left[ \frac{\sin \left(2 \sqrt{\mt} \right) }{2 \sqrt{\mt}} \right]_{ab} \partial_\mu \pi^b = \partial_\mu \pi^a - \tilde{\J}_\mu^{(2),a},\label{eq:cp2}
\eea
with
\bea
\tilde{\J}_\mu^{(2),a} = \sum_{k=1}^{\infty}  \frac{(-4)^k}{(2k+1)!} \left( \mt^k \right)_{ab} \partial_\mu \pi^b.
\eea
As we see from Eq. (\ref{eq:genrf}), each term in $\tilde{\J}_\mu^{(2),a}$ inserts the following single-trace vertex into $\R_\mu$, after we strip the flavor factors:
\bea
\V^{(2)}_{2k+1} (\mathbb{I}_{2k+1}) = \frac{-(-4)^k}{(2k+1)!f^{2k}}\sum_{j=0}^{2k} \left(\begin{array}{c}
2k\\
j
\end{array} \right) (-1)^{j} q \cdot p_{j+1},\label{eq:op2vi}
\eea
where $\mathbb{I}_{n} \equiv \{1,2, \cdots, n\}$ is the identity permutation for $n$ labels. As shown in Fig. \ref{fig:curt}, the legs of the above vertex are connected to semi-on-shell amplitudes, i.e. the Berends-Giele currents \cite{Berends:1987me}
\bea
J^{a_1 \cdots a_n, a} (p_1, \cdots ,p_n) \equiv \langle 0| \pi^a (0) | \pi^{a_1}(p_1) \cdots \pi^{a_n}(p_n)\rangle.\label{eq:bgcd}
\eea
These objects have one uncut off-shell leg of momentum $-\sum_i p_i$, which is connected to $\V$ in the single soft theorem, while all the other legs are on-shell. Just like the on-shell amplitudes, at $\ordr (p^2)$ the off-shell amplitudes can be ordered in a single trace basis, so that the  subleading single soft theorem for $\nlsm^{(2)}$ is
\bea
M_{n+1}^{(2)} (\mathbb{I}_{n+1}) &=& \tau \sum_{k=1}^{\lfloor n/2 \rfloor}  \frac{-(-4)^k}{(2k+1)!f^{2k}} \sum_{l}\sum_{j=1}^{2k-1} \left[ \left(\begin{array}{c}
2k\\
j
\end{array} \right) (-1)^{j} -1 \right] p_{n+1} \cdot q_{l_{j+1}}\non\\
&&\times \prod_{m=1}^{2k+1} J^{(2)} (l_{m-1}+1, \cdots , l_m) ,\label{eqnlsmosex}
\eea
where  we have taken $p_{n+1}$ to be soft. In the above $l$ is a way to split $\{1, 2, \cdots,n\}$ into $2k+1$ disjoint, ordered subsets $\{l_{m-1}+1, \cdots , l_m \}$, with $l_0=0$, $l_{2k+1} = n$ and $q_{l_{j+1}} = \sum_{i=l_j+1}^{l_{j+1}} p_i$.

It was first discovered in Ref. \cite{Cachazo:2016njl} using the CHY formalism  that an extended theory with the NGB's interacting with the bi-adjoint scalars  emerges in the soft theorem. Originally, NGB's generated by the coset of $\SU (N) \times \SU (N)/ \SU (N)$ were considered, so that $R$ is the adjoint representation of the unbroken $\SU (N)$ group, and the coset space is isomorphic to the unbroken $\SU (N)$. Then the generator $T^i_{ab}$ in the Lagrangian can be exchanged with the structure constant $-if^{iab}$ of $\SU(N)$: the difference between the broken and unbroken indices becomes non-existent. The bi-adjoint scalars $\phi^{a \ta}$, as the additional field content in the extended theory,  transform under both the original flavor group $\SU (N)$ and another flavor group $\SU (\tilde{N})$, and has the following cubic self-interaction:
\bea
-\frac{\lambda}{6} \phi^{a \ta} \phi^{b \tb} \phi^{c \tc} f^{abc} f^{\ta \tb \tc},\label{eq:phi3v}
\eea
characterized by the coupling constant $\lambda$. For the $(n+1)$-point (pt) amplitude, taking $p_{n+1}$ to be soft, we have
\bea
M_{n+1}^{(2)} (\mathbb{I}_{n+1}) = \frac{\tau}{\lambda f^2} \sum_{i=2}^{n-1}  s_{n+1,i} \ M_n^{\nlsm + \phi^3}(\mathbb{I}_{n}||1,n,i) + \ordr (\tau^2),\label{eq:sssstr}
\eea
where $s_{i,j} \equiv (p_i + p_j)^2$,  and $\nlsm + \phi^3$ denotes the extended theory\footnote{It should be understood that whenever it appears in the name of an mixed theory like $\nlsm + \phi^3$, ``NLSM'' means $\nlsm^{(2)}$.}. In the RHS of the above, ``$||$'' separates flavor structures of different flavor groups, which should not be confused with ``$|$'' in the multi-trace amplitudes in Eq. (\ref{eq:mttfo}) which separates traces of the generators of the same group. In $M_n^{\nlsm + \phi^3}(\mathbb{I}_{n}||1,n,i)$, the external states $1$, $n$ and $i$ are apparently bi-adjoint as they have two separate orderings, the left for $\SU(N)$ and the right for $\SU (\tilde{N})$; other external legs, which only have left orderings, then belong to the NGB's. In the DDM basis, the flavor factor for $M_n^{\nlsm + \phi^3}(\mathbb{I}_{n}||1,n,i)$ in Eq. (\ref{eq:sssstr}) is
\bea
(-1)^{(n-1)/2} f^{a_1 a_2 b_1} \left( \prod_{j=1}^{n-4} f^{b_j a_{j+2} b_{j+1}} \right) f^{b_{n-3} a_{n-1} a_n }  \tilde{f}^{\ta_1 \ta_n \ta_i},
\eea
where $f^{abc}$ and $\tilde{f}^{\ta \tb \tc}$ are structure constants of $\SU (N)$ and $\SU (\tilde{N})$, respectively.

Comparing with Eq. (\ref{eqnlsmosex}), one can identify all the new Feynman vertices in the extended theory that are relevant in Eq. (\ref{eq:sssstr}), in addition to ones already in $\nlsm^{(2)}$ \cite{Low:2017mlh,Low:2018acv,Yin:2018hht}:
\begin{itemize}
\item \textbf{Type I} vertices with two $\phi$ and an even number of $\pi$, which has to take exactly the same value as the vertices in $\nlsm^{(2)}$ with the same left ordering, i.e.
\bea
V_{2k}^{\nlsm + \phi^3} (\mathbb{I}_{2k} || i,j)  = V_{2k}^{(2)} (\mathbb{I}_{2k} ).
\eea
\item \textbf{Type II} vertices with three $\phi$, two of whose left orderings are adjacent, and an even number of $\pi$. These vertices are generated by the current $\J$, and to match Eq. (\ref{eq:sssstr}) we need
\bea
\V^{(2)}_{2k+1} (\mathbb{I}_{2k+1})  =\frac{1}{\lambda f^2} \sum_{i=2}^{2k} 2 q \cdot p_i  \  V_{2k+1}^{\nlsm + \phi^3} (\mathbb{I}_{2k+1} || 1,2k+1,j).\label{eq:type2r}
\eea
We know that $\V^{(2)} $ is linear in $q$, but for the above to hold the coefficients of $q \cdot p_1$ and $q\cdot p_{2k+1}$ need to vanish at the same time. Applying total momentum conservation to Eq. (\ref{eq:op2vi}), as well as the fact that whenever we use $\V^{(2)} $ we have the on-shell condition of $q^2 = 0$, we see that
\bea
V_{2k+1}^{\nlsm + \phi^3} (\mathbb{I}_{2k+1} || 1,2k+1,j)  =\frac{\lambda}{2} \frac{(-4)^k  }{(2k+1)!f^{2k-2}} \left[ 1- \left(\begin{array}{c}
2k\\
j-1
\end{array} \right) (-1)^{j} \right]\ .\label{eq:ev}
\eea
A special case is the 3-pt vertex
\bea
V_{2k+1}^{\nlsm + \phi^3} (1,2,3 || 1,3,2) = -\lambda,\label{eq:etp3}
\eea
which matches the $\phi^3$ interaction given by Eq. (\ref{eq:phi3v}).
\end{itemize}

\begin{figure}[t]
\centering
\includegraphics[width=0.8\textwidth]{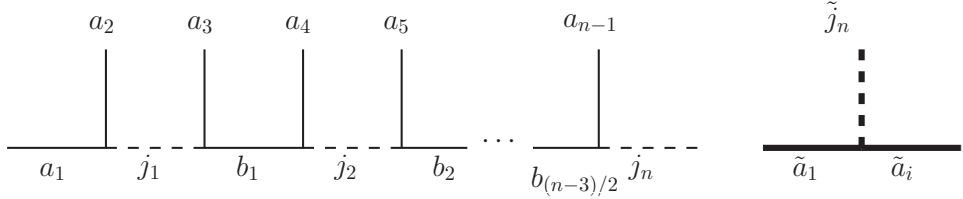}
\caption{The general flavor factor for $M_n^{\nlsm + \phi + \psi}(\mathbb{I}_{n}||1,n,i)$ in the DDM basis. Notice that these are not Feynman diagrams, but only represent the flavor structures of the associated ordered amplitudes. The thin lines are for the group $H$ of the left ordering, while the thick lines are for the group $\tilde{H}$ of the right ordering.\label{fig:etgd}}
\end{figure}

As discussed in Section \ref{sec:rb}, the flavor ordering works equally well for $\nlsm^{(2)}$ of a general coset $G/H$, implying that instead of restricting ourselves to $\SU (N) \times \SU (N)/ \SU (N)$, we can have a more general interpretation of the extended theory. Meanwhile, the isomorphism between the coset $\SU (N) \times \SU (N)/ \SU (N)$ and the group $\SU (N)$ is lost in a general coset, thus we need to discern the broken and unbroken indices. The flavor factor for  $M_n (\mathbb{I}_{n}||1,n,i)$ will need to be
\bea
T_{a_1 a_2}^{j_1} \left( \prod_{k=1}^{(n-5)/2} T^{j_k}_{a_{2k+1 } b_k } T^{j_{k+1}}_{b_k a_{2k+2}} \right) T^{j_{(n-3)/2}}_{a_{n-2} b_{(n-3)/2}} T^{j_n}_{b_{(n-3)/2} a_{n-1}} \tT^{\tdj_n}_{\ta_i \ta_1},\label{eq:gf}
\eea
where $T^{i}_{ab}$ and $\tT^{\tdi}_{\ta \tb}$ are generators of $H$ and $\tilde{H}$ in some representation $R$ and $\tilde{R}$, respectively. The flavor factor in Eq. (\ref{eq:gf}) can be presented graphically as in Fig. \ref{fig:etgd}. Notice that external states $1,2, \cdots n-1$ carry indices $a_k$ that furnish some representation $R$, while particle $n$ has indices in the adjoint. In other words, we have two different kinds of bi-index scalars: $\psi^{a \ta}$ in $R$ and $\tilde{R}$, and the bi-adjoint scalars $\phi^{j \tdj}$. We will denote such an extended theory as $\nlsm + \phi + \psi$. An example useful in the following will be $H = \SO (N)$ and $R$ is the fundamental representation, so that the amplitudes can also be expressed in the pair basis.

\section{The single soft theorem for $\bN$ of $\SO(N)$}
\label{sec:son}

In this section we consider NGB's furnishing $\bN$ of $\SO (N)$, focusing on  the leading $\ordr (p^2)$ in the EFT expansion. There are $N$ flavors in such a theory, and the minimal coset that realizes this is $\SO (N+1)/ \SO (N)$. The generators $T^i_{ab}$ satisfy the following completeness relation:
\bea
(T^i)_{ab } (T^i)_{cd} = \frac{1}{2}  (\delta^{ad} \delta^{bc} - \delta^{ac} \delta^{bd})\ ,\label{eq:comrel}
\eea
using which we can reduce the Lagrangian in Eq. (\ref{eq:lolg}) to \cite{Low:2019ynd}
\bea
\lag^{(2)} &=& \frac{1}{2} F_1^2 (r^2) \<\partial_\mu \pi| \partial^\mu \pi \> - \frac{1}{4f^2 r^2} \left[ F_1^2 (r^2) - 1 \right] \< \pi | \partial_\mu \pi \>^2,\label{eq:nlagso}
\eea
where we have adopted the bra-ket notation $(| \pi \>)_a \equiv \pi_a$, and $r \equiv \sqrt{ \< \pi | \pi \>/(2 f^2)}$. We see that in the vertices given by the above,  $\pi^a$ are pair-wise contracted, which implies that the flavor factor for the amplitudes are products of Kronecker deltas: we have
\bea
{\cal M}_n^{(2),a_1 \cdots a_n} &=& \sum_{\alpd \in P_n} \left( \prod_{j=1}^{n/2} \delta^{a_{\alpd (2j-1)} a_{\alpd (2j)}} \right) \non\\
&&\times M_n^{(2)} (\alpd (1),\alpd (2)|\alpd (3), \alpd(4)| \cdots | \alpd (2n-1), \alpd (2n)),\label{eq:fdpair}
\eea
where $P_n$ is all the distinct partitions of non-ordered set $\{1, 2, \cdots , n\}$ into $n/2$ subsets of two elements: $\{\alpd (1),\alpd (2)\}, \{\alpd (3), \alpd(4)\}, \cdots \{ \alpd (2n-1), \alpd (2n)\}$. The partial amplitude $M (\alpd) $ in Eq. (\ref{eq:fdpair}) contains $n/2$ non-ordered pairs of external particle indices. The RHS of Eq. (\ref{eq:fdpair}) is a sum of $(n-1)!!$ terms, and as the flavor factors in front of each term are completely independent of each other, they form a basis which we call the pair basis. As $\delta^{ab} = \tr \left( \gX^a \gX^b \right)$, the pair basis can also be understood as a multi-trace basis. The amplitude in the basis is invariant under exchanging the positions of different traces, as well as exchanging two labels in each trace.

When the multiplicity $n$ is large, we have $(n-1)!! \ll (n-3)!$, thus the pair basis is much smaller than the minimal BCJ basis of a general NLSM. A comparison of the size of the different bases is given in Table \ref{tab:compb}. Our ability to reduce to the $(n-1)!!$ basis depends on the special properties of the $\SO (N)$ fundamental representation, i.e. the completeness relation in Eq. (\ref{eq:comrel}).

\begin{table}[t]
\begin{center}
\begin{tabular}{c |c|c|c|c}
\hline           
Multiplicity & Single trace & DDM & BCJ & Pair\\
\hline
$n$ & $(n-1)!$ & $(n-2)!$ & $(n-3)!$ & $(n-1)!!$\\
\hline
$4$ & $6$ & $2$ & $1$ & $3$\\
\hline
$6$ & $120$ & $24$ & $6$ & $15$\\
\hline
$8$ & $5040$ & $720$ & $120$ & $105$\\
\hline
$10$ & $362880$ & $40320$ & $5040$ & $945$\\
\hline
\end{tabular}
\end{center}
\caption{\label{tab:compb} The size of different amplitude bases for NLSM.}
\end{table}

In the following, we explore the amplitude relations for the pair basis in Section \ref{sec:pbddmr}, which will be useful when we derive the subleading single soft theorem for the pair basis in Section \ref{sec:ssp}.

\subsection{Amplitude relations for the pair basis}

\label{sec:pbddmr}
\begin{figure}[t]
\bea
\begin{minipage}[c]{0.25\textwidth}
\begin{center}
\includegraphics[width=\textwidth]{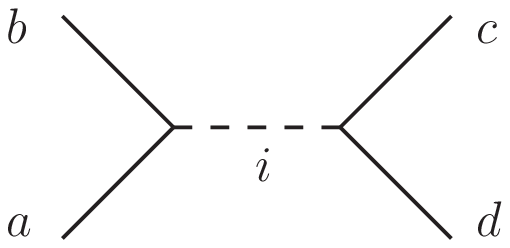} 
\end{center}
\end{minipage} = \frac{1}{2} \left( \qquad \begin{minipage}[c]{0.25\textwidth}
\begin{center}
\includegraphics[width=\textwidth]{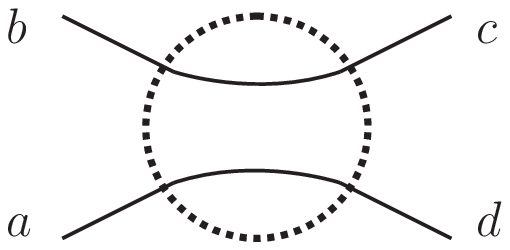} 
\end{center}
\end{minipage} - \qquad \begin{minipage}[c]{0.25\textwidth}
\begin{center}
\includegraphics[width=\textwidth]{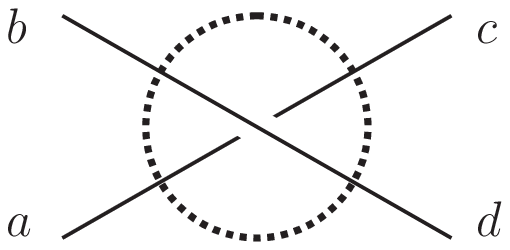} 
\end{center}
\end{minipage} \right) \non
\eea
\caption{\label{fig:comrel} The graphic representation of the completeness relation in Eq. (\ref{eq:comrel}). On the RHS, the solid lines do not intersect with each other, and each of them connects two fundamental indices of $\SO (N)$ and represents a Kronecker delta for the two indices. For a contraction of adjoint index $i$ on the LHS, there are two ways to contract the fundamental indices on the RHS, represented by a ``$\times$'' or an ``$=$'' in the dotted circle. The two choices have a sign difference.}
\end{figure}

\begin{figure}[t]
\centering
\includegraphics[width=0.7\textwidth]{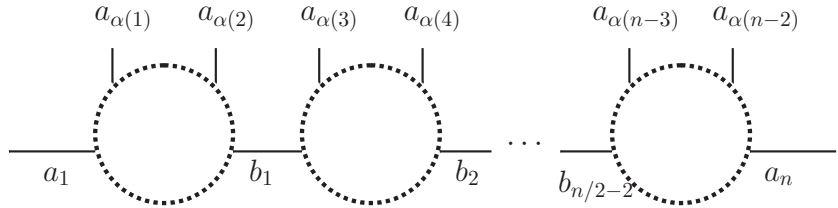}
\caption{\label{fig:pairld} The half-ladder after applying the completeness relation. Each dotted circle contains either a ``$\times$'' or an ``$=$''.}
\end{figure}

It turns out that the relation between partial amplitudes in the pair basis and the single-trace amplitudes is quite straightforward. Let us first look at the DDM basis given by Eq. (\ref{eq:ddm}),
the flavor factors of which are given by Fig. \ref{fig:ddm}. We would like to convert it to the pair basis, using the  completeness relation given by Eq. (\ref{eq:comrel}), which can be represented graphically as in Fig. \ref{fig:comrel}. Applying the relation to Fig. \ref{fig:ddm}, we arrive at flavor factors as in Fig. \ref{fig:pairld}.

Now let us consider the multi-trace partial amplitudes where indices $1$ and $n$ form a pair, e.g. $M_n^{(2)} (1,n|2,3|4,5| \cdots | n-2,n-1)$. By definition, it is the coefficient of the flavor factor
\bea
\delta^{a_1 a_n} \prod_{j=1}^{n/2-1} \delta^{a_{2j} a_{2j+1}}\label{eq:phlc}
\eea
in the full amplitude. As shown in Fig. \ref{fig:pairld}, the indices $a_1$ and $a_n$ are always on the two ends of the half-ladder. As explained in Figs. \ref{fig:comrel} and \ref{fig:pairld}, each dotted circle can be either a ``$\times $'' or an ``$=$'', but for $a_1$ and $a_n$ to be contracted to get $\delta^{a_1 a_n}$, all of the dotted circles must contain an ``$=$''. Therefore, the upper two indices in each of the dotted circles must be contracted as well. Then the coefficient of the flavor factor in Eq. (\ref{eq:phlc}) is
\bea
M_n^{(2)} (1,n|2,3|4,5| \cdots | n-2,n-1)=\frac{1}{2^{n/2-1}}   \sum_{\alpha \in \Ar_{n}} M_n^{(2)} (1, \alpha, n),\label{eq:pstr}
\eea
where $\Ar_{n}$ are permutations of $\{2, 3, \cdots, n-1\}$ so that for all pairs $\{2, 3\}$, $\{4,5\}$, $\cdots$, $\{n-2, n-1\}$, the two indices in the pair are adjacent to each other. The RHS of the above is a sum of $(n-2)!!$ terms. Other partial amplitudes containing the pair $\{1,n\}$ can be generated in exactly the same way, while partial amplitudes containing the pair $\{1, j\}$ where $j \ne n$ can be generated by starting with a DDM basis where $a_1$ and $a_j$ are at the two ends of the half-ladder. Apparently the formula in Eq. (\ref{eq:pstr}) has the correct relabeling symmetry. From the cyclic and reflection symmetries of $M_n (1, \alpha, n)$, we see that the RHS of Eq. (\ref{eq:pstr}) also has the correct permutation symmetry. The same relation can be derived for ordered vertices in exactly the same way\footnote{The practice of summing over all distinct planar Feynman diagrams still works in the pair basis, as the Kronecker deltas certainly satisfy the correct factorization property \cite{Low:2019ynd}.}. An example of Eq. (\ref{eq:pstr}) at 6-pt is
\bea
M_n^{(2)} (16|23|45) &=& \frac{1}{4} \left[ M_n^{(2)} (123456) + M_n^{(2)} (123546) +M_n^{(2)} (132456) + M_n^{(2)} (132546) \right. \non\\
&&\left.+ M_n^{(2)} (145236) + M_n^{(2)} (145326) + M_n^{(2)} (154236) + M_n^{(2)} (154326) \right].
\eea

In hindsight, the form of the RHS of Eq. (\ref{eq:pstr}) is very natural:  it has the correct pole structure, it satisfies the Adler zero condition, and it also has the correct mass dimension and permutation symmetry. The only thing non-trivial is that the RHS of Eq. (\ref{eq:pstr}) must also factorize correctly.

One should also recognize that Eq. (\ref{eq:pstr}) is not the unique way to write the multi-trace partial amplitudes in terms of single-trace ones: other representations can be easily generated by using the KK relations among the single-trace amplitudes.

An immediate consequence of Eq. (\ref{eq:pstr}) is that we can easily write down the CHY formula for the partial amplitudes in the pair basis. In the CHY representation, the tree-level amplitude for a scalar theory is in general written in the following form:
\bea
\M_n   = \oint d\mu_n \; \mathcal{I}_L (\{p, \sigma \}) \; \mathcal{I}_R (\{p, \sigma\}),\label{eq:chyg}
\eea
where $\{p \}$ are the on-shell external momenta, while $\{ \sigma \}$ are dimensionless variables satisfying the scattering equation
\bea
E_j \equiv \sum_{i \ne j} \frac{p_i \cdot p_j}{ \sigma_{ij} }=0,\label{eq:sceq}
\eea
with $\sigma_{ij} \equiv \sigma_i - \sigma_j$. This is enforced by the measure $d \mu_n$ of the integral:
\bea
d \mu_n &\equiv& ( \sigma_{ij}\sigma_{jk}\sigma_{ki}) (\sigma_{pq}\sigma_{qr}\sigma_{rp}) \prod_{a \neq i,j,k} E_a^{-1} \prod_{b \neq p,q,r} d\sigma_b.
\eea
Choosing $\{ i ,j ,k\}$ and $\{ p, q, r\}$ is called ``fixing the gauge'', and the measure $d \mu_n$ is actually gauge invariant, i.e. independent of the choice of $\{ i ,j ,k\}$ and $\{ p, q, r\}$. The integrands $\mathcal{I}_L$ and $\mathcal{I}_R$ are different among different theories.

For the general $\nlsm^{(2)}$, the single-trace partial amplitudes are given by, up to coupling constants, \cite{Cachazo:2014xea}
\bea
M_n^{(2)} (\alpha) =  \oint d\mu_n ~(\pf' \A_n)^2 ~\mathcal{C}_n  (\alpha),
\eea
where $\mathcal{C}_n (\alpha)$ is the Parke-Taylor factor given by
\bea
\mathcal{C}_n (\alpha) = \frac{1}{\sigma_{\alpha (1) \alpha (2)} \cdots \sigma_{\alpha (n-1) \alpha (n)} \sigma_{\alpha (n) \alpha (1)}},\label{eq:chynst}
\eea
and the anti-symmetric matrix $\A_n$ is given by
\bea
[\A_n]_{ab} =\left\{\begin{array}{ll} 	\dfrac{2 p_a \cdot p_b}{\sigma_{ab}}, & a \neq b, \\
	0, & a = b.\end{array} \right.
	\eea
The reduced Pfaffian $\pf'$ is defined as $\pf' \A_n = \frac{(-)^{a+b}}{\sigma_{ab}} \pf \A_n^{[a,b]}$, where $\A_n^{[a,b]}$ is the matrix $\A_n$ with rows and columns of labels $a$ and $b$ removed. It turns out such a definition does not depend on the choices of $\{a , b\}$.

An important observation is that in Eq. (\ref{eq:chynst}), both the measure $d\mu_n$ and the reduced Pfaffian $\pf' \A_n$ are independent of the ordering $\alpha$: the ordering information is only contained in the Parke-Taylor factor $\mathcal{C}_n (\alpha)$. Then using Eq. (\ref{eq:pstr}) we can easily arrive at the CHY formula for the partial amplitude in the pair basis:
\bea
M_n^{(2)} (1,n|2,3|4,5| \cdots | n-2,n-1) &=& \frac{1}{2^{n/2-1}}  \sum_{\alpha \in \Ar_{n}} \oint d\mu_n ~(\pf' \A_n)^2 ~\mathcal{C}_n (1, \alpha, n)\non\\
&=&  \oint d\mu_n ~(\pf' \A_n)^2 ~\frac{1}{2^{n/2-1}}  \sum_{\alpha \in \Ar_{n}}\mathcal{C}_n (1, \alpha, n).\label{eq:chypb}
\eea
In other words, we have
\bea
\mathcal{I}_L = (\pf' \A_n)^2,
\eea
which remains the same as the single-trace amplitudes, while 
\bea
\mathcal{I}_R = \frac{1}{2^{n/2-1}} \sum_{\alpha \in \Ar_{n}}\mathcal{C}_n (1, \alpha, n).\label{eq:pbir}
\eea

Another relation between the partial amplitudes in the pair basis can be easily proved using Eq. (\ref{eq:pstr}). Firstly, we know the $\U (1)$-decoupling relation between the single-trace partial amplitudes, which is the simplest kind of the KK relations:
\bea
M_n^{(2)} (1,2,3,\cdots,n-1, n) + M_n^{(2)}  (1,3,4,\cdots, n,2) + \cdots + M_n^{(2)} (1,n,2,\cdots,n-2,n-1) \non\\
= 0. \qquad \label{eq:u1dc}
\eea
Then
\bea
\sum_{\alpd \in P_n}  M_n^{(2)} (\alpd (1),\alpd (2)|\alpd (3), \alpd(4)| \cdots | \alpd (2n-1), \alpd (2n)) = 0,\label{eq:pasf}
\eea
as the LHS of the above can be expressed as $(n-2)!$ sums like the LHS of Eq. (\ref{eq:u1dc}). The relation given by Eq. (\ref{eq:pasf}) can also be easily understood from a physical perspective: we see from the definition of the pair basis in Eq. (\ref{eq:fdpair}), that
\bea
\M_n^{ (2),a a  \cdots a} &=& \sum_{\alpd \in P_n}  M_n^{ (2)} (\alpd (1),\alpd (2)|\alpd (3), \alpd(4)| \cdots | \alpd (2n-1), \alpd (2n)).
\eea
Namely, the LHS of Eq. (\ref{eq:pasf}) is actually the two-derivative full amplitude when all external particles are of a single flavor. The shift symmetry in  NLSM forbids any 2-derivative interactions for a single kind of scalar, and the corresponding leading order amplitude must vanish \cite{Low:2014nga}. Therefore, Eq. (\ref{eq:pasf}) just states the fact that at the two-derivative level, the NLSM amplitude between scalars of a single flavor vanishes.

\subsection{The single soft theorem in the pair basis}

\label{sec:ssp}

Now we are ready to discuss the subleading single soft theorem for the pair-basis amplitudes. In Section \ref{sec:st}, the subleading single soft theorem at $\ordr (p^2)$ for a general symmetric coset is presented in the single-trace basis. Then we can calculate the subleading single soft theorem in the pair basis, by using the relation between the pair basis and single-trace basis amplitudes. Applying Eq. (\ref{eq:pstr}) to Eq. (\ref{eq:sssstr}), we have
\bea
&&M_{n+1}^{(2)} (n+1,1|2,3|4,5|\cdots | n-1,n)\non\\
 &=& \frac{\tau}{\lambda f^2} \frac{1}{2^{(n-1)/2 }} \sum_{k=2}^{n}  s_{n+1,k} \ \sum_{\substack{j=2\\ j\ne k}}^n   \sum_{\alpha \in \Ar_{n-3} } M_n^{\nlsm + \phi + \psi}(1,\alpha, j_p, j||1,j,k) + \ordr (\tau^2),\label{eq:pbsti}
\eea
where $j_p \equiv j+ (-1)^j$, $\Ar_{n-3}$ are permutations of $\{2,3, \cdots, n\} \setminus \{ j, j_p \}$ so that for all the pairs $\{ m, m_p \}$ in the set, the two indices in the pair are adjacent to each other. Note that although $j \ne k$, the situation when $j_p = k$, i.e. $j$ and $k$ form a pair in the symmetrization of the LHS in the above, can still happen.

Next, we need to express the amplitudes of the extended theory in the pair basis as well. Using the procedure similar to Section \ref{sec:pbddmr}, we can derive a relation between the single-trace and the pair basis in the extended theory, the details and examples of which are shown in Appendix \ref{app:rpse}. We have
\bea
&&M_{n}^{\nlsm + \phi + \psi} (1,2,3|4,5| \cdots|n-1,n || 1^\psi, 2^\phi, i^\psi)  \non\\
&=& -\frac{1}{2^{(n-3)/2}} \sum_{\alpha \in \Ar_{n-3} } M_n^{\nlsm + \phi + \psi}(1,\alpha,3, 2 ||1^\psi, 2^\phi, i^\psi),\label{eq:ptret}
\eea
where we have identified the different bi-index scalars $\phi$ and $\psi$ in the right ordering. Plugging the above into Eq. (\ref{eq:pbsti}), we arrive at \cite{Low:2019wuv}
\bea
&&M_{n+1}^{(2)} (n+1,1|2,3|4,5|\cdots | n-1,n)\non\\
&=&  -\frac{\tau}{2\lambda f^2} \sum_{k=2}^n s_{k,n+1} \sum_{\substack{j=2\\j \ne k}}^{n} M_n^{\nlsm + \phi + \psi} (1,j,j_p|\alpd/j|| 1^\psi,j^\phi, k^\psi) + \ordr (\tau^2),\label{eq:slsspb}
\eea 
where $\alpd /j$ is the partition $\{2,3 |4,5| \cdots | n-1,n\}$ with the pair $\{j, j_p\}$ removed. Examples of Eq. (\ref{eq:slsspb}) are given in Appendix \ref{app:nlv}.

Similarly, we can also work out the vertices in the pair basis of $\nlsm + \phi + \psi$: the Type I vertices are given by
\bea
&&V_n^{\nlsm + \phi + \psi} (1,n|2,3|4,5| \cdots | n-2,n-1||i, j)\non\\
&=&\frac{1}{2^{n/2-1}}   \sum_{\alpha \in \Ar_{n}} V_n^{\nlsm + \phi + \psi} (1, \alpha, n||i,j) = V_n^{(2)} (1,n|2,3|4,5| \cdots | n-2,n-1),\label{eq:soeept}
\eea
while the Type II vertices are
\bea
&&V_{n}^{\nlsm + \phi + \psi} (1,2,3|4,5| \cdots|n-1,n || 1^\psi, 2^\phi, j^\psi) \non\\
&=& -\frac{1}{2^{(n-3)/2}} \sum_{\alpha \in \Ar_{n-3} } V_n^{\nlsm + \phi + \psi}(1,\alpha,3, 2||1, 2, j).
\eea
Plugging in the vertices in the single-trace basis given by Eq. (\ref{eq:ev}), we arrive at
\bea
V_{n}^{\nlsm + \phi + \psi} (1,2,3|4,5| \cdots|n-1,n || 1^\psi, 2^\phi, 4^\psi) = 0,\label{eq:vv}
\eea
while
\bea
V_{2n+1}^{\nlsm + \phi + \psi} (1,2,3|4,5| \cdots|2n,2n+1 || 1^\psi, 2^\phi, 3^\psi) =   \frac{\lambda}{2} \frac{-(-4)^{n} (n-1)!  }{ (2n)! f^{2n-2} },\label{eq:soevop}
\eea
with the 3-pt vertex given by
\bea
V_{2n+1}^{\nlsm + \phi + \psi} (1,2,3||1^\psi, 2^\phi, 3^\psi) = \lambda.
\eea

Note that although $V_{n}^{\nlsm + \phi + \psi} (1,2,3|4,5| \cdots  || 1^\psi, 2^\phi, 4^\psi) = 0$, the amplitude with the same ordering does not vanish: $M_{n}^{\nlsm + \phi + \psi} (1,2,3|4,5| \cdots  || 1^\psi, 2^\phi, 4^\psi) \ne 0$. This can be easily checked for the 5-pt amplitude $M_{5}^{\nlsm + \phi + \psi} (1,2,3|4,5|| 1^\psi, 2^\phi, 4^\psi)$, and it receives non-vanishing contributions from non-contact diagrams, as shown in Fig. (\ref{fig:soe5pt}). We give examples of $M_{n}^{\nlsm + \phi + \psi}$ in the pair basis in Appendix \ref{app:eet}.

\begin{figure}[t]
\centering
\includegraphics[width=0.8\textwidth]{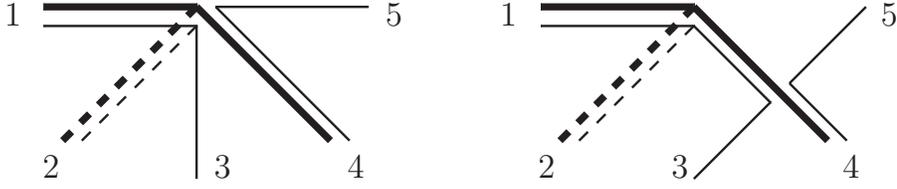}
\caption{\label{fig:soe5pt} Two types of Feynman diagrams that can contribute to $M_{5}^{\nlsm + \phi + \psi} (1,2,3|4,5|| 1^\psi, 2^\phi, 4^\psi)$. The solid lines represent fundamental indices, while the dashed lines represent adjoint indices. The thin lines are for the $\SO (N)$ flavor group of the NGB's, while the thick lines are for the other flavor group $\tilde{H}$ carried by $\phi$ and $\psi$. Although the contact term on the left vanishes, the diagram on the right still contribute, so that the amplitude is non-zero. The 4-pt vertex in the diagram on the right is given by Eq. (\ref{eq:soeept}).}
\end{figure}

We can work out the operators in the Lagrangian that give the vertices in Eq. (\ref{eq:soevop}):
\bea 
 \lambda T^i_{ab} T^{\tdi}_{\ta \tb} \phi^{i \tdi} \psi^{a \ta} \psi^{b \tb} \sum_{n=1}^\infty \frac{(-4)^{n-1} }{(2n)! } r^{2(n-1)} = \frac{ \lambda}{2} T^i_{ab} T^{\tdi}_{\ta \tb} \phi^{i \tdi} \psi^{a \ta} \psi^{b \tb}   F_1^2 (r^2) ,
\eea
where the cubic operator in the above is $(\lambda/2) T^i_{ab} T^{\tdi}_{\ta \tb} \phi^{i \tdi} \psi^{a \ta} \psi^{b \tb}$.

The results of Eqs. (\ref{eq:slsspb}) and (\ref{eq:soevop}) can also be confirmed by a direct calculation from the Ward identity. As we are still at $\ordr (p^2)$, the current is still given by Eq. (\ref{eq:cp2}), though it can be simplified using the completeness relations of the $\SO (N)$ fundamental generators given by Eq. (\ref{eq:comrel}):
\bea
\mathcal{J}^a_\mu &=&\partial_\mu \pi^a + \sum_{k=1}^\infty \frac{(-4)^k}{(2k+1)! } \left( r^{2k} \partial_\mu \pi^a - r^{2k-2} \frac{\<\pi | \partial_\mu \pi \>}{2f^2} \pi^a \right).
\eea
Such a current inserts the following flavor-ordered vertices to the soft theorem, i.e. the RHS of Eq. (\ref{eq:wia}):
\bea
&&\V_{2n+1} (1|2,3|4,5| \cdots | 2n, 2n+1) = - \left(\frac{-4}{f^2}\right)^n\frac{ (n-1)!}{2 (2n)!} q\cdot p_{1}\non\\
&=& -\frac{1}{2\lambda f^2} \sum_{k=2}^{2n+1} (2q \cdot p_k) V_{2n+1}^{\nlsm + \phi + \psi} (1,k,k_p|4,5| \cdots|2n,2n+1 || 1^\psi, k^\phi, k_p^\psi)\non\\
&=& -\frac{1}{2\lambda f^2} \sum_{k=2}^{2n+1} (2q \cdot p_k) \sum_{\substack{j=2\\j \ne k}}^{2n + 1} V_{2n+1}^{\nlsm + \phi + \psi} (1,j,j_p|\alpd/j|| 1^\psi,j^\phi, k^\psi) .
\eea
The second equality in the above utilizes total momentum conservation as well as the on-shell condition of $q$, while the last equality is a consequence of the vanishing contributions of vertices given by Eq. (\ref{eq:vv}). This directly leads to the soft theorem given by Eq. (\ref{eq:slsspb}).

\section{The single soft theorem at $\ordr (p^4)$}\label{sec4}

Now let us work out the subleading single soft theorem of NLSM at $\ordr (p^4)$ for a general group representation.  We will use the universal trace basis and focus on the 4 P-even operators given by Eq. (\ref{eq:o14}), which will always exist for a general spacetime dimension $d$. Unlike the case for the $\ordr (p^2)$ amplitudes as in Eq. (\ref{eqnlsmosex}), the single soft limit of $M^{(4)}$ receives two contributions: terms with the $\ordr (p^4)$ corrections to the current $\J$ of the shift symmetry, as well as terms with the $\ordr (p^4)$ corrections to one of the semi-on-shell amplitudes $J$. Taking the momentum $p_{n+1}$ to be soft, for the single trace amplitude we should have
\bea
M_{n+1}^{(4)} (\mathbb{I}_{n+1}) &=& \tau \sum_{k=1}^{\lfloor n/2 \rfloor} \sum_l \V^{(4)} (q_{l_{1}}, \cdots, q_{l_{2k+1}})   \prod_{m=1}^{2k+1} J^{(2)} (l_{m-1}+1, \cdots , l_m)\non\\
&&+\tau \sum_{k=1}^{\lfloor n/2 \rfloor} \sum_l \V^{(2)} (q_{l_{1}}, \cdots, q_{l_{2k+1}})  \sum_{i=1}^{2k+1} J^{(4)} (l_{i-1}+1, \cdots , l_i) \non\\
&&\times\prod_{\substack{m=1\\ m\ne i}}^{2k+1} J^{(2)} (l_{m-1}+1, \cdots , l_m) + \ordr (\tau^2)\ ,\label{eq:ssp4st}
\eea
where $\V^{(2)}$ is given by Eq. (\ref{eq:op2vi}), the choice of partitions $l$ is the same as in the single soft theorem for $\nlsm^{(2)}$ given by Eq. (\ref{eqnlsmosex}), and $\V^{(4)} (\cdots)$ comes from the single trace part in $\J^{(4)}$. Similarly, for the double trace amplitude we should have
\bea
&&M_{n+1}^{(4)} ( \mathbb{I}_{m}| m+1, m+2, \cdots, n+1) \non\\
&=& \tau \sum_{k,\gamma} \sum_{j'=1}^k \V^{(2)} (q_{\gamma_{1}},q_{\gamma_{2}},\cdots, q_{\gamma_{j'}} + q_{\mathbb{I}_{m}}, \cdots, q_{\gamma_{k}})  J^{(4)} (  \gamma_{j'}  |  \mathbb{I}_{m} )  \prod_{\substack{j=1\\ j\ne j'}}^{k} J^{(2)} ( \gamma_j )\non\\
&& + \tau  \sum_{k,i,\gamma,\gamma'} \V^{(4)} (q_{\gamma_1}, \cdots,q_{\gamma_{k}} |q_{\gamma'_{1}}, \cdots, q_{\gamma'_{i}})   \left[\prod_{j=1}^{k} J^{(2)} ( \gamma_j ) \right] \left[ \prod_{j'=1}^{i} J^{(2)} ( \gamma'_{j' }) \right] + \ordr (\tau^2)\ ,\label{eq:ssp4dt}
\eea
where $\{ \gamma_{1}, \cdots, \gamma_k \}$ are partitions of $\{ m+1, m+2, \cdots, n\}$, $\{ \gamma'_1, \gamma'_2, \cdots , \gamma'_i \}$ are partitions of any of the cyclic permutations of $\mathbb{I}_{m}$, and $q_{\alpha} \equiv \sum_{r \in \alpha} p_r$ for any sequence $\alpha$ . It should be understood in Eq. (\ref{eq:ssp4dt}) that  $n$ is odd, $m$ is even, $\V^{(2)}$ is non-vanishing only if $k$ is even (so that it has an odd number of arguments), $\V^{(4)}$ is non-vanishing only if $i$ is even  and $k$ is odd, $J^{(2)}$ is non-vanishing only if it has an odd number of arguments, and $J^{(4)} (\alpha | \beta)$ is non-vanishing only if one of the sets in $\{ \alpha, \beta \}$ has an even number of elements while the other has an odd number, with the off-shell leg in $J^{(4)} (\alpha | \beta)$ being in the same trace as the odd set.

Our goal then is to work out $\V^{(4)}$. For the single trace operators $O_3$ and $O_4$, it will be convenient to rewrite them as
\bea
C_3 O_3 + C_4 O_4 &=&2(C_3 + C_4) \tr ( d_\mu d_\nu d^\mu d^\nu ) - 2(C_3 - C_4) \tr ( d_\mu d^\mu d_\nu d^\nu )\non\\
&\equiv & C_{3'} O_{3'} + C_{4'} O_{4'},
\eea
with
\bea
O_{3'} = \tr ( d_\mu d_\nu d^\mu d^\nu ),\   O_{4'} = \tr ( d_\mu d^\mu d_\nu d^\nu ),\ C_{3'} = 2(C_3 + C_4),\  C_{4'} =  - 2(C_3 - C_4).
\eea

The contribution of $O_{3'}$ to the current is
\bea
\left(\J^{(4),3'}\right)_\mu^a &=&\frac{4f}{\Lambda^2}  \left[\cos \sqrt{\mt} \right]_{ab} \tr \left(  \gX^b d_\nu d_\mu d^\nu \right),\label{eq:o3po}
\eea
with
\bea
\cos \sqrt{\mt} = \sum_{n=0}^\infty \frac{(-1)^n}{(2n)! } \mt^n.
\eea
Using the Lie algebra in Eq. (\ref{eq:crg}), one can  show that
\bea
d_\mu = - \frac{i}{2} \xi^\dagger \partial_\mu U \xi^\dagger = \frac{i}{2} \xi \partial_\mu U^\dagger \xi,\qquad \xi \left( \left[\cos \sqrt{\mt} \right]_{ab}  \gX^b \right) \xi = \frac{1}{2} \left\{ \gX^a, U \right\},
\eea
where
\bea
\xi  = e^{i \Pi}, \qquad \Pi = \frac{\pi^a \gX^a}{f}, \qquad  U = \xi^2.
\eea
Then
\bea
\left(\J^{(4),3'}\right)_\mu^a &=& \frac{if}{4\Lambda^2}   \tr \left(  \left\{ \gX^a, U \right\} \partial_\nu U^\dagger \partial_\mu U \partial^\nu U^\dagger \right) \non\\
 &=& \sum_{n=1}^\infty \sum_{l_1 = 0}^{2n +1} \sum_{l_2 = 0}^{2n +1 -l_1} \sum_{l_3 =0 }^{2n +1 -l_1- l_2} \frac{f}{\Lambda^2} \frac{(-4)^{n } (-1)^{l_1 + l_3}}{l_1! l_2! l_3!(2n+1-l_1 - l_2 - l_3)!}\non\\
&&\times \tr \left(   \gX^a \Pi^{l_1}  \partial_\nu \Pi^{l_2} \partial_\mu \Pi^{l_3} \partial^\nu \Pi^{2n+1-l_1 - l_2 -l_3} \right)\non\\
&=& \frac{4 f}{ \Lambda^2} \tr \left(  \gX^a \partial_\nu \Pi \partial_\mu \Pi 
\partial^\nu \Pi \right) + \frac{4f}{3 \Lambda^2} \tr \left[  \gX^a  \left( -6 \Pi^2 \partial_\nu \Pi \partial_\mu \Pi \partial^\nu \Pi\right. \right.\non\\
&& + 6\Pi \partial_\nu \Pi^2 \partial_\mu \Pi 
\partial^\nu \Pi -6 \Pi \partial_\nu \Pi \partial_\mu \Pi^2
\partial^\nu \Pi +6 \Pi \partial_\nu \Pi \partial_\mu \Pi 
\partial^\nu \Pi^2 \non\\
&& - 2\partial_\nu \Pi^3 \partial_\mu \Pi 
\partial^\nu \Pi - 2\partial_\nu \Pi \partial_\mu \Pi^3
\partial^\nu \Pi -2\partial_\nu \Pi \partial_\mu \Pi 
\partial^\nu \Pi^3 \non\\
&&\left. \left. +3\partial_\nu \Pi^2 \partial_\mu \Pi^2 
\partial^\nu \Pi + 3\partial_\nu \Pi \partial_\mu \Pi^2 
\partial^\nu \Pi^2 -3\partial_\nu \Pi^2 \partial_\mu \Pi 
\partial^\nu \Pi^2 \right) \right]\non\\
&& + \cdots.\label{eq:c3pc}
\eea
Similarly, the contribution of $O_{4'}$ to the current is 
\bea
\left( \J^{(4),4'} \right)_\mu^a &=& \sum_{n=1}^\infty \sum_{l_1 = 0}^{2n +1} \sum_{l_2 = 0}^{2n +1 -l_1} \sum_{l_3 =0 }^{2n +1 -l_1- l_2} \frac{f }{2 \Lambda^2} \frac{(-4)^{n } (-1)^{l_1 + l_3}}{l_1! l_2! l_3!(2n+1-l_1 - l_2 - l_3)!}\non\\
&&\times \tr \left[  \gX^a \Pi^{l_1}  \left(\partial_\mu \Pi^{l_2} \partial_\nu \Pi^{l_3} \partial^\nu \Pi^{2n+1-l_1 - l_2 -l_3}  +\partial_\nu \Pi^{l_2} \partial^\nu \Pi^{l_3} \partial_\mu \Pi^{2n+1-l_1 - l_2 -l_3} \right) \right].\ \ 
\eea
The corresponding vertices inserted into $\R_\mu$, after stripping the single-trace flavor factors, are
\bea
\V^{(4),3'}_{2n+1} (\mathbb{I}_{2n+1}) &=&  \sum_{l_1 = 0}^{2n -2} \sum_{l_2 = 1}^{2n -1 -l_1} \sum_{l_3 =1 }^{2n  -l_1- l_2} \frac{1}{f^{2n} \Lambda^2} \frac{(-4)^{n } (-1)^{l_1 + l_3}}{l_1! l_2! l_3!(2n+1-l_1 - l_2 - l_3)!}\non\\
&&\times q\cdot p_{l_1+ l_2 +1;l_1 + l_2 + l_3} \ p_{l_1+1;l_1+l_2} \cdot p_{l_1 + l_2 +l_3 + 1; 2n+1},\label{eq:c3pjv}\\
\V^{(4),4'}_{2n+1} (\mathbb{I}_{2n+1}) &=&  \sum_{l_1 = 0}^{2n -2} \sum_{l_2 = 1}^{2n -1 -l_1} \sum_{l_3 =1 }^{2n  -l_1- l_2} \frac{1}{2 f^{2n} \Lambda^2} \frac{(-4)^{n } (-1)^{l_1 + l_3}}{l_1! l_2! l_3!(2n+1-l_1 - l_2 - l_3)!}\non\\
&&\times \left( q\cdot p_{l_1+1;l_1+l_2} \  p_{l_1+ l_2 +1;l_1 + l_2 + l_3} \cdot p_{l_1 + l_2 +l_3 + 1; 2n+1} \right.\non\\
&&+  \left. q\cdot p_{l_1 + l_2 +l_3 + 1; 2n+1} \  p_{l_1+1;l_1+l_2} \cdot   p_{l_1+ l_2 +1;l_1 + l_2 + l_3}  \right),
\eea
where
\bea
p_{i;j} \equiv \sum_{k=i}^{j} p_k
\eea
and $q = - \sum_{i=1}^{2n+1} p_i$.

On the other hand, the contributions of $O_1$  and $O_2$ to the current are
\bea
\left(\J^{(4),1} \right)^a_\mu &=& \sum_{n=0}^{\infty} \sum_{m=1}^\infty \sum_{l_1 = 0}^{2n+1} \sum_{l_2 = 0}^{2m}  \frac{1}{f^{2(m+n)} \Lambda^2} \frac{(-4)^{(m+n)} (-1)^{l_1 + l_2}}{l_1! l_2! (2n+1-l_1)!(2m-l_2)!}\non\\
&&\times \tr \left(  X^a \Pi^{l_1}  \partial_\mu \Pi^{2n+1-l_1}\right) \tr \left( \partial_\nu \Pi^{l_2} \partial^\nu \Pi^{2m- l_2} \right),\\
\left(\J^{(4),2} \right)^a_\mu &=& \sum_{n=0}^{\infty} \sum_{m=1}^\infty \sum_{l_1 = 0}^{2n+1} \sum_{l_2 = 0}^{2m}  \frac{1}{f^{2(m+n)} \Lambda^2} \frac{(-4)^{(m+n)} (-1)^{l_1 + l_2}}{l_1! l_2! (2n+1-l_1)!(2m-l_2)!}\non\\
&&\times \tr \left(  X^a \Pi^{l_1}  \partial_\nu \Pi^{2n+1-l_1}\right) \tr \left( \partial_\mu \Pi^{l_2} \partial^\nu \Pi^{2m- l_2} \right),
\eea
and the corresponding double-trace vertices are
\bea
&&\V^{(4),1}_{2m+2n+1} (1, \cdots, 2m|2m+1, \cdots, 2m+2n+1 ) \non\\
&=& \sum_{l_1 = 0}^{2n} \sum_{l_2 = 1}^{2m-1} \sum_{i=1}^{2m} \frac{1}{f^{2(m+n)} \Lambda^2} \frac{(-4)^{(m+n)} (-1)^{l_1 + l_2}}{l_1! l_2! (2n+1-l_1)!(2m-l_2)!}\non\\
&&\times q\cdot p_{2m+l_1+1;2m+ 2n+1}\  p_{m|1;l_2|i} \cdot p_{m|l_2+1;2m|i},\\
&&\V^{(4),2}_{2m+2n+1} (1, \cdots, 2m|2m+1, \cdots, 2m+2n+1 ) \non\\
&=& \sum_{l_1 = 0}^{2n} \sum_{l_2 = 1}^{2m-1} \sum_{i=1}^{2m} \frac{1}{f^{2(m+n)} \Lambda^2} \frac{(-4)^{(m+n)} (-1)^{l_1 + l_2}}{l_1! l_2! (2n+1-l_1)!(2m-l_2)!}\non\\
 &&\times q\cdot p_{m|1;l_2|i} \ p_{m|l_2+1;2m|i} \cdot p_{2m+l_1+1;2m+ 2n+1}, 
\eea
where
\bea
p_{m|i;j|k} = \sum_{l = \mod (i+k,2m)+1}^{\mod (j+k,2m)+1} p_l.
\eea
Then we have
\bea
\V^{(4)} (\alpha) =C_{3'} \V^{(4),3'} (\alpha) +C_{4'} \V^{(4),4'} (\alpha) ,\quad \V^{(4)} (\alpha | \beta) = C_1 \V^{(4),1} (\alpha | \beta)+ C_2 \V^{(4),2} (\alpha | \beta).
\eea
We give low-pt examples of these vertices in Appendix \ref{app:p4cv}.

The next question to ask is: can we interpret the RHS of Eqs. (\ref{eq:ssp4st}) and (\ref{eq:ssp4dt}) as given by the amplitudes of some extended theory? We will first focus on a special case where we have a definite answer, and then proceed to the more general case. To avoid complicated flavor labels, we will assume that the NGB's furnish the adjoint representation, so that the extended theory of $\nlsm^{(2)}$ is just $\nlsm + \phi^3$, though the results can be straightforwardly reinterpreted for a general group representation.

\subsection{The $d_2$ case}

As demonstrated in Ref. \cite{Cachazo:2016njl}, the extended theory can be identified in a concrete manner when the amplitudes of the original theory have a CHY representation and admit a double copy structure. In Ref. \cite{Low:2020ubn}  a special case of the NLSM up to $\ordr (p^4)$, dubbed $\nlsm^{d_2}$, was observed to demonstrate these properties. In such a theory  the Wilson coefficients are fixed to be
\bea
C_1 =  \frac{1}4, \qquad C_2 =- \frac{1}2,\qquad C_3 = C_4 = 0.\label{eq:defd2}
\eea
The corresponding $\ordr (p^4)$ amplitude $M^{(4),d_2}$ always has a double trace ordering, and is a component of an EFT named the extended Dirac-Born-Infeld (DBI) theory \cite{Cachazo:2014xea}, also called $\dbi + \nlsm$ \cite{Chiodaroli:2017ngp}, which is a double copy of $\nlsm^{(2)}$ and a gauged version of the bi-adjoint scalar theory called $\ym + \phi^3$ \cite{Chiodaroli:2014xia}, or generalized YM scalar \cite{Cachazo:2014xea}. We can write
\bea
\nlsm^{d_2} \subset \dbi + \nlsm = \nlsm^{(2)} \stackrel{\rm KLT}{\otimes} \left({\textrm{YM}+\phi^3}\right),
\eea
where $\stackrel{\rm KLT}{\otimes}$ indicates a Kawai-Lewellen-Tye (KLT) relation \cite{Kawai:1985xq}  between the ordered amplitudes of two theories, which we will use extensively below to show that in the single soft limit of $\nlsm^{d_2}$ there is indeed an extended theory.

\subsubsection{From the double copy}

The amplitudes for $\nlsm^{d_2}$ can be expressed using the KLT formula as
\bea
&&M^{(4),d_2}_n (n,\mathbb{I}_{l}|\alpha) \non\\
&=& \sum_{a,b=2}^{n-2} \sum_{\beta_L, \beta_R \in S_{n-4}} M^{(2)}_n (1,\beta_L,n-1,a,n) K_n(1,\beta_L,n-1,a,n || 1,\beta_R,n-1,n,b)\non\\
&& \times M^{ \ym + \phi^3}_n (n,\mathbb{I}_l|\alpha ||1,\beta_R,n-1,n,b),\label{eq:kltd2}
\eea
where $l$ is odd and $1\le l \le n-3$, $M^{ \ym + \phi^3}$ is the amplitude for $\ym + \phi^3$ where all external states are scalars, and $K_n$ is the KLT kernel for the $n$-pt amplitude satisfying \cite{Cachazo:2013iea}
\bea
K_n = \left[M_n^{\phi^3} \right]^{-1},\label{eq:kltp3}
\eea
with $M_n^{\phi^3}$ being the doubly ordered amplitudes for the bi-adjoint scalar theory. In Eq. (\ref{eq:kltp3}) $K_n$ and $M_n^{\phi^3}$ are understood as $(n-3)! \times (n-3)!$ square matrices, whose rows and columns correspond to $(n-3)!$ ways of the left and right ordering, respectively. For example, in Eq. (\ref{eq:kltd2}) the positions of $1$, $n-1$ and $n$ are fixed in both the left and the right orderings of $K_n$, so that the remaining $(n-3)!$ left and right orderings define the square KLT matrix.

\begin{figure}[t]
\centering
\includegraphics[width=0.18\textwidth]{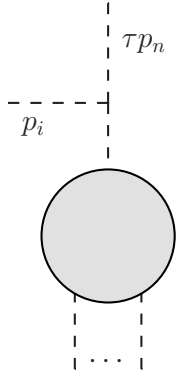}
\caption{\label{fig:p3p} The ``pole diagrams'' giving $\ordr (\tau^{-1})$ contributions in the $\phi^3$ theory, where the soft leg and another external leg are attached to the same 3-pt vertex, leading to a pole in $\tau$ given by the propagator $i/(2\tau p_{i} \cdot p_n )$.}
\end{figure}

Now let us take the single soft limit of state $n$ in Eq. (\ref{eq:kltd2}). The  $\nlsm^{(2)}$ amplitude $M^{(2)}$ on the RHS of Eq. (\ref{eq:kltd2}) is expanded using Eq. (\ref{eq:sssstr}).
The $\phi^3$ amplitudes start at $\ordr (\tau^{-1})$, and similar to what we have seen in Section \ref{sec:st}, the leading contributions in the soft limit have to be given by ``pole diagrams'' shown in Fig. \ref{fig:p3p}. Then a simple calculation yields
\bea
&&M_n^{\phi^3} (1,\beta_L,n-1,a,n || 1,\beta_R,n-1,n,b) \non\\
& =&  \frac{ \lambda}{\tau } \frac{\delta_{ab}}{s_{an}}  M_{n-1}^{\phi^3} (1,\beta_L,n-1,a || 1,\beta_R,n-1,b) + \ordr (\tau^0),
\eea
so that the KLT kernel in Eq. (\ref{eq:kltd2}) is given by
\bea
\left[M_n^{\phi^3} \right]^{-1} =  \frac{\tau}\lambda \delta_{ab} s_{an} \left[M_{n-1}^{\phi^3} \right]^{-1}+\ordr (\tau^2).\label{eq:d2kl}
\eea

\begin{figure}[t]
\centering
\includegraphics[width=0.18\textwidth]{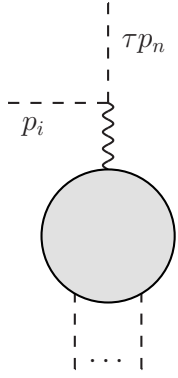}
\caption{\label{fig:ymp3p} The additional pole diagram  in  $\ym + \phi^3$, where the internal line that gives the $1/\tau $ pole is a gauge boson.}
\end{figure}

Similar to the $\phi^3$ amplitudes, we have $M_n^{\ym + \phi^3} = \ordr (\tau^{-1})$, but the pole diagrams here not only contain ones shared with the $\phi^3$ theory as shown in Fig. \ref{fig:p3p}, but more diagrams where the internal leg giving the $1/\tau$ pole belongs to a gauge boson, as shown in Fig. \ref{fig:ymp3p}. However, these additional diagrams are proportional to
\bea
\frac{1}{\tau s_{in}} p_i^\mu M^{\ym + \phi^3}_{\mu, n-1} (i^g) + \ordr (\tau^0),
\eea
where $\epsilon_i^\mu M^{\ym + \phi^3}_{\mu, n-1} (i^g)$ is the on-shell amplitude of $\ym + \phi^3$ with all external states being scalar except for the state $i$, which is a gauge boson with polarization vector $\epsilon_i$. Then
\bea
p_i^\mu M^{\ym + \phi^3}_{\mu, n-1} (i^g) = 0
\eea
because of gauge invariance, so that Fig. \ref{fig:ymp3p} starts at $\ordr (\tau^0)$ after all.
Therefore, the only $\ordr (\tau^{-1})$ contributions are still given by Fig. \ref{fig:p3p} as in the $\phi^3$ theory, and
\bea
M_n^{\ym + \phi^3} (n,\mathbb{I}_l|\alpha||1,\beta_R,n-1,n,b) = \frac{\lambda}{\tau}  \frac{\delta_{bl}}{ s_{bn}} M^{\ym + \phi^3}_{n-1} (\mathbb{I}_l |\alpha||1,\beta_R,n-1,b) + \ordr (\tau^0).\label{eq:d2yp}
\eea
Notice that in the above, the $\ordr (\tau^{-1})$ term vanishes when $l=1$. 

Combining Eq. (\ref{eq:kltd2}) with Eqs. (\ref{eq:sssstr}), (\ref{eq:d2kl}) and (\ref{eq:d2yp}), we arrive at the   single soft  theorem of $\nlsm^{d_2}$:
\bea
M^{(4),d_2}_n (n,\mathbb{I}_{l}|\alpha) =\frac{\tau}{\lambda f^2} \sum_{\substack{i=2 \\ i \ne l}}^{n-1}  s_{n,i} M_{n-1}^{d_2 + \phi^3 } (\mathbb{I}_{l}|\alpha||1,l,i) + \ordr (\tau^2).\label{eq:sttd2}
\eea
In the above, the amplitudes of the extended theory $d_2 + \phi^3$ are given by the following double copy formula:
\begin{align}
M_{n}^{d_2 + \phi^3 } (\mathbb{I}_{l}|\alpha||1,l,i) &\equiv&  \sum_{\beta_L, \beta_R\in S_{n-3}} M^{\nlsm+\phi^3}_n (\beta_L||1,l,i) K_n(\beta_L || \beta_R)    M^{ \ym + \phi^3}_n (\mathbb{I}_l|\alpha||\beta_R),\ \ 
\end{align}
where $\beta_L$ and $\beta_R$ are permutations of $\{ 1,2, \cdots, n\}$ with the positions of three arbitrary elements fixed. Again, the above does not exist when $l = 1$, thus
\bea
M^{(4),d_2}_n (n,1|\alpha) = \ordr (\tau^2).\label{eq:ssd2t2}
\eea
This can be easily checked for the 6-pt amplitude $M^{(4),d_2}_6 (6,1|2,3,4,5)$, which we show in Appendix \ref{app:nlv}.

The soft theorem given by Eq. (\ref{eq:sttd2}) is very similar to the single soft theorem of $\nlsm^{(2)}$ as in Eq. (\ref{eq:sssstr}): 3 legs are given additional ordering in the extended theory, where two of them, i.e. $1$ and $l$ in  Eq. (\ref{eq:sttd2}), are adjacent to the soft leg $n$ in the original amplitude, while the third leg $i$ ranges over all the hard legs non-adjacent to $n$ and enters  the simple the soft factor $s_{n,i}$.

Just like the theory $\nlsm^{d_2}$ is part of $\dbi + \nlsm$, the extended theory $d_2 + \phi^3$ in Eq. (\ref{eq:sttd2}) can be identified as part of a theory called $\dbi +\ym+ \nlsm   + \phi^3$ \cite{Chiodaroli:2017ngp}, which is the double copy of $\nlsm + \phi^3$ and $\ym + \phi^3$:
\bea
d_2 + \phi^3 \subset \dbi +\ym+ \nlsm   + \phi^3 = \left(\nlsm + \phi^3 \right) \stackrel{\rm KLT}{\otimes} \left( \ym + \phi^3 \right)
\eea
Actually, by the same argument as above one can easily calculate the single soft theorem for a general $n$-pt amplitude of $\dbi + \nlsm$, taking $p_n$ to be soft:
\bea
&&M^{\dbi + \nlsm}_n (\alpha_1 | \alpha_2 |\cdots|\alpha_m, n) \non\\
&=& \frac{\tau}{\lambda f^2} \sum_{\substack{i=1\\i\ne j_L, j_R}}^{n-1}  s_{n,i} M^{\dbi +\ym+ \nlsm   + \phi^3}_{n-1} (\alpha_1 | \alpha_2 |\cdots|\alpha_m ||j_R,j_L,i)  + \ordr (\tau^2),\label{eq:stdn}
\eea
where $j_L$ and $j_R$ are the legs left/right adjacent to the leg $n$ in the ordering of the original amplitude, i.e. the last and first element in the sequence $\alpha_m$.

\subsubsection{Matching to the Ward identity}
\label{sec:d2ev}

Now let us compare Eq. (\ref{eq:sttd2}) with what we know from the Ward identity, i.e. Eq. (\ref{eq:ssp4dt}), and work out the vertices in $d_2 + \phi^3$. As Eq. (\ref{eq:sttd2}) has essentially the same form as the $\ordr (p^2)$ soft theorem of Eq. (\ref{eq:sssstr}), we again have the Type I and Type II vertices: Type I are vertices with two $\phi$ that equals the $\nlsm^{d_2}$ ones with the same left ordering, while Type II  comes from vertices $\V$ generated by the current. We then need to compute the Type II vertices in the following. In $\nlsm^{d_2}$ the Wilson coefficients of the $\ordr (p^4)$ operators take the fixed values of Eq. (\ref{eq:defd2}), so that the relevant vertex $\V^{(4)}$ is
\bea
\V^{(4),d_2} (\alpha|\beta) = \frac{1}{4} \V^{(4),1} (\alpha|\beta)  - \frac{1}{2}\V^{(4),2} (\alpha|\beta) .
\eea

The first thing one may want to check is the special case of Eq. (\ref{eq:ssd2t2}), where the $\ordr (\tau)$ term vanishes. In such a case, Eq. (\ref{eq:ssp4dt}) is reduced to
\bea
M_{n+1}^{(4)} ( n,1| \alpha) =  \tau  \sum_{i,\gamma'} \V^{(4),d_2} (1|q_{\gamma'_{1}}, \cdots,q_{\gamma'_{i}} )   \prod_{j=1}^{i} J^{(2)} ( \gamma_j ) + \ordr (\tau^2)\ 
\eea
with $p_n$ being soft, where $\{ \gamma'_1, \cdots, \gamma'_i \}$ are partitions of any of the cyclic permutations of $\alpha$. It is clear that vertices $\V^{(4),d_2} (j|\alpha) $ with only one leg $j$ in one of their traces appear in all the $\ordr (\tau)$ terms in the above.

Therefore, to satisfy Eq. (\ref{eq:ssd2t2}) we need total cancellations of all terms involving $\V^{(4),d_2} (j|\alpha) $. This, however, does not necessarily imply that $\V^{(4),d_2} (j|\alpha) = 0$. It is easy to show that
\bea
\V^{(4),d_2} (1,2|3)  = \frac{2}{f^2 \Lambda^2} \left( q\cdot p_1\ p_2^2 + q \cdot p_2\ p_1^2 \right),\label{eq:3pd2o}
\eea
which does not vanish, but its effect can be moved to higher-pt vertices. For example, if $p_{1}$ in $\V^{(4),d_2} (1,2|3)$ is an internal momentum in a Feynman diagram, the term with $p_{1}^2$ will cancel the propagator and effectively resulting in a higher-pt vertex. This suggests that the EoM may be needed here. Indeed, let us directly look at the current: the vertices $\V^{(4),d_2} (j|\alpha) $ come from
\bea
\left( \J^{(4),d_2} \right)^a_\mu &=& \frac{f}{\Lambda^2}  \left[\cos \sqrt{\mt} \right]_{ab} \left[ \tr \left( X^b d_\mu \right) \tr \left( d_\nu d^\nu \right) - 2 \tr \left(  X^b d_\nu  \right) \tr \left(d_\mu d^\nu \right) \right]\non\\
&\supset & \frac{1}{\Lambda^2}   \left[ \tr \left( X^a \partial_\mu \Pi \right) \tr \left( d_\nu d^\nu \right) - 2 \tr \left(  X^a \partial_\nu \Pi  \right) \tr \left(d_\mu d^\nu \right) \right]\non\\
&\simeq & -\frac{2}{\Lambda^2} \tr \left( X^a  \Pi \right) \left[ 2 \tr \left( d^\nu \nabla_{[\mu} d_{\nu]} \right) - \tr \left( d_\mu \nabla_\nu d^\nu \right) \right]\non\\
&\simeq &0.
\eea
In the second line of the above we select all terms that generate $\V^{(4),d_2} (j|\alpha) $, in the third line we drop total derivatives, which give a contribution of $\ordr (\tau^2)$ in the soft theorem, while in the last line we apply an identity: $\nabla_{[\mu} d_{\nu]}  \equiv (\nabla_{\mu} d_{\nu} -\nabla_{\nu} d_{\mu} )/2= 0$, as well as the EoM $\nabla_\mu d^\mu = \ordr (\partial^4)$, and dropped terms beyond $\ordr (\partial^3)$. We provide a proof of these relations in Appendix \ref{app:eom}. This indicates that all contributions from the vertices like $\V^{(4),d_2} (j|\alpha)$ will be cancelled out in the final amplitude.

The vertex that we need to consider is then
\bea
&&\V^{(4),d_2}_{2m+2n+1} (\mathbb{I}_{2m}|2m+1, \cdots, 2m+2n+1 )\non\\
&=& \frac{1}{\Lambda^2} \left(-\frac{4}{f^2}\right)^{m+n} \sum_{l_1 = 0}^{2n} \sum_{l_2 = 1}^{2m-1} \sum_{i=1}^{2m}  \frac{ (-1)^{l_1 + l_2}}{l_1! l_2! (2n+1-l_1)!(2m-l_2)!}\non\\
&&\times \left( \frac{1}4 q\cdot p_{2m+l_1+1;2m+ 2n+1}\  p_{m|1;l_2|i} \cdot p_{m|l_2+1;2m|i}  \right. \non\\
 &&\left.- \frac{1}2 q\cdot p_{m|1;l_2|i} \ p_{m|l_2+1;2m|i} \cdot p_{2m+l_1+1;2m+ 2n+1} \right)\non\\
 &=& \frac{1}{\lambda f^2} \sum_{\substack{i=1 \\ i \ne 2m+1}}^{2m+2n} 2 q \cdot p_i  \  V_{2k+1}^{d_2 + \phi^3} (\mathbb{I}_{2k+1} || 2m,2m+2n+1,j).
\eea
Similar to what we see in Eq. (\ref{eq:type2r}), the above dictates that terms involving $q \cdot p_{2m+1}$ and $q \cdot p_{2m+2n+1}$ in $\V^{(4),d_2}$ need to be eliminated at the same time. The final form of the vertices for the extended theory would be
\bea
&&V^{d_2   + \phi^3}_{2m+2n+1} (\mathbb{I}_{2m}|2m+1,\cdots,2m+2n+1||2m+1,2m+2n+1,j)\non\\
&=& \frac{\lambda  }{2\Lambda^2}  \left(-\frac{4}{f^2}\right)^{m+n-1}   \frac{ 1 }{ (2m)!(2n+1) !}  \sum_{i=1}^{2m}  \left[ \sum_{l_2 = 1}^{2m-1} (-1)^{ l_2} \left(\begin{array}{c} 
2m\\
l_2
\end{array}\right)p_{m|1;l_2|i} \cdot p_{m|l_2+1;2m|i} \right. \non\\
&&   \left. + 2 \mathsf{F} (m, |i-j|) \sum_{l_1 = 0}^{2n}  (-1)^{l_1}   \left(\begin{array}{c}
2n+1\\
l_1
\end{array}\right)  p_i \cdot p_{2m+l_1+1;2m+ 2n+1} \right]
\eea
for $1 \le j \le 2m$, and
\bea
&&V^{d_2   + \phi^3}_{2m+2n+1}(\mathbb{I}_{2m}|2m+1,\cdots,2m+2n+1||2m+1,2m+2n+1,j)\non\\
&=& \frac{\lambda  }{2\Lambda^2}  \left(-\frac{4}{f^2}\right)^{m+n-1} \sum_{l_2 = 1}^{2m-1} \sum_{i=1}^{2m}  \frac{ (-1)^{ l_2}}{ (2m)!(2n+1) !} \left(\begin{array}{c}
2m\\
l_2
\end{array}\right) \non\\
&&\times\left[   1 +  (-1)^{j}  \left(\begin{array}{c}
2n\\
j-2m-1
\end{array}\right) \right] p_{m|1;l_2|i} \cdot p_{m|l_2+1;2m|i} 
\eea
for $ 2m+2 \le j \le 2m+2n$, with
\bea
\mathsf{F} (m , k ) &\equiv& -2m \min ( k, 2m-k) \non\\
&&+ 
  \frac{\left(1- \delta_{1,m}\right) (-1)^k k (2m-k) \left[ \min (k, 2m-k) - 1 \right]}{2(m-1 + \delta_{1,m})(2m-1)} \left(\begin{array}{c}
2m\\
k
\end{array}\right) .
\eea
Examples of the vertices and amplitudes for $d_2 + \phi^3$ are presented in Appendix \ref{app:eet}.

\subsection{The general case}

It is still unknown whether the most general $\ordr (p^4)$ amplitudes of NLSM, with arbitrary Wilson coefficients $C_i$, have a double copy structure or a CHY representation. Without such input it is hard to answer definitely whether the soft theorems of Eqs. (\ref{eq:ssp4st}) and (\ref{eq:ssp4dt}) have an interpretation of extended theories. If they do, as they are for amplitudes of the higher derivative corrections to $\nlsm^{(2)}$, it is natural to assume that they are higher derivative corrections to the $\nlsm^{(2)}$ soft theorem, given by Eq. (\ref{eq:sssstr}). However, the correction can either enter the extended theory or the soft factor. Schematically, we can have
\bea
M^{(4)}_{n+1} = \tau \S^{(2)} M^{(4)+ }_n + \tau \S^{(4)} M^{\nlsm +\phi^3 }_n + \ordr (\tau^2),\label{eq:gsssch}
\eea
where $\S^{(2)}$ is the known soft factor in the soft theorem of $\nlsm^{(2)}$, $(4)+$ is some higher derivative corrections to the extended theory $\nlsm +\phi^3$, and $\S^{(4)}$ a new soft factor at $\ordr (p^4)$. One needs to determine which part of the RHS of Eqs. (\ref{eq:ssp4st}) and (\ref{eq:ssp4dt}) enters the first/second term on the RHS of Eq. (\ref{eq:gsssch}).

Here we make the first steps to answer these questions. We observe that in the special case of $\nlsm^{d_2}$, the form of the soft theorem is very similar to that of $\nlsm^{(2)}$: the soft factor is not modified at all, while all the higher derivative corrections go into the extended theory. One can then ask: is it possible to do this for the general amplitudes of NLSM at $\ordr (p^4)$? To be more concrete, we would like to ask whether the single trace $\ordr (p^4)$ soft theorem given by Eq. (\ref{eq:ssp4st}) can be interpreted as
\bea
M_{n+1}^{(4)} (\mathbb{I}_{n+1}) = \frac{\tau}{\lambda f^2} \sum_{i=2}^{n-1}  s_{n+1,i} \ M_n^{(4) + \phi^3}(\mathbb{I}_{n}||1,n,i) + \ordr (\tau^2),\label{eq:sstast}
\eea
and similarly in the double trace case, whether Eq. (\ref{eq:ssp4dt}) can be interpreted as
\bea
M_{n+1}^{(4)  } (n+1,\mathbb{I}_{l}|\sigma )  = \frac{\tau}{\lambda f^2} \sum_{\substack{i=2 \\ i \ne l}}^{n}  s_{n+1,i} M_{n}^{(4) + \phi^3 } (\mathbb{I}_{l}|\sigma||1,l,i) + \ordr (\tau^2).\label{eq:sstadt}
\eea
Notice that Eqs. (\ref{eq:sstast}) and (\ref{eq:sstadt}) are much more stringent constraints than Eq. (\ref{eq:gsssch}): we require that at $\ordr (\tau)$, the coefficient of $s_{n+1,i}$, with $n+1$ labeling the soft leg, vanishes if $i$ is adjacent to $n+1$ in the ordering of the original amplitude. Furthermore, $(4) + \phi^3$ is understood as a higher derivative correction to $\nlsm + \phi^3$, and the coefficient of $s_{n+1,i}$ is an amplitude where only three of the $n$ external legs are bi-adjoint scalars, while all the other external states are still NGB's. Then the extended theory amplitudes should satisfy the Adler zero condition for all the external legs, except for $\{1,n,i\}$ in Eq. (\ref{eq:sstast}) and $\{1,l,i \}$ in Eq. (\ref{eq:sstadt}).

One can then use low-pt amplitudes of $\nlsm$ to test whether such interpretations are possible. For the double trace amplitudes, satisfying Eq. (\ref{eq:sstadt}) leads to a constraint of the Wilson coefficients: $C_2 = -2 C_1$. This effectively selects the amplitudes of $\nlsm^{d_2}$. On the other hand, we have tested up to 8-pt that the interpretation of Eq. (\ref{eq:sstast}) is possible for the single trace amplitudes if $C_{4'} = 0$. 

In the $\nlsm^{d_2}$ case the result is somewhat expected, given that these amplitudes can be built by a double copy with $\nlsm^{(2)}$ amplitudes, which themselves satisfy this type of single soft theorem. The single trace case  on the other hand is surprising, as these $\mathcal{O}(p^4)$ amplitudes are not known to have any direct connection to the $\mathcal{O}(p^2)$ amplitudes. We will call the NLSM with  $\ordr (p^4)$ Wilson coefficients $C_{3'} = 1$, $C_{4'} = C_1 = C_2 = 0$ as $\nlsm^{C_{3'}}$, and the corresponding extended theory $C_{3'} + \phi^3$. The soft theorem for the $\ordr (p^4)$ amplitudes of $\nlsm^{C_{3'}}$ is then
\bea
M_{n+1}^{(4),3'} (\mathbb{I}_{n+1}) = \frac{\tau}{\lambda f^2} \sum_{i=2}^{n-1}  s_{n+1,i} \ M_n^{C_{3'}  + \phi^3}(\mathbb{I}_{n}||1,n,i) + \ordr (\tau^2).\label{eq:sstc3p}
\eea

Again, as in the case of $d_2 + \phi^3$, there should be Type I and Type II vertices in $C_{3'} + \phi^3$, Type I taking the same value the $\nlsm^{C_{3'}}$ vertices with identical left orderings, while Type II vertices should be given by
\bea
\V^{(4),3'}_{2k+1} (\mathbb{I}_{2k+1})  =\frac{1}{\lambda f^2} \sum_{i=2}^{2k} 2 q \cdot p_i  \  V_{2k+1}^{C_{3'} + \phi^3} (\mathbb{I}_{2k+1} || 1,2k+1,j).
\eea
We see from Eq. (\ref{eq:c3pjv}) that $\V^{(4),3'}_{2k+1} (\mathbb{I}_{2k+1})$ can indeed be put in a form without $2 q \cdot  p_1$ and $2q \cdot p_{2k+1}$. However, when $k=1$, $j$ has to be $2$, and the coefficient of $q \cdot p_2$ is proportional to
\bea
 p_1 \cdot p_3,\label{eq:v3po}
\eea
which cannot work as $V^{C_{3'}+\phi^3}_3 ( 1,2,3|| 1,3,2)$, because the 3-pt $\phi^3$ vertex need to be invariant under cyclic permutations. This problem persists in higher-pt vertices as well: we need
\bea
V^{C_{3'}+\phi^3}_{2k+1} ( \mathbb{I}_{2k+1}|| 1,2k+1,2) = V^{C_{3'}+\phi^3}_{2k+1} ( 1,2k+1,2k,2k-1,\cdots, 2|| 1,2,2k+1),\label{eq:c3pcst}
\eea
the two sides of which are related by a reversion of both the left and the right orderings. For example, the coefficient of $2q \cdot p_2$ for $\V^{(4),3'}_{5} (\mathbb{I}_{5})$ in Eq. (\ref{eq:c3pjv}) is
\bea
 \frac{2}{3 f^4 \Lambda^2 }  p_1 \cdot (-2 p_3 + p_4 - p_5),
\eea
which is not invariant under $2 \leftrightarrow 5$, $3 \leftrightarrow 4$.

At 3-pt, we see that
\bea
p_1 \cdot p_3 = - \left( p_1 \cdot p_3 + p_1 \cdot p_2 + p_2 \cdot p_3 + p_1^2 + p_3^2 \right).\label{eq:c3p3om}
\eea
In other words, the difference between Eq. (\ref{eq:v3po}) and a cyclic form is associated with $p_i^2$. Similar to what we have seen in $\V^{(4),d_2}(1,2|3)$ as in Eq. (\ref{eq:3pd2o}), this amounts to corrections to higher-pt vertices. Similar to Section \ref{sec:d2ev}, it would be easier to fix the problem at the level of the current using the EoM. The suitably modified vertices given by the current is
\bea
\hat{\V}^{(4),3'}_{2k+1} (1,2,\cdots, 2k+1) &=& -\frac{1}{\Lambda^2 f^{2k} } \sum_{l_1 = 0}^{2k-2} \sum_{l_2=1}^{2k-1-l_1} \sum_{l_3=1}^{2k-l_1 - l_2} \frac{(-4)^k (-1)^{l_1 + l_3}}{l_1! l_2 ! l_3 ! (2k+1 - l_1 - l_2 - l_3)! }\non\\
 && \left( q \cdot p_{l_3 + 1;l_3 + l_2} \ p_{l_3 + 1;l_3 + l_2} \cdot p_{l_1 + l_2 + l_3 + 1; 2k+1}\right.\non\\
 &&\left. + q \cdot p_{l_1+l_2 + 1;l_1 + l_2 + l_3} \ p_{l_1 + 1;l_1 + l_2} \cdot p_{l_1 + l_2  + 1; 2k+1}  \right)\non\\
 &=&\frac{1}{\lambda f^2} \sum_{i=2}^{2k} 2 q \cdot p_i  \  V_{2k+1}^{C_{3'} + \phi^3} (\mathbb{I}_{2k+1} || 1,2k+1,j),\label{eq:v3pm}
\eea
and the details of the derivation of the above are presented in Appendix \ref{app:c3pfc}. Then the Type II vertices are given by
\bea
&&V^{C_{3'} + \phi^3}_{2k+1} (1,2,\cdots, 2k+1||1,2k+1,j) \non\\
&=& -\frac{\lambda (-4)^k}{2\Lambda^2 f^{2k-2} } \sum_{l_1 = 1}^{j-1} \left\{ \sum_{l_2=j-l_1}^{2k-l_1} \sum_{l_3=0}^{2k-l_1 - l_2} \frac{ (-1)^{l_1 + l_3}}{l_1! l_2 ! l_3 ! (2k+1 - l_1 - l_2 - l_3)! } \right.\non\\
&&\times p_{l_1 + 1;l_1 + l_2} \cdot   p_{l_1 + l_2 + l_3 + 1; 2k+1} \non\\
&&+ \sum_{l_2=1}^{j-l_1} \left[ 1 +  (-1)^j  \left( \begin{array}{c}
2k+1-l_1 - l_2 \\
2k+1 - j
\end{array} \right) \right] \non\\
&& \left. \phantom{\sum_j^{l_2}} \times  \frac{ (-1)^{l_2 }}{(l_1-1)! l_2 ! (2k+2-l_1 - l_2)! } p_{l_1 ;l_1 + l_2 -1} \cdot p_{l_1 + l_2  ; 2k+1} \right\},
\eea
which satisfy Eq. (\ref{eq:c3pcst}). We provide examples in Appendix \ref{app:eet}.

\section{The double soft theorem at $\ordr (p^\infty)$}\label{sec5}

It was shown in Ref. \cite{Low:2015ogb} that the first two non-vanishing orders of the double-soft theorem for NLSM can be straightforwardly derived for the $\ordr (p^2)$, flavor-dressed full amplitudes $\M^{(2)}$. Using the same method, we can generalize such results to all orders in the derivative expansion for the tree amplitude $\M^\nlsm$.

\begin{figure}[t]
\centering
\includegraphics[width=0.18\textwidth]{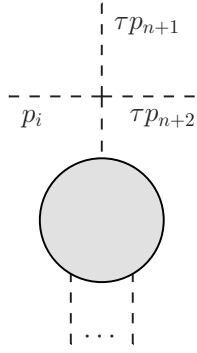}
\caption{\label{fig:dousopol} The pole diagram in the double soft limit.}
\end{figure}

Let us consider the $(n+2)$-pt tree amplitude $\M_{n+2}$, and take $p_{n+1}$ and $p_{n+2}$ to be soft. The leading non-vanishing term in the double soft limit is $\ordr (\tau^0)$, instead of  $\ordr (\tau)$ which the Adler zero in the single soft limit may suggest. This is because when two of the external legs are soft, there are Feynman diagrams where a pole in the soft parameter  $\tau $ develops: this happens when both of the soft legs are attached to a single 4-pt vertex, and a third external leg with momentum $p_i$ is attached to the vertex as well, as shown in Fig. (\ref{fig:dousopol}). The pole in $\tau$ then appears in the propagator attached to the vertex:
\bea
\frac{i}{\tau (s_{i,n+1} + s_{i,n+2} + \tau s_{n+1,n+2})}.
\eea

Therefore, we are able to classify all Feynman diagrams into two groups: the pole diagrams with such a pole in $\tau$, summed to $\M_{\pole}$; all the other diagrams, which  are called ``gut diagrams'' and sum to $\M_\gut$. It is straightforward to calculate $\M_{\pole}$ in the double soft limit, while the contributions of the gut diagrams are then fixed by applying symmetry constraints on the full amplitude $\M = \M_{\pole} + \M_\gut$, including flavor symmetry as well as the shift symmetry, which manifests as the Adler zero condition in the single soft limit. We will be able to calculate the double soft limit up to $\ordr (\tau)$ in this manner.

The pole diagrams for $\M_{n+2}$ are given by
\bea
&&\M_{n+2,\ \pole}^{a_1 \cdots a_{n+2}}  (p_1, \cdots, p_n, \tau p_{n+1}, \tau p_{n+2}) \non\\ &=& \sum_{i=1}^n \frac{- V^{a_i a_{n+1} a_{n+2} b}_4 (p_i, \tau p_{n+1}, \tau p_{n+2}, -\tilde{p_i})}{\tau (s_{i,n+1} + s_{i,n+2} + \tau s_{n+1,n+2})}   \tilde{\M}_n^{a_1 \cdots a_{i-1} b a_{i+1} \cdots n} (p_1, \cdots, p_{i-1},\tilde{p_i}, p_{i+1},\cdots , p_n) \non\\
&& + \ordr (\tau^2),\label{eq:dspei}
\eea
where $V_4$ is the 4-pt vertex attached by the two soft legs as shown in Fig. (\ref{fig:dousopol}), while
\bea
\tilde{\M}_n^{a_1 \cdots a_{i-1} b a_{i+1} \cdots a_n} (p_1, \cdots, p_{i-1},\tilde{p_i}, p_{i+1},\cdots , p_n) 
\eea
is the $n$-pt amplitude with leg $i$ off-shell, with momentum  $\tilde{p_i} = p_i + \tau (p_{n+1} + p_{n+2})$. Here $\tilde{\M}$ is clearly equivalent to the Berends-Giele current given by Eq. (\ref{eq:bgcd}), except that the off-shell leg is also cut in $\tilde{\M}$ but not in $J$.  Of course, in the fully on-shell limit we have $\tilde{\M}_n (p_i) = \M_n ( p_i)$. However, for a general off-shell momentum $\tilde{p_i}$, $\tilde{\M}_n (\tilde{p_i}) = \M_n (\tilde{p_i})$ is not guaranteed. Therefore, to relate $\tilde{\M}$ and $\M$ in a concrete manner, it is necessary to express  $\tilde{\M}_n (\tilde{p_i})$  as
\bea
&&\tilde{\M}_n^{a_1 \cdots  b  \cdots a_n} ( \cdots,\tilde{p_i},\cdots)\non\\
&  =& \M_n^{a_1 \cdots b \cdots a_n} (\cdots,\tilde{p_i},\cdots ) + \tilde{p_i}^2 X^{a_1 \cdots b \cdots a_n} (p_1, \cdots,\tilde{p_i},\cdots , p_n),
\eea
where $X(\tilde{p_i})$ is some unknown function of the momenta and must be non-singular in the limit of  $\tilde{p_i} \to p_i$. The $X$ piece is usually neglected in the literature, but should be included in a rigorous derivation. As it clearly contains off-shell data, it should vary under local redefinitions of the field variables, but its contributions need to vanish in any on-shell results. Then $\tilde{\M}_n (\tilde{p_i})$ in Eq. (\ref{eq:dspei}) can be expanded in  $\tau$ as
\bea
&&\tilde{\M}_n^{a_1 \cdots  b  \cdots a_n} ( \cdots,\tilde{p_i},\cdots)\non\\
&=&\tilde{\M}_n^{a_1 \cdots  b  \cdots a_n} ( \cdots,p_i,\cdots) \non\\
&&+ \tau (p_{n+1} + p_{n+2}) \cdot \left.\left[\frac{\partial}{\partial \tilde{p_i} }\tilde{M}_n^{a_1 \cdots  b  \cdots a_n} ( \cdots,\tilde{p_i},\cdots) \right]\right|_{\tilde{p_i} = p_i} + \ordr (\tau^2)\non\\
&=&  \M_n^{a_1 \cdots  b  \cdots a_n} ( \cdots,p_i,\cdots) + \tau (p_{n+1} + p_{n+2}) \cdot  \frac{\partial}{\partial p_i }\M_n^{a_1 \cdots  b  \cdots a_n} ( \cdots,p_i,\cdots) \non\\
&&+ 2\tau p_i \cdot (p_{n+1} + p_{n+2}) X^{a_1 \cdots b \cdots a_n} ( \cdots,p_i,\cdots) + \ordr (\tau^2).\label{eq:dstme}
\eea

As we would like to calculate the double soft limit up to $\ordr (\tau)$, from  Eq. (\ref{eq:dspei}) we know that we need $V_4$ up to $\ordr (\tau^2)$, which turns out to be completely fixed by the Lagrangian up to $\ordr (p^4)$. This is because for a general, off-shell 4-pt scalar vertex $V_4$ with momenta $p_1 ,\cdots, p_4$ satisfying total momentum conservation, it can only be a polynomial of 6 independent momentum invariants, for example
\bea
s_{12},\ s_{13}, \ s_{23}, \ p_1^2, \ p_2^2, \ p_3^2.
\eea
However, in $V^{a_i a_{n+1} a_{n+2} b}_4 (p_i, \tau p_{n+1}, \tau p_{n+2}, -\tilde{p_i})$ three of the momenta are on-shell, reducing the number of independent momentum invariants to 3, e.g.
\bea
\tau s_{i, n+1}, \ \tau s_{i,n+2},\ \tau^2 s_{n+1,n+2},
\eea
which are at least at $\ordr (\tau)$. The contribution of the $\ordr (p^{2n})$ Lagrangian then consists of $n$ powers of these invariants, which are at least $\ordr (\tau^n)$. Therefore, to calculate  $V_4$ in Eq. (\ref{eq:dspei}) up to $\ordr (\tau^2)$, we only need the Lagrangian up to  $\ordr (p^4)$; moreover, as the Lagrangian starts at $\ordr (p^2)$, $V_4$ in Eq. (\ref{eq:dspei}) starts at $\ordr (\tau)$, thus we only need to calculate $\tilde{\M}_n (\tilde{p_i})$ up to $\ordr (\tau)$.

From the Lagrangian  given by Eq. (\ref{eq:nlsml4}) we can then expand $V_4$ as
 \bea
&&V^{a_i a_{n+1} a_{n+2} b}_4 (p_i, \tau p_{n+1}, \tau p_{n+2}, p_I)\non\\
 &=&-\tau \frac{1}{3f^2} \left[     T^j_{a_i a_{n+1}} T^j_{a_{n+2} b} (2s_{i,n+2} - s_{i,n+1})   - T^j_{a_i a_{n+2}} T^j_{ a_{n+1} b} (2s_{i,n+1} - s_{i,{n+2}})  \right] \non\\
&&-\tau^2\left[ \frac{1}{3f^2}  \left( T^j_{a_i a_{n+1}} T^j_{a_{n+2} b} + T^j_{a_i a_{n+2}} T^j_{ a_{n+1} b} \right) s_{n+1,n+2}\right.\non\\
&&\left. +  \frac{1}{f^2 \Lambda^2} \ff^{(4),a_i a_{n+1} a_{n+2} b}  s_{i,n+1} s_{i,n+2}   \right] + \ordr (\tau^3),\label{eq:dsv4e}
\eea
where
\bea
\quad \ff^{(4),abcd}   & =&(2C_1 + C_2) \left(\delta^{ac } \delta^{bd} + \delta^{ab} \delta^{cd} \right)  +2 C_2\ \delta^{ad} \delta^{bc}   \non\\
&&+  \left( C_3 + \frac{2}{3} C_4 \right) \left( T^j_{ab } T^j_{ cd} + T^j_{ac } T^j_{bd} \right)   + 16 C_4 \gd^{abcd}
\eea
is the flavor factor associated with the $\ordr (p^4)$ contribution in the Lagrangian, and the totally symmetric flavor tensor $\gd^{abcd}$ is defined as
\bea
\gd^{a_1 a_2 a_3 a_4} = \frac{1}{3!} \sum_{\alpha \in S_3} \tr \left( \gX^{a_{\alpha (1)}} \gX^{a_{\alpha (2)}} \gX^{a_{\alpha (3)}} \gX^{a_{4}} \right).
\eea
Notice that in Eq. (\ref{eq:dsv4e}) we have used the closure condition Eq. (\ref{eq:clocon}) for the generators $T^i_{ab}$ of the flavor group.

Now that we have expanded $\M_{\pole}$ to  $\ordr (\tau)$, it is time to do the same for $\M_\gut$:
\bea
&&\M_{n+2,\ \gut}^{a_1 \cdots a_{n+2}} (p_1, \cdots , p_n, \tau p_{n+1}, \tau p_{n+2})\non\\
& = & \M_{n+2,\ \gut}^{a_1 \cdots a_{n+2}} (p_1, \cdots , p_n, 0, 0)  \non\\
&&+\tau p_{n+1} \cdot \left. \left[ \frac{\partial}{\partial p_{n+1}} \M_{n+2,\ \gut}^{a_1 \cdots a_{n+2}} (p_1, \cdots , p_n,  p_{n+1}, 0) \right] \right|_{p_{n+1} = 0}\non\\
&&+\tau p_{n+2} \cdot \left. \left[ \frac{\partial}{\partial p_{n+2}} \M_{n+2,\ \gut}^{a_1 \cdots a_{n+2}} (p_1, \cdots , p_n, 0,  p_{n+2}) \right] \right|_{p_{n+2} = 0}+ \ordr (\tau^2).\ \label{eq:gutoe} 
\eea
Apparently, we need to calculate  the leading contributions of $\M_\gut$ in both the single and the double soft limits.

Let us first consider the single soft limit with $p_{n+1} = 0$:
\bea
&&\M_{n+2,\ \gut}^{a_1 \cdots a_{n+2}}(p_1, \cdots, p_n , 0, p_{n+2})\non\\
& =&M_{n+2}^{a_1 \cdots a_{n+2}} (p_1, \cdots, p_n , 0, p_{n+2}) - \M_{n+2, \ \pole}^{a_1 \cdots a_{n+2}}(p_1, \cdots, p_n , 0, p_{n+2})\non\\
&=&- \M_{n+2, \ \pole}^{a_1 \cdots a_{n+2}}(p_1, \cdots, p_n , 0, p_{n+2}),\label{eq:gutsmpole}
\eea
where the last equality is dictated by the Adler zero condition. Eq. (\ref{eq:gutoe}) requires expanding the above to the linear order in soft $p_{n+2}$. Therefore, let us make the usual replacement of $p_{n+2} \to \tau p_{n+2}$, and expand $\tau $ around $0$; we have
\bea
&&\M_{n+2,\ \pole}^{a_1 \cdots a_{n+2}}  (p_1, \cdots, p_n, 0, \tau p_{n+2}) \non\\
&=& \sum_{i=1}^n \frac{- V^{a_i a_{n+1} a_{n+2} b}_4 (p_i, 0,  \tau p_{n+2}, -p_i - \tau p_{n+2})}{ \tau  s_{i,n+2} }  \non\\
&&\times \tilde{\M}^{a_1 \cdots a_{i-1} b a_{i+1} \cdots n} (p_1, \cdots, p_{i-1}, p_i + \tau p_{n+2}, p_{i+1},\cdots , p_n)  .
\eea
Notice that in the above, $V_4 $ must be a polynomial of  $\tau s_{i,n+2}$, which is the only non-vanishing momentum invariant left. Therefore, the contribution of  $\ordr (p^{2n})$ terms in the Lagrangian must be proportional to  $\tau^n s^{n}_{i, n+2}$. From the Lagrangian in Eq. (\ref{eq:nlsml4}) we see that the $\ordr (p^4)$ contribution actually vanishes, thus
\bea
&&V^{a_i a_{n+1} a_{n+2} b}_4 (p_i, 0,  \tau p_{n+2}, -p_i - \tau p_{n+2})\label{eq:v4at3pt}\non \\ &=&- \frac{\tau}{3f^2} \left( T^j_{a_i b} T^j_{a_{n+1} a_{n+2}}    - T^j_{a_i a_{n+1}} T^j_{a_{n+2} b} \right)  s_{i, n+2} + \ordr (\tau^3),
\eea
and Eq. (\ref{eq:gutsmpole}) becomes
\bea
&&\M_{n+2,\ \gut}^{a_1 \cdots a_{n+2}}(p_1, \cdots, p_n , 0, \tau p_{n+2}) \label{eq:gutsins1} \non\\
 &=&  \frac{1}{3f^2} \sum_{i=1}^n \left( T^j_{a_i b} T^j_{a_{n+1} a_{n+2}}    - T^j_{a_i a_{n+1}} T^j_{a_{n+2} b} \right)  \tilde{\M}_n^{a_1 \cdots  b  \cdots a_n} ( \cdots, p_i + \tau p_{n+2},\cdots) + \ordr (\tau^2) ,\non\\
 \eea
where the expansion of $\tilde{\M}$ in terms of $\tau p_{n+2}$ in the above is given by Eq. (\ref{eq:dstme}). Specifically, if we then set $\tau = 0$ in the above, we get
\bea
\M_{n+2,\ \gut}^{a_1 \cdots a_{n+2}} (p_1, \cdots, p_n , 0, 0)=  -\frac{1}{3f^2} \sum_{i=1}^n  T^j_{a_i a_{n+2}} T^j_{a_{n+1} b}  \M_n^{a_1 \cdots  b  \cdots a_n} (p_1, \cdots, p_n),\label{eq:gut00}
\eea
where again we have used the closure condition in Eq. (\ref{eq:clocon}), as well as the flavor symmetry of the full on-shell amplitude:
\bea
\sum_{i=1}^n  T^j_{a_i b}  \M_n^{a_1 \cdots  b  \cdots a_n} ( \cdots, p_i ,\cdots) = 0.\label{eq:onslw}
\eea

Similarly, we have
\bea
&&\M_{n+2,\ \gut}^{a_1 \cdots a_{n+2}}(p_1, \cdots, p_n ,  \tau p_{n+1}, 0)= - \M_{n+2,\ \pole}^{a_1 \cdots a_{n+2}}(p_1, \cdots, p_n ,  \tau p_{n+1}, 0)\non\\
 &=&  \frac{1}{3f^2} \sum_{i=1}^n \left( T^j_{a_i b} T^j_{a_{n+2} a_{n+1}}    - T^j_{a_i a_{n+2}} T^j_{a_{n+1} b} \right)  \tilde{\M}_n^{a_1 \cdots  b  \cdots a_n} ( \cdots, p_i + \tau p_{n+1},\cdots) + \ordr(\tau^2),\non\\
 \label{eq:gutsins2} 
 \eea
and if we set $\tau = 0$ in the above, we get back to Eq. (\ref{eq:gut00}). Then putting everything together into Eq. (\ref{eq:gutoe}) we have
\bea
&&\M_{n+2,\ \gut}^{a_1 \cdots a_{n+2}} (p_1, \cdots , p_n, \tau p_{n+1}, \tau p_{n+2})\non\\
& = & -\frac{1}{3f^2} \sum_{i=1}^n \left\{ T^j_{a_i a_{n+2}} T^j_{a_{n+1} b}  \M_n^{a_1 \cdots  b  \cdots a_n} (p_1, \cdots, p_n) \right.\non\\
&&-\tau\left[    \left( T^j_{a_i b} T^j_{a_{n+2} a_{n+1}}    - T^j_{a_i a_{n+2}} T^j_{a_{n+1} b} \right) \right.  p_{n+1} \cdot  \frac{\partial}{\partial p_{i}} \M_n^{a_1 \cdots  b  \cdots a_n} ( p_1, \cdots, p_n)  \non\\
&&+     \left( T^j_{a_i b} T^j_{a_{n+1} a_{n+2}}    - T^j_{a_i a_{n+1}} T^j_{a_{n+2} b} \right)  p_{n+2} \cdot  \frac{\partial}{\partial p_{i}} \M_n^{a_1 \cdots  b  \cdots a_n} (p_1, \cdots, p_n)\non\\
 && +     \left( T^j_{a_i b} T^j_{a_{n+2} a_{n+1}}    - T^j_{a_i a_{n+2}} T^j_{a_{n+1} b} \right)  s_{i, n+1} X^{a_1 \cdots b \cdots a_n} (p_1, \cdots, p_n)\non\\
  && \left. \left.+     \left( T^j_{a_i b} T^j_{a_{n+1} a_{n+2}}    - T^j_{a_i a_{n+1}} T^j_{a_{n+2} b} \right)  s_{i, n+2} X^{a_1 \cdots b \cdots a_n} (p_1, \cdots, p_n) \right]\right\}\non\\
  &&+ \ordr (\tau^2).\label{eq:finalgut}
\eea

Notice that  the result of $\M_\gut$ in Eq. (\ref{eq:finalgut}) is exactly the same as in  Ref. \cite{Low:2015ogb}, apart from pieces containing the unknown function $X$, although Ref. \cite{Low:2015ogb} only considered the $\ordr (p^2)$ Lagrangian while the result here is valid in all orders of the derivative expansion.  This is because  the only direct input from the Lagrangian needed in the above calculations is for the 4-pt vertices like in Eq. (\ref{eq:v4at3pt}), where the  $\ordr (p^4)$ contributions vanishes, while terms beyond  $\ordr (p^4)$ are included in the $\ordr (\tau^2)$ terms in Eqs. (\ref{eq:gutsins1}) and (\ref{eq:gutsins2}), which are eliminated when being plugged into Eq. (\ref{eq:gutoe}).

Now we are ready to calculate the full amplitude up to $\ordr (\tau)$:
\bea
&&\M^{a_1 \cdots a_{n+2}}_{n+2} (p_1, \cdots , p_n, \tau p_{n+1}, \tau p_{n+2}) \non \\
&=& \M_{n+2, \ \pole}^{a_1 \cdots a_{n+2}}  (p_1, \cdots, p_n, \tau p_{n+1}, \tau p_{n+2})  + \M_{n+2,\ \gut}^{a_1 \cdots a_{n+2}}  (p_1, \cdots, p_n, \tau p_{n+1}, \tau p_{n+2}),
\eea
thus the double soft theorem is
\bea
&&\M^{a_1 \cdots a_{n+2}}_{n+2} (p_1, \cdots , p_n, \tau p_{n+1}, \tau p_{n+2}) \\
&=& - \frac{1}{2f^2} \sum_{i=1}^n  \frac{1}{s_{i,n+1} + s_{i,n+2} + \tau s_{n+1,n+2}}    \left\{     T^j_{a_i b } T^j_{a_{n+1} a_{n+2} } (s_{i,n+2} - s_{i,n+1})  \phantom{\frac{2}{ \Lambda^2}}\right. \non\\
&& + \tau\left[ \left( T^j_{a_i a_{n+1}} T^j_{a_{n+2} b} + T^j_{a_i a_{n+2}} T^j_{ a_{n+1} b} \right) s_{n+1,n+2}   -  \frac{2}{ \Lambda^2} \ff^{(4),a_i a_{n+1} a_{n+2} b}  s_{i,n+1} s_{i,n+2}  \right.  \non\\
&& \left.\left. \phantom{\frac{2}{ \Lambda^2}}+4 T^j_{a_i b} T^j_{a_{n+1} a_{n+2}}    p_{n+1}^\nu p_{n+2}^\mu J_{i,\mu \nu} \right] \right\} \M_n^{a_1 \cdots b \cdots a_n} (p_1, \cdots, p_n)+ \ordr (\tau^2),\label{eq:ds}
\eea
where
\bea
J_{i, \mu \nu} = p_{i,\mu} \frac{\partial}{ \partial p_i^\nu} - p_{i,\nu} \frac{\partial}{ \partial p_i^\mu}
\eea
is the angular momentum operator for external state $i$. Notice that the off-shell $X$ terms are cancelled nicely. The only difference between the $\ordr (p^2)$ result in Ref. \cite{Low:2015ogb} and our all-order result is the $\ordr (\tau)$ contribution associated with $\ff^{(4)}$. In other words, the $\ordr (p^2)$ Lagrangian completely fixes the $\ordr (\tau^0)$ double soft theorem, while the $\ordr (\tau)$ soft factor is fixed by the Lagrangian up to $\ordr (p^4)$. The soft theorem is unaffected by the possible WZW term in $4$ spacetime dimensions, as its interaction starts at $5$-pt while the only input we need here are $4$-pt vertices.

\section{Conclusion}\label{sec6}
In this work we have explored NLSM amplitudes beyond leading order in the derivative expansion, as well as the usual flavor structure of $\SU (N) \times \SU (N) / \SU (N)$. We have demonstrated  the universality of the trace ordering for a general group representation up to $\ordr (p^4)$. This enables us to reinterpret the extended theory $\nlsm + \phi^3$, in the previously known $\ordr (p^2)$ single soft theorem of Eq. (\ref{eq:sssstr}), as a more general theory $\nlsm + \phi + \psi$.

In particular, for NGB's furnishing $\bN$ of $\SO (N)$, the alternative flavor decomposition of the pair basis is convenient, and we derive the corresponding single soft theorem of $\nlsm^{(2)}$ in Eq. (\ref{eq:slsspb}) where the extended theory is presented in this basis as well. To achieve this, we have studied the pair basis in detail, uncovering novel amplitude relations given in Eqs. (\ref{eq:pstr}), (\ref{eq:pasf}) and (\ref{eq:ptret}), as well as the CHY formula Eq. (\ref{eq:chypb}) for the pair basis $\nlsm^{(2)}$ amplitudes.

As the pair basis can be regarded as a multi-trace basis where each trace only contains two external states, it appears in amplitudes of other theories as well, such as the multi-flavor DBI scalars as well as the YM scalar theory of a single flavor but multiple colors. There may be connections between these theories and the $\nlsm^{(2)}$ for $\bN$ of $\SO (N)$, such as in the web of theories of Ref. \cite{Cheung:2017ems}. Amplitude relations for these theories are also worth exploring. Moreover, the DBI scalar amplitudes start at $\ordr (\tau^2)$ in the single soft limit, which can be seen as a special case of the $\dbi + \nlsm$ soft theorem given by Eq. (\ref{eq:stdn}). The leading non-vanishing terms at $\ordr (\tau^2)$ have been calculated using the Ward identity \cite{Yin:2018hht}, and it remains to be seen whether the double copy structure of the DBI amplitudes leads to extended theories in this context.

Using the Ward identity corresponding to the shift symmetry, we have also extended the subleading single soft theorem to $\ordr (p^4)$, given by Eqs. (\ref{eq:ssp4st}) and (\ref{eq:ssp4dt}). We found that for at least two specific choices of the Wilson coefficients for the $\ordr (p^4)$ operators, which result in $\nlsm^{d_2}$ and $\nlsm^{C_{3'}}$, the single soft theorems can be interpreted as generating extended theory amplitudes, similar to the $\nlsm^{(2)}$. For the $\nlsm^{d_2}$ case, as seen in Eq. (\ref{eq:sttd2}), this fact is expected as the theory is known to have a double copy construction in terms of $\nlsm^{(2)}$, but the emergence of $C_{3'} + \phi^3$ in Eq. (\ref{eq:sstc3p}) can only be understood from the perspective of the Ward identity. This may give us a handle to explore the possible double copy structure or CHY representation of $\nlsm^{C_{3'}}$.

Finally, we computed the general double soft theorem to the subleading order of the soft expansion, which is valid to all orders in the EFT expansion. As shown in Eq. (\ref{eq:ds}), the only higher order corrections compared to the known $\nlsm^{(2)}$ results enter in the subleading soft order, and only $\ordr (p^4)$ operators give non-vanishing contributions. Similar situations, where only a limited number of EFT operators modify soft theorems, are known in gauge theory and gravity \cite{Elvang:2016qvq}. It would be interesting to see how these patterns extend to higher orders in the soft expansion, as well as the multi-soft limits studied in Refs. \cite{Du:2016njc,Low:2018acv}. The investigation of the multi-soft limits has previously relied heavily on manipulating Feynman diagrams in the Cayley parameterization, which is only available for the adjoint $\U (N)$ $\nlsm$ \cite{Kampf:2013vha}. Although the resulting amplitudes in the trace basis are universal, it would still be beneficial to develop other techniques that are more physically intuitive.

We have seen clearly that the higher derivative operators at $\ordr (p^4)$ generate amplitudes with properties similar to what we have seen in $\nlsm^{(2)}$, but only for different specific choices of the Wilson coefficients. These include:
\begin{itemize}
\item A double copy construction and a CHY formula for $\nlsm^{d_2}$.
\item The same subleading single soft theorem in $\nlsm^{C_{3'}}$ and $\nlsm^{d_2}$ as in $\nlsm^{(2)}$, with emergent extended theories.
\item The double soft theorem of $\nlsm^{(2)}$ does not change when we only add the WZW term.
\item The double soft theorem, after flavor ordering and with two adjacent legs taken to be soft, is not modified by higher derivative operators if $C_3 = -3C_4$, $C_1 = C_2 = 0$.
\item The KK relation is satisfied when $C_1 = C_2 = C_4 = C_- = 0$.
\end{itemize}
On the other hand, no combination of the Wilson coefficients generates  amplitudes that satisfy BCJ relations or appear in the Z-theory. It is evident that the higher derivative amplitudes in NLSM exhibit a wide variety of behaviors, and is an ideal testing ground to study the origin of the numerous properties known in $\nlsm^{(2)}$ as well as in gauge theory and gravity. It is also well-known that causality enforces positivity constraints on the Wilson coefficients of the chiral Lagrangian \cite{Pham:1985cr,Adams:2006sv}. This suggests a potential positive geometry which exists in the EFT expansion of NLSM that may be worth investigating as well, using the approach recently developed in \cite{Arkani-Hamed:2020blm}, including for the closely related Z-theory \cite{Huang:2020nqy}.

As for the WZW term, which is not featured prominently in this work, although it does not modify the double soft theorem of $\nlsm^{(2)}$ at all, it definitely will contribute to the subleading single soft theorem. Calculating its contribution using the Ward identity would be straightforward, but the result is more useful if we have a closed-form expression of the WZW operator at the beginning.

For applications to phenomenological models such as the chiral perturbation theory or the composite Higgs models, more features need to be added to the idealized NLSM, including explicit symmetry breaking which gives the pseudo NGB's masses, as well as interactions with fermions and vector bosons. As long as the explicit symmetry breaking is soft, the Ward identity, which is our main tool for deriving the soft theorems, will still hold at the leading order. It is certainly desirable to have purely on-shell techniques such as the CHY formalism generalized to these realistic cases as well. Further investigations in this direction are left for future work.

Lastly, there is the question of loop corrections. Generically, these can modify the soft theorems, whether at integrand or integrated level \cite{Bern:2014oka,He:2014bga,Bianchi:2014gla}, and it would be interesting to also explore these issues in the context of the NLSM.

\begin{acknowledgments}
The authors would like to thank Ian Low for collaborations at the early stage of this project, as well as on other closely related projects. The authors also thank Mattias Sj\"{o} for providing expressions for high-point NLSM amplitudes, and John Joseph M. Carrasco for enlightening discussions.  ZY would like to thank Henrik Johansson and Jaroslav Trnka for useful discussions. ZY is supported in part by the Knut and Alice Wallenberg Foundation under grants KAW 2018.0116 and KAW 2018.0162. LR is supported by Taiwan Ministry of Science and Technology Grant No. 109-2811-M-002-523.

\end{acknowledgments}

\begin{appendix}
\section{Derivation details}
\label{app:der}

\subsection{Pair basis of the extended theory $\nlsm + \phi + \psi$}
\label{app:rpse}
In this section all the amplitudes discussed are for the theory of $\nlsm + \phi + \psi$, so that the superscript for $M^{\nlsm + \phi + \psi}$ is omitted. Consider the amplitude $M_{n} (1,2,3|4,5| \cdots |n-1,n || 1^\psi, 2^\phi, 4^\psi)$, which has the following coefficient in the flavor decomposition of the pair basis:
\bea
T^{j_2}_{a_1 a_{3}} \left[ \prod_{l=2}^{(n-1)/2} \delta^{a_{2l} a_{2l+1}}  \right]T^{\tdj_2}_{\ta_1 \ta_{4}}.\label{eq:etff1}
\eea
Note that for the left ordering of such an amplitude, shuffling the positions of the traces still leaves the amplitude invariant--this include the three component trace $\{1, 2,3 \}$. The left and the right three component traces are separately invariant under cyclic permutations, but generate a minus sign when two indices are exchanged in one of the two traces.

\begin{figure}[t]
\centering
\includegraphics[width=\textwidth]{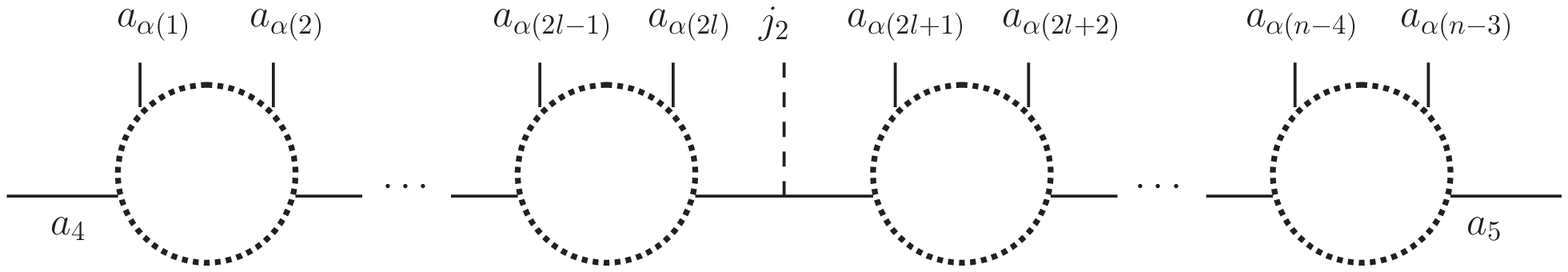}
\caption{\label{fig:etf1} The first kind of flavor factor in Eq. (\ref{eq:etddm}), which does not generate $\delta^{a_4 a_{5}}$. As explained in Fig. (\ref{fig:comrel}), each of the dotted circles can contain either an ``$=$'' or a ``$\times$''.}
\end{figure}

\begin{figure}[t]
\centering
\includegraphics[width=0.9\textwidth]{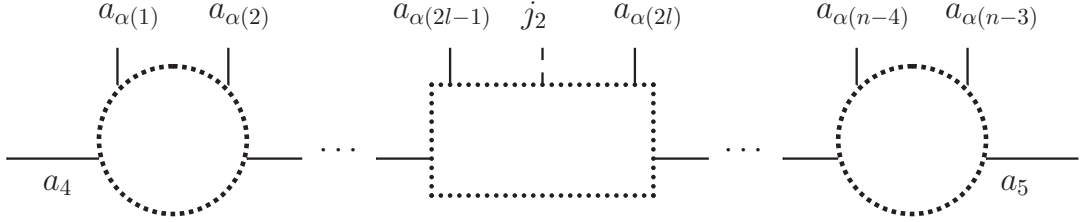}
\caption{\label{fig:etf2} The second kind of flavor factor in Eq. (\ref{eq:etddm}), where the structure inside the dotted box is given by Fig. \ref{fig:comr2}.}
\end{figure}

\begin{figure}[t]
\centering
\bea
&&\begin{minipage}[c]{0.33\textwidth}
\begin{center}
\includegraphics[height=2.7cm]{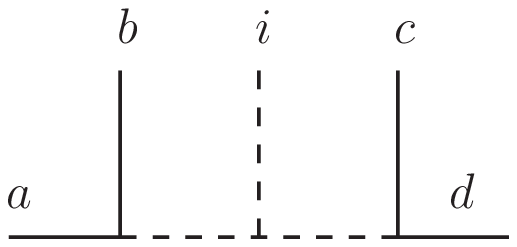} 
\end{center}
\end{minipage} \non\\
&=& \frac{1}{2} \left( \quad \begin{minipage}[c]{0.33\textwidth}
\begin{center}
\includegraphics[height=3cm]{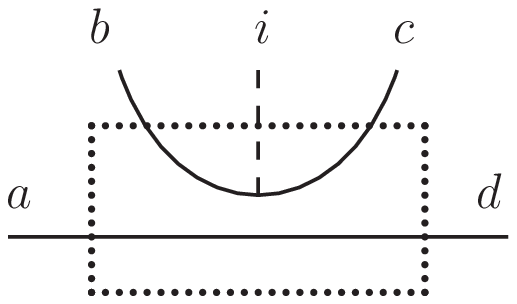} 
\end{center}
\end{minipage} \quad+ \qquad \begin{minipage}[c]{0.33\textwidth}
\begin{center}
\includegraphics[height=3cm]{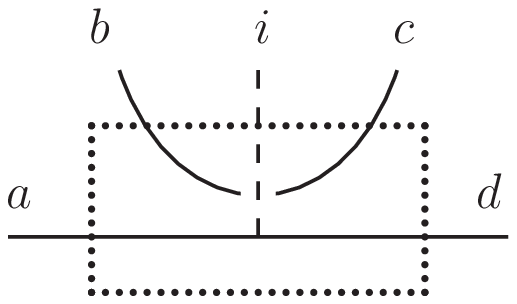} 
\end{center}
\end{minipage}\right.\non\\
&&\left.   - \begin{minipage}[c]{0.33\textwidth}
\begin{center}
\includegraphics[height=3cm]{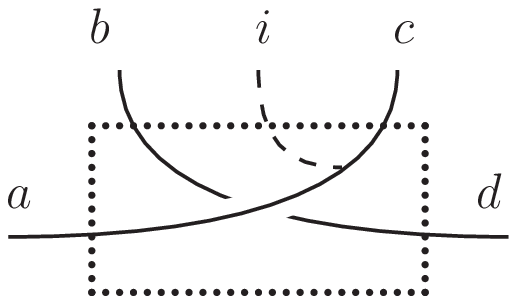} 
\end{center}
\end{minipage}\qquad -  \begin{minipage}[c]{0.33\textwidth}
\begin{center}
\includegraphics[height=3cm]{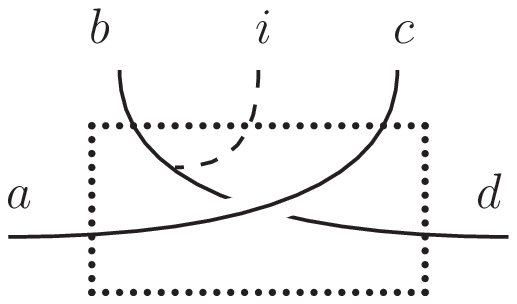} 
\end{center}
\end{minipage}\qquad \right) \non
\eea
\caption{\label{fig:comr2} Graphical representation of Eq. (\ref{eq:comr2}). Only the first term on the RHS generates $\delta^{ad}$.}
\end{figure}

To express $M_{n} (1,2,3|4,5| \cdots |n-1,n || 1^\psi, 2^\phi, 4^\psi)$ as a sum of single-trace amplitudes, we decompose the full amplitude in the DDM basis, setting $4$ and $5$ as the two ends of the half-ladders:
\bea
&&{\cal M}_n^{a_1 j_2 a_3 \cdots  a_{n} ; \ta_1 \ta_{4} \tdj_2 } (p_1, \cdots, p_n)\non\\
&=& \sum_{\alpha \in S_{n-3}}  T^{i_1}_{a_4 a_{\alpha (1)}  } \left( \prod_{m=1}^{(n-5)/2} T^{i_{2m}}_{ a_{ \alpha (2m)} b_{2m-1}} T^{i_{2m+1}}_{b_{2m} a_{\alpha (2m+1)} } \right)  T^{i_{n-3}}_{ a_{\alpha (n-3)} a_{5}}\non\\
&&\times \left[ \left( \prod_{ m=1 }^{(n-3)/2} \delta^{i_{2m-1} i_{2m}}  \right) \sum_{l=1}^{(n-5)/2}  \left( \prod_{\substack{m=1 \\ m \ne l}}^{(n-5)/2} \delta^{b_{2m-1} b_{2m}} \right) T^{j_2}_{b_{2l-1} b_{2l}} \right.  \non\\
&&\left. + \left( \prod_{ m=1 }^{(n-5)/2} \delta^{b_{2m-1} b_{2m}}  \right) \sum_{l=1}^{(n-3)/2}  \left( \prod_{\substack{m=1 \\ m \ne l}}^{(n-3)/2} \delta^{i_{2m-1} i_{2m}} \right) i f^{i_{2l-1} j_2 i_{2l}}  \right]\non\\
&&\times   T^{\tdj_2}_{\ta_1 \ta_{4}}  M_n (4,\alpha, 5 ||1^\psi,2^\phi,4^\psi),\label{eq:etddm}
\eea
where $\alpha$ is a permutation of $\{1,2, \cdots, n\} \setminus \{4,5 \}$. As demonstrated in above equation, there are two kinds of flavor factors for the left group $\SO (N)$, corresponding to the graphs in Figs. \ref{fig:etf1} and \ref{fig:etf2}, respectively. It turns out that the first kind cannot contribute to Eq. (\ref{eq:etff1}), because it cannot generate $\delta^{a_4 a_{5}}$. For the second kind, we need to work out what happens in the dotted box in Fig. \ref{fig:etf2}: we have
\bea
T^j_{ab} \ if^{jik} T^k_{cd} = T^j_{ab} [T^j, T^i]_{cd} = \frac{1}{2} \left( \delta^{ad} T^i_{bc} + \delta^{bc} T^i_{ad} - \delta^{bd} T^i_{ac} - \delta^{ac} T^i_{bd} \right),\label{eq:comr2}
\eea
which is demonstrated in Fig. \ref{fig:comr2}.

Therefore, the contribution to Eq. (\ref{eq:etff1}) is a sum of all single-trace amplitudes where in the left ordering, $4$ and $5$ are fixed at two ends, all other pairs $\ \{6,7\}, \ \cdots ,  \ \{n-1, n\}$ are adjacent, while $1$ and $3$ sandwich $2$. The only complication is that there is an extra minus sign for terms with $\{3, 2, 1\}$  instead of $\{1, 2, 3\}$. We have
\bea
 M_{n} (1,2,3|4,5| \cdots|n-1,n || 1^\psi, 2^\phi, 4^\psi) = \frac{1}{2^{(n-3)/2}} \sum_{\alpha \in \Ar'_{n-2}} \sgn (\alpha) M_n (4, \alpha, 5||1, 2, 4),\label{eq:cr1}
\eea
where $\Ar'_{n-2}$ is for any permutation of $\{ 1, 2, \cdots, n\} \setminus \{4, 5 \}$ where all pairs $\{6,7\}, \ \cdots ,  \ \{n-1, n\}$ are adjacent and also contains the sequence $\{1,2,3\}$ or $\{3,2,1\}$. The signature of $\alpha$ is defined as
\bea
\sgn (\cdots, 1,2,3, \cdots ) =1,\qquad \sgn (\cdots, 3,2,1, \cdots )= -1.
\eea
Now, as the left ordering in the RHS of Eq. (\ref{eq:cr1}) is a single trace with $n$ labels, a reversal of the left ordering will give a factor $(-1)^n = -1$. Then it is very easy to see that Eq. (\ref{eq:cr1}) is equivalent to
\bea
M_{n} (1,2,3|4,5| \cdots|n-1,n || 1^\psi, 2^\phi, 4^\psi)  = -\frac{1}{2^{(n-3)/2}} \sum_{\alpha \in \Ar_{n-3} } M_n (1,\alpha,3, 2 ||1^\psi, 2^\phi, 4^\psi).
\eea
Similarly, we have
\bea
M_{n}  (1,2,3|4,5| \cdots|n-1,n || 1^\psi, 2^\phi,3^\psi)  = -\frac{1}{2^{(n-3)/2}} \sum_{\alpha \in \Ar_{n-3} } M_n (1,\alpha,3, 2 ||1^\psi,2^\phi, 3^\psi).
\eea
For example,
\bea
M_{5} (123|45|67 || 1^\psi  2^\phi  i^\psi)  &=& -\frac{1}{4} \left[ M_5 (1456732 ||12i) + M_5 (1546732 ||12i) \right.\non\\
&&+M_5 (1457632 ||12i) + M_5 (1547632 ||12i)\non\\
&&+M_5 (1674532 ||12i) + M_5 (1764532 ||12i)\non\\
&&\left.+M_5 (1675432 ||12i) + M_5 (1765432 ||12i)\right],
\eea
where we have omitted the labels $\phi$ and $\psi$ on the RHS, and $i \ne 1 $ or $2$.

\subsection{Useful relations involving the NGB field operators}

\label{app:eom}

Here we briefly prove relations used in this work that involve the NGB field operator $\pi$. The building blocks of the NLSM Lagrangian, $d_\mu$ and $E_\mu$, come from the Maurer-Cartan 1-form
\bea
-i \xi^\dagger \partial_\mu \xi = d_\mu^a \gX^a + E_\mu^i \gT^i = d_\mu + E_\mu,
\eea
where
\bea
\xi (\pi) = e^{i \Pi},\qquad \Pi \equiv \frac{\pi^a \gX^a}{f}.
\eea
The Lie algebra of a symmetric coset, given by Eq. (\ref{eq:crg}), leads to an automorphism $\text{Aut}$ under which the broken generators change sign, namely
\bea
\text{Aut} (\gX^a) = -\gX^a.\label{eq:aut}
\eea
As $\text{Aut} (\xi) = \xi^\dagger$, we have 
\bea
d_\mu &=& \frac{1}{2}[-i \xi^\dagger \partial_\mu \xi + i \text{Aut}(\xi^\dagger \partial_\mu \xi)] =-\frac{i}{2} [\xi^\dagger \partial_\mu \xi -\xi \partial_\mu \xi^\dagger ],\label{eq:gdefd}\\
E_\mu &=&-\frac{i}{2} [\xi^\dagger \partial_\mu \xi + \xi \partial_\mu \xi^\dagger ],\label{eq:gdefe}
\eea
while the covariant derivative is given by $\nabla_{\mu} d_{\nu} = \partial_\mu d_\nu + i[E_\mu, d_\nu]$, thus
\bea
\nabla_{[\mu} d_{\nu] } &=& 0 \label{eq:covd0}
\eea
and
\bea
\nabla_\mu d^\mu = -\frac{i}{2} \left[ \partial_\mu \left( \xi^\dagger \partial^\mu U \xi^\dagger \right) + \partial_\mu \xi^\dagger U  \partial^\mu \xi^\dagger - \partial_\mu \xi U^\dagger  \partial^\mu \xi \right].
\eea

Now let us work out the EoM of the NLSM \cite{Panico:2015jxa}. From the completely fixed form of $\lag^{(2)}$ given in Eq. (\ref{eq:nlsml4}), we calculate the EoM as
\bea
D_\eom^{a,(2)} = \frac{\delta \lag^{(2)}}{\delta \pi^a}  - \partial_\nu  \frac{\delta \lag^{(2)}}{\delta \partial_\nu \pi^a}  = \ordr (\partial^4).\label{eq:l2eom}
\eea
An important observation is
\bea
\partial_\mu F(\Pi) = \frac{\delta F(\Pi) }{\delta \pi^a} \partial_\mu \pi^a,\label{eq:varf1}
\eea
where  $F(\Pi)$ can be any function of $\Pi$. The above implies
\bea
\frac{\delta \partial_\mu F(\Pi)}{\delta \partial_\nu \pi^a} = \delta^\nu_{\mu} \frac{\delta F(\Pi) }{\delta \pi^a}.\label{eq:varf2}
\eea
Combining Eqs. (\ref{eq:l2eom}), (\ref{eq:varf1}) and (\ref{eq:varf2}), one can show that
\bea
\tr \left( \frac{ \delta d_\nu}{ \delta \partial_\nu \pi^a} \nabla_\mu d^\mu \right) = \frac{1}{f^2} D_\eom^{a,(2)} = \ordr (\partial^4).
\eea
The explicit form of $d_\mu$, given by Eq. (\ref{eq:dexpo}), tells us that
\bea
\tr \left( \frac{ \delta d_\nu}{ \delta \partial_\nu \pi^a} \nabla_\mu d^\mu \right) = \frac{1}{f} [ F_1 (\mt) ]_{ a b} (\nabla_\mu d^\mu)^b.
\eea
We know that $F_1(\mt)$ is invertible, thus
\bea
(\nabla_\mu d^\mu)^a = f \left\{ [ F_1(\mt) ]^{-1} \right\}_{ab} \tr \left\{ \frac{ \delta d_\nu}{ \delta \partial_\nu \pi^b} \nabla_\mu d^\mu \right\} = \ordr (\partial^4).\label{eq:eom}
\eea

\subsection{Preparing the current for the theory $C_{3'}+\phi^3$}
\label{app:c3pfc}

The current associated with the coefficient $C_{3'}$, as we originally derive it in Eq. (\ref{eq:c3pc}), is
\bea
\left(\J^{(4),3'}\right)_\mu^a = \frac{if}{4\Lambda^2}   \tr \left( \left\{ \gX^a, U \right\} \partial_\nu U^\dagger \partial_\mu U \partial^\nu U^\dagger \right) 
= \frac{4f}{\Lambda^2  } \tr \left(   \gX^a  \partial_\nu \Pi \partial_\mu \Pi \partial^\nu \Pi \right) + \cdots.
\eea
We need to rewrite the above so that the resulting vertices of the extended theory $C_{3'}+\phi^3$ satisfy the correct symmetry condition of Eq. (\ref{eq:c3pcst}). As we see from the above, the current is neatly written in terms of $U$. We will need to use the EoM given in Eq. (\ref{eq:eom}), so it would be convenient to rewrite it using $U$ as well: one can show that
\bea
 \frac{i}{2} \partial_\mu \left(\partial^\mu U^\dagger  U \right)  = -\frac{i}{2} \partial_\mu \left( U^\dagger \partial^\mu U \right) = \xi^\dagger \nabla_\mu d^\mu \xi = \ordr (\partial^4).\label{eq:eomu1}
\eea
The existence of the automorphism given by Eq. (\ref{eq:aut}) implies that
\bea
\frac{i}{2} \partial_\mu \left(\partial^\mu U  U^\dagger \right)  = -\frac{i}{2} \partial_\mu \left( U \partial^\mu U^\dagger \right)  = \ordr (\partial^4)\label{eq:eomu2}
\eea
as well.

For the rest of this section, whenever we rewrite the current, we freely drop any terms involving the form of the EoM as in Eqs. (\ref{eq:eomu1}) and (\ref{eq:eomu2}), as well as any terms that are total derivatives, because  the former lead to terms of higher orders in the derivative expansion, while the latter only contribute $\ordr (\tau^2)$ effects in the soft expansion.

Let us take the inspiration from the 3-pt vertex. The manipulation on the current corresponding to Eq. (\ref{eq:c3p3om}) is
\bea
\tr \left(   \gX^a  \partial_\nu \Pi \partial_\mu \Pi \partial^\nu \Pi \right) &=&- \tr \left(   \gX^a  \partial_\nu \Pi \partial_\mu \Pi \partial^\nu \Pi + \gX^a   \Pi  \partial_\mu \partial_\nu \Pi \partial^\nu \Pi \right.\non\\
&& \left. +\gX^a  \partial_\nu \Pi \partial_\mu \partial^\nu \Pi \Pi  + \gX^a   \Pi  \partial_\mu  \Pi \DAl \Pi + \gX^a  \DAl \Pi \partial_\mu  \Pi \Pi \right), \ \ 
\eea
where the terms with $\DAl \Pi$ can be converted to higher-pt vertices using the EoM.

Similarly, we can use the following relations:
\bea
 U \partial_\nu U^\dagger \partial_\mu U \partial^\nu U^\dagger = \partial_\nu U \partial_\mu U^\dagger  U \partial^\nu U^\dagger &=& - U \partial_\mu \partial_\nu U^\dagger  U \partial^\nu U^\dagger,\\
 \partial_\mu \partial_\nu U^\dagger  U \partial^\nu U^\dagger &=& 0,\\
U\partial_\nu U^\dagger \partial_\mu U \partial^\nu U^\dagger + U\partial_\nu U^\dagger \partial_\mu \partial^\nu U  U^\dagger &=& 0,
\eea
to rewrite the current as
\bea
\left(\J^{(4),3'}\right)_\mu^a &=& -\frac{if}{4\Lambda^2}   \tr \left\{  \gX^a \left[ (U-1) \partial_\mu \partial_\nu U^\dagger  U \partial^\nu U^\dagger    +U\partial_\nu U^\dagger \partial_\mu U \partial^\nu U^\dagger \right.\right.\non\\
&&\left.\left.+ U\partial_\nu U^\dagger \partial_\mu \partial^\nu U  (U^\dagger -1) \right] - U \leftrightarrow U^\dagger \right\}\non\\
&=& -\frac{f}{\Lambda^2} \sum_{l_1 = 0}^{2n-2} \sum_{l_2=1}^{2n-1-l_1} \sum_{l_3=1}^{2n-l_1 - l_2} \frac{(-4)^n (-1)^{l_1 + l_3}}{l_1! l_2 ! l_3 ! (2n+1 - l_1 - l_2 - l_3)! }\non\\
&&\times \tr \left[ \Pi^{l_3} \partial_\mu \partial_\nu \Pi^{l_2} \Pi^{l_1} \partial^\nu \Pi^{2n+1 - l_1 - l_2 - l_3} \right.\non\\
&& \left.+ \Pi^{l_1}  \partial_\nu \Pi^{l_2} \partial^\nu \left( \partial_\mu \Pi^{l_3}  \Pi^{2n+1 - l_1 - l_2 - l_3}  \right) \right],\label{eq:c3pcr}
\eea
which generates the vertices given by Eq. (\ref{eq:v3pm}).

\section{Examples in the single soft theorems}
\label{app:lpv}

Here we give low-pt examples of various quantities appearing in the single soft theorems.

\subsection{The NLSM}
\label{app:nlv}

The flavor-ordered 4-pt vertices of NLSM in the general trace basis, up to $\ordr (p^4)$:
\bea
V (1,2,3,4) &=& V^{(2)} (1,2,3,4) +\frac{2}{f^2 \Lambda^2} \left[ 2 C_{3' } p_1 \cdot p_3\ p_2 \cdot p_4 \right.\non\\
&&\left. + C_{4'} \left(p_1 \cdot p_2\ p_3 \cdot p_4 + p_1 \cdot p_4 \ p_2 \cdot p_3 \right)  \right] + \ordr \left(\frac{1}{\Lambda^4} \right),\\
V (1,2|3,4) &=& \frac{4}{f^2 \Lambda^2} \left[ 2C_1 p_1 \cdot p_2 \ p_3 \cdot p_4 \right.\non\\
&&\left.+ C_2 ( p_1 \cdot p_3 \ p_2 \cdot p_4+ p_1 \cdot p_4 \ p_2 \cdot p_3 ) \right] + \ordr \left(\frac{1}{\Lambda^4} \right),
\eea
with the $\ordr (p^2)$ vertex given by
\bea
V^{(2)} (1,2,3,4) = \frac{1}{f^2} \left[ s_{12} + s_{23} - \frac{2}{3} \left(p_1^2 + p_2^2 + p_3^2 + p_4^2\right) \right].
\eea

The 4-pt vertex in the pair basis for $\nlsm^{(2)}$:
\bea
V^{(2)} (1,2|3,4) = \frac{1}{6f^2} \left[2 p_1 \cdot p_2 + 2 p_3 \cdot p_4 + (p_1 + p_2)^2\right].
\eea

The 4-pt amplitudes in the trace basis up to $\ordr (p^4)$, with $p_4$ as the soft momentum, are
\bea
M_4(1,2,3,4) &=&  -\frac{\tau}{f^2} s_{24} + \ordr (\tau^2)  = \frac{\tau}{\lambda f^2} s_{24} M^{\nlsm + \phi^3}_3 (1,2,3||1,3,2) + \ordr (\tau^2),\label{eq:nlsmst4}\\
M_4(1,2|3,4) &=&   \ordr (\tau^2),
\eea
with
\bea
M^{\nlsm + \phi^3}_3 (1,2,3||1,3,2) =  -\lambda,
\eea
given by the cubic vertex in Eq. (\ref{eq:etp3}).

The 4-pt amplitude in the pair basis for $\nlsm^{(2)}$, with $p_4$ as the soft momentum:
\bea
&&M_4^{(2)} (1,2|3,4)  =  \frac{\tau}{2f^2} s_{34} + \ordr (\tau^2)  \non\\
&=& -\frac{\tau}{2\lambda f^2} \left[ s_{14} M^{\nlsm + \phi^3}_3 (3,2,1||3^\psi,2^\phi,1^\psi) + s_{24} M^{\nlsm + \phi^3}_3 (3,1,2||3^\psi,1^\phi,2^\psi) \right] + \ordr (\tau^2).\non\\
\eea

The 6-pt amplitude in the pair basis for $\nlsm^{(2)}$, with $p_6$ as the soft momentum:
\bea
M_6^{(2)} (61|23|45) &=&- \frac{1}{4} \left[s_{23} s_{16} \left( \frac{1}{P^2_{235}} + \frac{1}{P^2_{234}}\right)  + s_{23} s_{45} \left( \frac{1}{P^2_{236}} + \frac{1}{P^2_{123}}\right)\right.\non\\
&&\left. + s_{45} s_{16} \left( \frac{1}{P^2_{245}} + \frac{1}{P^2_{345}}\right)  -s_{23} - s_{45} - s_{16} \right]\non\\
&=&-\frac{\tau}{2\lambda f^2 }  \left\{ s_{26} \left[ M_5 (132|45||1 3 2) +M_5 (145|23||1 4 2) + M_5 (154|23||1 5 2) \right]\right.\non\\
&&+ s_{36} \left[ M_5 (123|45||1 2 3) +M_5 (145|23||1 4 3) + M_5 (154|23||1 5 3) \right]\non\\
&&+s_{46} \left[ M_5 (123|45||1 2 4) +M_5 (132|45||1 3 4) + M_5 (154|23||1 5 4) \right]\non\\
&&\left.+ s_{56} \left[ M_5 (123|45||1 2 5) +M_5 (132|45||1 3 5) + M_5 (145|23||1 4 5) \right] \right\}\non\\
&&+ \ordr (\tau^2),\label{eq:6ptst}
\eea
where $P_{ijk} \equiv p_i +p_j + p_k$, and we have omitted the superscript $\nlsm + \phi + \psi$ as well as particle labels for the bi-index scalars in the 5-pt amplitudes, whose second state in the right ordering should be understood as $\phi$, while the other two are $\psi$.

The 6-pt amplitude of $\nlsm^{d_2}$:
\bea
M^{(4),d_2}_6 (6,1|2,3,4,5) &=& \frac{1}{\Lambda^2f^4} \left[ s_{35} \left(\frac{s_{26} s_{12}}{P^2_{126}} + \frac{s_{46} s_{14}}{P^2_{146}}  \right)  + s_{24} \left(\frac{s_{36} s_{13}}{P^2_{136}} + \frac{s_{56} s_{15}}{P^2_{156}}  \right) \right.\non\\
&&\left.\phantom{\frac{ s_{12}}{P^2_{126}}}- \left(s_{46} + s_{26}) (s_{14} + s_{12}\right) + s_{16} s_{24}  \right] ,
\eea
which is clearly $\ordr (\tau^2)$ when state $1$ or $6$ is taken to be soft. On the other hand, taking $p_5$ to be soft,
\bea
&&M^{(4),d_2}_6 (6,1|2,3,4,5) \non\\
&=& \frac{\tau}{\Lambda^2f^4} \left[s_{35} \left(\frac{ s_{12} s_{26}}{s_{34}} +\frac{ s_{13} s_{36}}{s_{24}}+ \frac{ s_{14} s_{46}}{ s_{23}} + s_{16}  \right)   + s_{15} \left( \frac{ s_{13} s_{36}}{ s_{24}} - s_{36}\right) \right.\non\\
&&\left.+ s_{56} \left( \frac{ s_{13} s_{36}}{ s_{24}} - s_{13}\right) \right] + \ordr (\tau^2)\non\\
&=& \frac{\tau}{\lambda f^2} \left[ s_{56}M^{d_2 + \phi^3}_5 (6,1|2,3,4||2,4,6) + s_{15}M^{d_2 + \phi^3}_5 (6,1|2,3,4||2,4,1) \right.\non\\
&&\left. + s_{35}M^{d_2 + \phi^3}_5 (6,1|2,3,4||2,4,3) \right]  + \ordr (\tau^2).\label{eq:ed46ps}
\eea

The 6-pt amplitude of $\nlsm^{C_3'}$:
\bea
M_6^{(4),3'} (1,2,3,4,5,6)  &=&   \frac{1}{\Lambda^2f^4} \left[\frac{s_{13}  s_{46}(s_{13} + s_{46})}{P^2_{123}} + \frac{s_{24}  s_{15}(s_{24} + s_{15}) }{P^2_{234}}+ \frac{s_{35} s_{26} ( s_{35} + s_{26} ) }{P^2_{345}} \right. \non\\
&&\left.\phantom{\frac{(s_{13} )}{P^2_{123}}}- \left( P^2_{135} \right)^2 - s_{13} s_{46} - s_{24} s_{15} - s_{35} s_{26}  \right],
\eea
so that when we take the soft limit of external state $6$,
\bea
&&M_6^{(4),3'} (1,2,3,4,5,6)\non\\
  &=&   \frac{\tau}{\Lambda^2f^4}\left[ s_{26} \left( \frac{s_{35}^2}{s_{12}} + \frac{s_{24}^2}{s_{15}} - s_{24} - s_{35} \right) + s_{36} \left(  \frac{s_{24}^2}{s_{15}} + s_{24} \right)\right.\non\\
&&\left. +s_{46} \left( \frac{s_{13}^2}{s_{45}} + \frac{s_{24}^2}{s_{15}} - s_{24} - s_{13} \right)\right] + \ordr (\tau^2)\non\\
&=&   \frac{\tau}{\lambda f^2}\left[ s_{26} M^{C_{3'} + \phi^3}_5 (1,2,3,4,5||1,5,2) + s_{36} M^{C_{3'} + \phi^3}_5 (1,2,3,4,5||1,5,3) \right.\non\\
&&\left. +s_{46} M^{C_{3'} + \phi^3}_5 (1,2,3,4,5||1,5,4) \right] + \ordr (\tau^2).\label{eq:c3pst6p}
\eea

\subsection{The current}
\label{app:p4cv}

Below we present the vertices $\V$ generated by the current.

3-pt vertices:
\bea
\V^{(4),1}_3 (1,2|3) &=& \frac{8}{f^2 \Lambda^2} q\cdot p_3 \ p_1 \cdot p_2,  \label{eq:tso1}\\
\V^{(4),2}_3 (1,2|3) &= & \frac{4}{f^2 \Lambda^2} \left( q\cdot p_2 \ p_1 \cdot p_3+ q \cdot p_1 \ p_2 \cdot p_3 \right),\\
\V^{(4),3'}_3 (1,2,3) &=& \frac{4}{f^2 \Lambda^2}  q \cdot p_2 \  p_1 \cdot p_3,\\
\V^{(4),4'}_3 (1,2,3) &=& \frac{2}{f^2 \Lambda^2} (q \cdot p_1 \ p_2 \cdot p_3 + q \cdot p_3 \ p_1 \cdot p_2).
\eea

5-pt vertices:
\bea
\V^{(4),1}_5 (1,2,3,4|5) &=&  \frac{8}{3f^4 \Lambda^2}q \cdot p_5 \left[ p_1\cdot (p_2 - p_3) + p_2\cdot (p_3 - p_4)   \right.\non\\
&&\left.+ p_3\cdot (p_4 - p_1) + p_4\cdot (p_1 - p_2) \right],\\
\V^{(4),2}_5 (1,2,3,4|5) &=&  \frac{4}{3f^4 \Lambda^2} \left[q \cdot p_1 \ p_5 \cdot (p_2 -2 p_3 + p_4) + q \cdot p_2 \ p_5 \cdot (p_3 -2 p_4 + p_1)\right.\non\\
&& \left.+q \cdot p_3 \ p_5 \cdot (p_4 -2 p_1 + p_2) + q \cdot p_4 \ p_5 \cdot (p_1 -2 p_2 + p_3)\right],\\
\V^{(4),1}_5 (1,2|3,4,5)  &=&  -\frac{16}{3f^4 \Lambda^2} q \cdot  (p_3 -2 p_4 + p_5) \ p_1 \cdot p_2,\\
\V^{(4),2}_5 (1,2|3,4,5) &=&  -\frac{8}{3f^4 \Lambda^2} \left[q \cdot p_1 \ p_2 \cdot (p_3 -2 p_4 + p_5) + q \cdot p_2 \ p_1 \cdot (p_3 -2 p_4 + p_5)\right],\\
 \V^{(4),3'}_5 (1,2,3,4,5) &=& \frac{4 }{3 f^4 \Lambda^2 } \left[ q\cdot p_2 \ p_1 \cdot (-2 p_3 + p_4 - p_5) + q\cdot p_4 \ p_5 \cdot (-p_1 + p_2 - 2 p_3 ) \right.\non\\
&&\left. +q\cdot p_3 \ (p_1 \cdot p_5 + 3 p_2 \cdot p_4) \right],\\
 \V^{(4),4'}_5 (1,2,3,4,5) &=&\frac{2 }{3 f^4 \Lambda^2 } \left\{ q\cdot p_1 \left[ p_2 \cdot (p_4 - 2p_3 ) + p_5 \cdot (p_1+ 2 p_3 ) \right]\  +q \cdot p_2 \ p_4 \cdot (3p_3 + p_5) \right.\non\\
&&-2 q\cdot p_3 \ (p_1 \cdot p_2 + p_4 \cdot p_5) + q\cdot p_4 \ p_2 \cdot (p_1 + 3 p_3)  \non\\
&&\left. + q\cdot p_5 \left[p_1 \cdot (2p_3 + p_5 ) +p_4 \cdot (p_2 - 2 p_3) \right] \right\}.
\eea

\subsection{The extended theories}
\label{app:eet}

The 5-pt amplitudes of $\nlsm + \phi + \pi$ in the pair basis:
\bea
M_5(1,2,3|4,5||1^\psi ,2^\phi ,3^\psi)  &=& -\frac{\lambda}{2 f^2} s_{45} \left( \frac{1}{s_{12}} + \frac{1}{s_{23}} \right),\\
M_5(1,2,3|4,5||1^\psi ,2^\phi ,4^\psi) &=&  -\frac{\lambda}{2 f^2}  \left( \frac{s_{45}}{s_{12}} - \frac{s_{25}}{s_{14}} \right),
\eea
which match Eq. (\ref{eq:6ptst}).

For the extended theory $d_2 + \phi^3$, as $\V^{(4),d_2} (j | \alpha)$ does not contribute $\ordr (\tau)$ terms in the soft theorem, there are no 3-pt vertices at $\ordr (1/\Lambda^2)$, though there is of course the $\ordr (1/\Lambda^0)$ vertex of the $\phi^3$ theory given by Eq. (\ref{eq:etp3}). Similarly, all the higher-pt vertices at $\ordr (1/\Lambda^0) $ are the same as in $\nlsm + \phi^3$. The 5-pt vertices at $\ordr (1/\Lambda^2)$ are given by
\bea
V^{d_2 + \phi^3}_{5} (1,2 | 3,4,5 ||3,5,1) &=& -\frac{2\lambda }{3\Lambda^2 f^2}  p_2  \cdot \left( p_2 + 3 p_4  \right) ,\\
V^{d_2 + \phi^3}_{5} (1,2 | 3,4,5 ||3,5,4) &=& \frac{2\lambda }{\Lambda^2 f^2} p_1 \cdot p_2.
\eea
One can then calculate the amplitudes of the extended theory to be
\bea
M^{d_2 + \phi^3}_{5} (1,2 | 3,4,5 ||3,5,1) &=& \frac{\lambda }{\Lambda^2 f^2} \left[\frac{s_{14} s_{24}}{s_{35}} - s_{24} \right],\\
M^{d_2 + \phi^3}_{5} (1,2 | 3,4,5 ||3,5,4) &=& \frac{\lambda }{\Lambda^2 f^2} \left[\frac{s_{13} s_{23}}{s_{45}} + \frac{s_{14} s_{24}}{s_{35}} + \frac{s_{15} s_{25}}{s_{34}} + s_{12} \right],
\eea
which matches Eq. (\ref{eq:ed46ps}).

For $C_{3'} + \phi^3$, first we present the 3-pt and 5-pt vertices $\V^{(4),3'}$ after we rewrite the current  $\J^{(4),3'}$ as in Eq. (\ref{eq:c3pcr}):
\bea
\hat{\V}^{C_{3'} + \phi^3}_3 (1,2,3) &=& -\frac{4}{\Lambda^2 f^2}  q \cdot p_2 \left( p_1 \cdot p_2 + p_1 \cdot p_3 + p_2 \cdot p_3 \right),\\
\hat{\V}^{C_{5'} + \phi^3}_5 (1,2,3,4,5) &=&   \frac{4}{3\Lambda^2 f^4} \left\{ q \cdot p_2 \left[ -p_1^2 + p_2 \cdot \left(2 p_3 - p_4 + p_5 \right) + p_3 \cdot \left(3p_4 - p_5 \right) + 2p_4 \cdot p_5 \right]\ \right.\non\\
&&+q\cdot p_3 \left( p_1^2 + p_1 \cdot p_5 + 3 p_2 \cdot p_4 + p_5^2 \right) \non\\
&&\left.+ q\cdot p_4 \left[ 2 p_1 \cdot p_2 + p_3 \cdot \left( - p_1 +3 p_2 \right)  + p_4 \cdot \left( p_1 - p_2 + 2 p_3 \right) - p_5^2 \right]\right\}.
\eea
The 5-pt vertices of $C_{3'} + \phi^3$ at $\ordr (1/\Lambda^2)$ are given by
\bea
V^{C_{3'} + \phi^3}_{3} (1,2 , 3 ||1,3,2) &=& - \frac{2\lambda }{\Lambda^2 } \left( p_1 \cdot p_2 + p_1 \cdot p_3 + p_2 \cdot p_3 \right),\\
V^{C_{3'} + \phi^3}_{5} (1,2 , 3,4,5 ||1,5,2) &=& -\frac{2\lambda }{3\Lambda^2 f^2}  \left[ p_2^2 + p_3^2 + p_4^2 + p_5^2 + 3\left( p_2 \cdot p_4 + p_3 \cdot p_5 \right) \right.\non\\
&&\left. \phantom{p_2^2}+ p_2 \cdot p_5 - p_3 \cdot p_4 \right] ,\\
V^{C_{3'} + \phi^3}_{5} (1,2 , 3,4,5 ||1,5,3) &=& \frac{2\lambda }{3\Lambda^2 f^2} \left( p_1^2  + p_1 \cdot p_5 + 3 p_2 \cdot p_4 + p_5^2 \right) .
\eea
We see explicitly that $V^{C_{3'} + \phi^3}_{3} (1,2 , 3 ||1,3,2)$ and $V^{C_{3'} + \phi^3}_{5} (1,2 , 3,4,5 ||1,5,2)$ has the correct relabeling property of Eq. (\ref{eq:c3pcst}).

Although we have a 3-pt vertex at $\ordr (1/\Lambda^2)$, the corresponding 3-pt amplitude still vanishes, so that it contributes nothing to the soft theorem in Eq. (\ref{eq:nlsmst4}). On the other hand, we can calculate the 5-pt amplitudes to be
\bea
M^{C_{3'} + \phi^3}_{5} (1,2 , 3,4,5 ||1,5,2) &=& \frac{\lambda }{\Lambda^2 f^2}   \left( \frac{s_{35}^2}{s_{12}} + \frac{s_{24}^2}{s_{15}} - s_{24} - s_{35} \right)  ,\\
M^{C_{3'} + \phi^3}_{5} (1,2 , 3,4,5 ||1,5,3) &=& \frac{\lambda }{\Lambda^2 f^2} \left(  \frac{s_{24}^2}{s_{15}} + s_{24} \right) ,
\eea
which agrees with Eq. (\ref{eq:c3pst6p}).

\end{appendix}

\bibliography{draft}

\begin{thebibliography}{81}%
\makeatletter
\providecommand \@ifxundefined [1]{%
 \@ifx{#1\undefined}
}%
\providecommand \@ifnum [1]{%
 \ifnum #1\expandafter \@firstoftwo
 \else \expandafter \@secondoftwo
 \fi
}%
\providecommand \@ifx [1]{%
 \ifx #1\expandafter \@firstoftwo
 \else \expandafter \@secondoftwo
 \fi
}%
\providecommand \natexlab [1]{#1}%
\providecommand \enquote  [1]{``#1''}%
\providecommand \bibnamefont  [1]{#1}%
\providecommand \bibfnamefont [1]{#1}%
\providecommand \citenamefont [1]{#1}%
\providecommand \href@noop [0]{\@secondoftwo}%
\providecommand \href [0]{\begingroup \@sanitize@url \@href}%
\providecommand \@href[1]{\@@startlink{#1}\@@href}%
\providecommand \@@href[1]{\endgroup#1\@@endlink}%
\providecommand \@sanitize@url [0]{\catcode `\\12\catcode `\$12\catcode
  `\&12\catcode `\#12\catcode `\^12\catcode `\_12\catcode `\%12\relax}%
\providecommand \@@startlink[1]{}%
\providecommand \@@endlink[0]{}%
\providecommand \url  [0]{\begingroup\@sanitize@url \@url }%
\providecommand \@url [1]{\endgroup\@href {#1}{\urlprefix }}%
\providecommand \urlprefix  [0]{URL }%
\providecommand \Eprint [0]{\href }%
\providecommand \doibase [0]{http://dx.doi.org/}%
\providecommand \selectlanguage [0]{\@gobble}%
\providecommand \bibinfo  [0]{\@secondoftwo}%
\providecommand \bibfield  [0]{\@secondoftwo}%
\providecommand \translation [1]{[#1]}%
\providecommand \BibitemOpen [0]{}%
\providecommand \bibitemStop [0]{}%
\providecommand \bibitemNoStop [0]{.\EOS\space}%
\providecommand \EOS [0]{\spacefactor3000\relax}%
\providecommand \BibitemShut  [1]{\csname bibitem#1\endcsname}%
\let\auto@bib@innerbib\@empty
\bibitem [{\citenamefont {Gell-Mann}\ and\ \citenamefont
  {Levy}(1960)}]{GellMann:1960np}%
  \BibitemOpen
  \bibfield  {author} {\bibinfo {author} {\bibfnamefont {M.}~\bibnamefont
  {Gell-Mann}}\ and\ \bibinfo {author} {\bibfnamefont {M.}~\bibnamefont
  {Levy}},\ }\href {\doibase 10.1007/BF02859738} {\bibfield  {journal}
  {\bibinfo  {journal} {Nuovo Cim.}\ }\textbf {\bibinfo {volume} {16}},\
  \bibinfo {pages} {705} (\bibinfo {year} {1960})}\BibitemShut {NoStop}%
\bibitem [{\citenamefont {Coleman}\ \emph {et~al.}(1969)\citenamefont
  {Coleman}, \citenamefont {Wess},\ and\ \citenamefont
  {Zumino}}]{Coleman:1969sm}%
  \BibitemOpen
  \bibfield  {author} {\bibinfo {author} {\bibfnamefont {S.~R.}\ \bibnamefont
  {Coleman}}, \bibinfo {author} {\bibfnamefont {J.}~\bibnamefont {Wess}}, \
  and\ \bibinfo {author} {\bibfnamefont {B.}~\bibnamefont {Zumino}},\ }\href
  {\doibase 10.1103/PhysRev.177.2239} {\bibfield  {journal} {\bibinfo
  {journal} {Phys. Rev.}\ }\textbf {\bibinfo {volume} {177}},\ \bibinfo {pages}
  {2239} (\bibinfo {year} {1969})}\BibitemShut {NoStop}%
\bibitem [{\citenamefont {Callan}\ \emph {et~al.}(1969)\citenamefont {Callan},
  \citenamefont {Coleman}, \citenamefont {Wess},\ and\ \citenamefont
  {Zumino}}]{Callan:1969sn}%
  \BibitemOpen
  \bibfield  {author} {\bibinfo {author} {\bibfnamefont {C.~G.}\ \bibnamefont
  {Callan}, \bibfnamefont {Jr.}}, \bibinfo {author} {\bibfnamefont {S.~R.}\
  \bibnamefont {Coleman}}, \bibinfo {author} {\bibfnamefont {J.}~\bibnamefont
  {Wess}}, \ and\ \bibinfo {author} {\bibfnamefont {B.}~\bibnamefont
  {Zumino}},\ }\href {\doibase 10.1103/PhysRev.177.2247} {\bibfield  {journal}
  {\bibinfo  {journal} {Phys. Rev.}\ }\textbf {\bibinfo {volume} {177}},\
  \bibinfo {pages} {2247} (\bibinfo {year} {1969})}\BibitemShut {NoStop}%
\bibitem [{\citenamefont {Weinberg}(1979)}]{Weinberg:1978kz}%
  \BibitemOpen
  \bibfield  {author} {\bibinfo {author} {\bibfnamefont {S.}~\bibnamefont
  {Weinberg}},\ }\href {\doibase 10.1016/0378-4371(79)90223-1} {\bibfield
  {journal} {\bibinfo  {journal} {Physica A}\ }\textbf {\bibinfo {volume}
  {96}},\ \bibinfo {pages} {327} (\bibinfo {year} {1979})}\BibitemShut
  {NoStop}%
\bibitem [{\citenamefont {Gasser}\ and\ \citenamefont
  {Leutwyler}(1984)}]{Gasser:1983yg}%
  \BibitemOpen
  \bibfield  {author} {\bibinfo {author} {\bibfnamefont {J.}~\bibnamefont
  {Gasser}}\ and\ \bibinfo {author} {\bibfnamefont {H.}~\bibnamefont
  {Leutwyler}},\ }\href {\doibase 10.1016/0003-4916(84)90242-2} {\bibfield
  {journal} {\bibinfo  {journal} {Annals Phys.}\ }\textbf {\bibinfo {volume}
  {158}},\ \bibinfo {pages} {142} (\bibinfo {year} {1984})}\BibitemShut
  {NoStop}%
\bibitem [{\citenamefont {Gasser}\ and\ \citenamefont
  {Leutwyler}(1985)}]{Gasser:1984gg}%
  \BibitemOpen
  \bibfield  {author} {\bibinfo {author} {\bibfnamefont {J.}~\bibnamefont
  {Gasser}}\ and\ \bibinfo {author} {\bibfnamefont {H.}~\bibnamefont
  {Leutwyler}},\ }\href {\doibase 10.1016/0550-3213(85)90492-4} {\bibfield
  {journal} {\bibinfo  {journal} {Nucl. Phys. B}\ }\textbf {\bibinfo {volume}
  {250}},\ \bibinfo {pages} {465} (\bibinfo {year} {1985})}\BibitemShut
  {NoStop}%
\bibitem [{\citenamefont {Susskind}\ and\ \citenamefont
  {Frye}(1970)}]{Susskind:1970gf}%
  \BibitemOpen
  \bibfield  {author} {\bibinfo {author} {\bibfnamefont {L.}~\bibnamefont
  {Susskind}}\ and\ \bibinfo {author} {\bibfnamefont {G.}~\bibnamefont
  {Frye}},\ }\href {\doibase 10.1103/PhysRevD.1.1682} {\bibfield  {journal}
  {\bibinfo  {journal} {Phys. Rev. D}\ }\textbf {\bibinfo {volume} {1}},\
  \bibinfo {pages} {1682} (\bibinfo {year} {1970})}\BibitemShut {NoStop}%
\bibitem [{\citenamefont {Ellis}(1970)}]{Ellis:1970nn}%
  \BibitemOpen
  \bibfield  {author} {\bibinfo {author} {\bibfnamefont {J.~R.}\ \bibnamefont
  {Ellis}},\ }\href {\doibase 10.1016/0550-3213(70)90516-X} {\bibfield
  {journal} {\bibinfo  {journal} {Nucl. Phys. B}\ }\textbf {\bibinfo {volume}
  {21}},\ \bibinfo {pages} {217} (\bibinfo {year} {1970})}\BibitemShut
  {NoStop}%
\bibitem [{\citenamefont {Low}(2015{\natexlab{a}})}]{Low:2014nga}%
  \BibitemOpen
  \bibfield  {author} {\bibinfo {author} {\bibfnamefont {I.}~\bibnamefont
  {Low}},\ }\href {\doibase 10.1103/PhysRevD.91.105017} {\bibfield  {journal}
  {\bibinfo  {journal} {Phys. Rev. D}\ }\textbf {\bibinfo {volume} {91}},\
  \bibinfo {pages} {105017} (\bibinfo {year} {2015}{\natexlab{a}})},\ \Eprint
  {http://arxiv.org/abs/1412.2145} {arXiv:1412.2145 [hep-th]} \BibitemShut
  {NoStop}%
\bibitem [{\citenamefont {Low}(2015{\natexlab{b}})}]{Low:2014oga}%
  \BibitemOpen
  \bibfield  {author} {\bibinfo {author} {\bibfnamefont {I.}~\bibnamefont
  {Low}},\ }\href {\doibase 10.1103/PhysRevD.91.116005} {\bibfield  {journal}
  {\bibinfo  {journal} {Phys. Rev. D}\ }\textbf {\bibinfo {volume} {91}},\
  \bibinfo {pages} {116005} (\bibinfo {year} {2015}{\natexlab{b}})},\ \Eprint
  {http://arxiv.org/abs/1412.2146} {arXiv:1412.2146 [hep-ph]} \BibitemShut
  {NoStop}%
\bibitem [{\citenamefont {Liu}\ \emph {et~al.}(2018)\citenamefont {Liu},
  \citenamefont {Low},\ and\ \citenamefont {Yin}}]{Liu:2018vel}%
  \BibitemOpen
  \bibfield  {author} {\bibinfo {author} {\bibfnamefont {D.}~\bibnamefont
  {Liu}}, \bibinfo {author} {\bibfnamefont {I.}~\bibnamefont {Low}}, \ and\
  \bibinfo {author} {\bibfnamefont {Z.}~\bibnamefont {Yin}},\ }\href {\doibase
  10.1103/PhysRevLett.121.261802} {\bibfield  {journal} {\bibinfo  {journal}
  {Phys. Rev. Lett.}\ }\textbf {\bibinfo {volume} {121}},\ \bibinfo {pages}
  {261802} (\bibinfo {year} {2018})},\ \Eprint
  {http://arxiv.org/abs/1805.00489} {arXiv:1805.00489 [hep-ph]} \BibitemShut
  {NoStop}%
\bibitem [{\citenamefont {Liu}\ \emph {et~al.}(2019)\citenamefont {Liu},
  \citenamefont {Low},\ and\ \citenamefont {Yin}}]{Liu:2018qtb}%
  \BibitemOpen
  \bibfield  {author} {\bibinfo {author} {\bibfnamefont {D.}~\bibnamefont
  {Liu}}, \bibinfo {author} {\bibfnamefont {I.}~\bibnamefont {Low}}, \ and\
  \bibinfo {author} {\bibfnamefont {Z.}~\bibnamefont {Yin}},\ }\href {\doibase
  10.1007/JHEP05(2019)170} {\bibfield  {journal} {\bibinfo  {journal} {JHEP}\
  }\textbf {\bibinfo {volume} {05}},\ \bibinfo {pages} {170} (\bibinfo {year}
  {2019})},\ \Eprint {http://arxiv.org/abs/1809.09126} {arXiv:1809.09126
  [hep-ph]} \BibitemShut {NoStop}%
\bibitem [{\citenamefont {Liu}\ \emph {et~al.}(2020)\citenamefont {Liu},
  \citenamefont {Low},\ and\ \citenamefont {Vega-Morales}}]{Liu:2019rce}%
  \BibitemOpen
  \bibfield  {author} {\bibinfo {author} {\bibfnamefont {D.}~\bibnamefont
  {Liu}}, \bibinfo {author} {\bibfnamefont {I.}~\bibnamefont {Low}}, \ and\
  \bibinfo {author} {\bibfnamefont {R.}~\bibnamefont {Vega-Morales}},\ }\href
  {\doibase 10.1140/epjc/s10052-020-8244-8} {\bibfield  {journal} {\bibinfo
  {journal} {Eur. Phys. J. C}\ }\textbf {\bibinfo {volume} {80}},\ \bibinfo
  {pages} {829} (\bibinfo {year} {2020})},\ \Eprint
  {http://arxiv.org/abs/1904.00026} {arXiv:1904.00026 [hep-ph]} \BibitemShut
  {NoStop}%
\bibitem [{\citenamefont {Kaplan}\ and\ \citenamefont
  {Georgi}(1984)}]{Kaplan:1983fs}%
  \BibitemOpen
  \bibfield  {author} {\bibinfo {author} {\bibfnamefont {D.~B.}\ \bibnamefont
  {Kaplan}}\ and\ \bibinfo {author} {\bibfnamefont {H.}~\bibnamefont
  {Georgi}},\ }\href {\doibase 10.1016/0370-2693(84)91177-8} {\bibfield
  {journal} {\bibinfo  {journal} {Phys. Lett. B}\ }\textbf {\bibinfo {volume}
  {136}},\ \bibinfo {pages} {183} (\bibinfo {year} {1984})}\BibitemShut
  {NoStop}%
\bibitem [{\citenamefont {Agashe}\ \emph {et~al.}(2005)\citenamefont {Agashe},
  \citenamefont {Contino},\ and\ \citenamefont {Pomarol}}]{Agashe:2004rs}%
  \BibitemOpen
  \bibfield  {author} {\bibinfo {author} {\bibfnamefont {K.}~\bibnamefont
  {Agashe}}, \bibinfo {author} {\bibfnamefont {R.}~\bibnamefont {Contino}}, \
  and\ \bibinfo {author} {\bibfnamefont {A.}~\bibnamefont {Pomarol}},\ }\href
  {\doibase 10.1016/j.nuclphysb.2005.04.035} {\bibfield  {journal} {\bibinfo
  {journal} {Nucl. Phys. B}\ }\textbf {\bibinfo {volume} {719}},\ \bibinfo
  {pages} {165} (\bibinfo {year} {2005})},\ \Eprint
  {http://arxiv.org/abs/hep-ph/0412089} {arXiv:hep-ph/0412089} \BibitemShut
  {NoStop}%
\bibitem [{\citenamefont {Panico}\ and\ \citenamefont
  {Wulzer}(2016)}]{Panico:2015jxa}%
  \BibitemOpen
  \bibfield  {author} {\bibinfo {author} {\bibfnamefont {G.}~\bibnamefont
  {Panico}}\ and\ \bibinfo {author} {\bibfnamefont {A.}~\bibnamefont
  {Wulzer}},\ }\href {\doibase 10.1007/978-3-319-22617-0} {\emph {\bibinfo
  {title} {{The Composite Nambu-Goldstone Higgs}}}},\ Vol.\ \bibinfo {volume}
  {913}\ (\bibinfo  {publisher} {Springer},\ \bibinfo {year} {2016})\ \Eprint
  {http://arxiv.org/abs/1506.01961} {arXiv:1506.01961 [hep-ph]} \BibitemShut
  {NoStop}%
\bibitem [{\citenamefont {Arkani-Hamed}\ \emph {et~al.}(2010)\citenamefont
  {Arkani-Hamed}, \citenamefont {Cachazo},\ and\ \citenamefont
  {Kaplan}}]{ArkaniHamed:2008gz}%
  \BibitemOpen
  \bibfield  {author} {\bibinfo {author} {\bibfnamefont {N.}~\bibnamefont
  {Arkani-Hamed}}, \bibinfo {author} {\bibfnamefont {F.}~\bibnamefont
  {Cachazo}}, \ and\ \bibinfo {author} {\bibfnamefont {J.}~\bibnamefont
  {Kaplan}},\ }\href {\doibase 10.1007/JHEP09(2010)016} {\bibfield  {journal}
  {\bibinfo  {journal} {JHEP}\ }\textbf {\bibinfo {volume} {09}},\ \bibinfo
  {pages} {016} (\bibinfo {year} {2010})},\ \Eprint
  {http://arxiv.org/abs/0808.1446} {arXiv:0808.1446 [hep-th]} \BibitemShut
  {NoStop}%
\bibitem [{\citenamefont {Kampf}\ \emph
  {et~al.}(2013{\natexlab{a}})\citenamefont {Kampf}, \citenamefont {Novotny},\
  and\ \citenamefont {Trnka}}]{Kampf:2012fn}%
  \BibitemOpen
  \bibfield  {author} {\bibinfo {author} {\bibfnamefont {K.}~\bibnamefont
  {Kampf}}, \bibinfo {author} {\bibfnamefont {J.}~\bibnamefont {Novotny}}, \
  and\ \bibinfo {author} {\bibfnamefont {J.}~\bibnamefont {Trnka}},\ }\href
  {\doibase 10.1103/PhysRevD.87.081701} {\bibfield  {journal} {\bibinfo
  {journal} {Phys. Rev. D}\ }\textbf {\bibinfo {volume} {87}},\ \bibinfo
  {pages} {081701} (\bibinfo {year} {2013}{\natexlab{a}})},\ \Eprint
  {http://arxiv.org/abs/1212.5224} {arXiv:1212.5224 [hep-th]} \BibitemShut
  {NoStop}%
\bibitem [{\citenamefont {Kampf}\ \emph
  {et~al.}(2013{\natexlab{b}})\citenamefont {Kampf}, \citenamefont {Novotny},\
  and\ \citenamefont {Trnka}}]{Kampf:2013vha}%
  \BibitemOpen
  \bibfield  {author} {\bibinfo {author} {\bibfnamefont {K.}~\bibnamefont
  {Kampf}}, \bibinfo {author} {\bibfnamefont {J.}~\bibnamefont {Novotny}}, \
  and\ \bibinfo {author} {\bibfnamefont {J.}~\bibnamefont {Trnka}},\ }\href
  {\doibase 10.1007/JHEP05(2013)032} {\bibfield  {journal} {\bibinfo  {journal}
  {JHEP}\ }\textbf {\bibinfo {volume} {05}},\ \bibinfo {pages} {032} (\bibinfo
  {year} {2013}{\natexlab{b}})},\ \Eprint {http://arxiv.org/abs/1304.3048}
  {arXiv:1304.3048 [hep-th]} \BibitemShut {NoStop}%
\bibitem [{\citenamefont {Chen}\ and\ \citenamefont {Du}(2014)}]{Chen:2013fya}%
  \BibitemOpen
  \bibfield  {author} {\bibinfo {author} {\bibfnamefont {G.}~\bibnamefont
  {Chen}}\ and\ \bibinfo {author} {\bibfnamefont {Y.-J.}\ \bibnamefont {Du}},\
  }\href {\doibase 10.1007/JHEP01(2014)061} {\bibfield  {journal} {\bibinfo
  {journal} {JHEP}\ }\textbf {\bibinfo {volume} {01}},\ \bibinfo {pages} {061}
  (\bibinfo {year} {2014})},\ \Eprint {http://arxiv.org/abs/1311.1133}
  {arXiv:1311.1133 [hep-th]} \BibitemShut {NoStop}%
\bibitem [{\citenamefont {Chen}\ \emph {et~al.}(2015)\citenamefont {Chen},
  \citenamefont {Du}, \citenamefont {Li},\ and\ \citenamefont
  {Liu}}]{Chen:2014dfa}%
  \BibitemOpen
  \bibfield  {author} {\bibinfo {author} {\bibfnamefont {G.}~\bibnamefont
  {Chen}}, \bibinfo {author} {\bibfnamefont {Y.-J.}\ \bibnamefont {Du}},
  \bibinfo {author} {\bibfnamefont {S.}~\bibnamefont {Li}}, \ and\ \bibinfo
  {author} {\bibfnamefont {H.}~\bibnamefont {Liu}},\ }\href {\doibase
  10.1007/JHEP03(2015)156} {\bibfield  {journal} {\bibinfo  {journal} {JHEP}\
  }\textbf {\bibinfo {volume} {03}},\ \bibinfo {pages} {156} (\bibinfo {year}
  {2015})},\ \Eprint {http://arxiv.org/abs/1412.3722} {arXiv:1412.3722
  [hep-th]} \BibitemShut {NoStop}%
\bibitem [{\citenamefont {He}\ \emph {et~al.}(2016)\citenamefont {He},
  \citenamefont {Liu},\ and\ \citenamefont {Wu}}]{He:2016vfi}%
  \BibitemOpen
  \bibfield  {author} {\bibinfo {author} {\bibfnamefont {S.}~\bibnamefont
  {He}}, \bibinfo {author} {\bibfnamefont {Z.}~\bibnamefont {Liu}}, \ and\
  \bibinfo {author} {\bibfnamefont {J.-B.}\ \bibnamefont {Wu}},\ }\href
  {\doibase 10.1007/JHEP07(2016)060} {\bibfield  {journal} {\bibinfo  {journal}
  {JHEP}\ }\textbf {\bibinfo {volume} {07}},\ \bibinfo {pages} {060} (\bibinfo
  {year} {2016})},\ \Eprint {http://arxiv.org/abs/1604.02834} {arXiv:1604.02834
  [hep-th]} \BibitemShut {NoStop}%
\bibitem [{\citenamefont {Du}\ and\ \citenamefont {Fu}(2016)}]{Du:2016tbc}%
  \BibitemOpen
  \bibfield  {author} {\bibinfo {author} {\bibfnamefont {Y.-J.}\ \bibnamefont
  {Du}}\ and\ \bibinfo {author} {\bibfnamefont {C.-H.}\ \bibnamefont {Fu}},\
  }\href {\doibase 10.1007/JHEP09(2016)174} {\bibfield  {journal} {\bibinfo
  {journal} {JHEP}\ }\textbf {\bibinfo {volume} {09}},\ \bibinfo {pages} {174}
  (\bibinfo {year} {2016})},\ \Eprint {http://arxiv.org/abs/1606.05846}
  {arXiv:1606.05846 [hep-th]} \BibitemShut {NoStop}%
\bibitem [{\citenamefont {Chen}\ \emph {et~al.}(2016)\citenamefont {Chen},
  \citenamefont {Li},\ and\ \citenamefont {Liu}}]{Chen:2016zwe}%
  \BibitemOpen
  \bibfield  {author} {\bibinfo {author} {\bibfnamefont {G.}~\bibnamefont
  {Chen}}, \bibinfo {author} {\bibfnamefont {S.}~\bibnamefont {Li}}, \ and\
  \bibinfo {author} {\bibfnamefont {H.}~\bibnamefont {Liu}},\ }\href@noop {} {\
   (\bibinfo {year} {2016})},\ \Eprint {http://arxiv.org/abs/1609.01832}
  {arXiv:1609.01832 [hep-th]} \BibitemShut {NoStop}%
\bibitem [{\citenamefont {Cheung}\ \emph
  {et~al.}(2018{\natexlab{a}})\citenamefont {Cheung}, \citenamefont {Shen},\
  and\ \citenamefont {Wen}}]{Cheung:2017ems}%
  \BibitemOpen
  \bibfield  {author} {\bibinfo {author} {\bibfnamefont {C.}~\bibnamefont
  {Cheung}}, \bibinfo {author} {\bibfnamefont {C.-H.}\ \bibnamefont {Shen}}, \
  and\ \bibinfo {author} {\bibfnamefont {C.}~\bibnamefont {Wen}},\ }\href
  {\doibase 10.1007/JHEP02(2018)095} {\bibfield  {journal} {\bibinfo  {journal}
  {JHEP}\ }\textbf {\bibinfo {volume} {02}},\ \bibinfo {pages} {095} (\bibinfo
  {year} {2018}{\natexlab{a}})},\ \Eprint {http://arxiv.org/abs/1705.03025}
  {arXiv:1705.03025 [hep-th]} \BibitemShut {NoStop}%
\bibitem [{\citenamefont {Cheung}\ \emph
  {et~al.}(2018{\natexlab{b}})\citenamefont {Cheung}, \citenamefont {Remmen},
  \citenamefont {Shen},\ and\ \citenamefont {Wen}}]{Cheung:2017yef}%
  \BibitemOpen
  \bibfield  {author} {\bibinfo {author} {\bibfnamefont {C.}~\bibnamefont
  {Cheung}}, \bibinfo {author} {\bibfnamefont {G.~N.}\ \bibnamefont {Remmen}},
  \bibinfo {author} {\bibfnamefont {C.-H.}\ \bibnamefont {Shen}}, \ and\
  \bibinfo {author} {\bibfnamefont {C.}~\bibnamefont {Wen}},\ }\href {\doibase
  10.1007/JHEP04(2018)129} {\bibfield  {journal} {\bibinfo  {journal} {JHEP}\
  }\textbf {\bibinfo {volume} {04}},\ \bibinfo {pages} {129} (\bibinfo {year}
  {2018}{\natexlab{b}})},\ \Eprint {http://arxiv.org/abs/1709.04932}
  {arXiv:1709.04932 [hep-th]} \BibitemShut {NoStop}%
\bibitem [{\citenamefont {Mizera}\ and\ \citenamefont
  {Skrzypek}(2018)}]{Mizera:2018jbh}%
  \BibitemOpen
  \bibfield  {author} {\bibinfo {author} {\bibfnamefont {S.}~\bibnamefont
  {Mizera}}\ and\ \bibinfo {author} {\bibfnamefont {B.}~\bibnamefont
  {Skrzypek}},\ }\href {\doibase 10.1007/JHEP10(2018)018} {\bibfield  {journal}
  {\bibinfo  {journal} {JHEP}\ }\textbf {\bibinfo {volume} {10}},\ \bibinfo
  {pages} {018} (\bibinfo {year} {2018})},\ \Eprint
  {http://arxiv.org/abs/1809.02096} {arXiv:1809.02096 [hep-th]} \BibitemShut
  {NoStop}%
\bibitem [{\citenamefont {Carrillo~Gonz\'alez}\ \emph
  {et~al.}(2018)\citenamefont {Carrillo~Gonz\'alez}, \citenamefont {Penco},\
  and\ \citenamefont {Trodden}}]{Carrillo-Gonzalez:2018pjk}%
  \BibitemOpen
  \bibfield  {author} {\bibinfo {author} {\bibfnamefont {M.}~\bibnamefont
  {Carrillo~Gonz\'alez}}, \bibinfo {author} {\bibfnamefont {R.}~\bibnamefont
  {Penco}}, \ and\ \bibinfo {author} {\bibfnamefont {M.}~\bibnamefont
  {Trodden}},\ }\href {\doibase 10.1007/JHEP11(2018)065} {\bibfield  {journal}
  {\bibinfo  {journal} {JHEP}\ }\textbf {\bibinfo {volume} {11}},\ \bibinfo
  {pages} {065} (\bibinfo {year} {2018})},\ \Eprint
  {http://arxiv.org/abs/1809.04611} {arXiv:1809.04611 [hep-th]} \BibitemShut
  {NoStop}%
\bibitem [{\citenamefont {Bjerrum-Bohr}\ \emph {et~al.}(2019)\citenamefont
  {Bjerrum-Bohr}, \citenamefont {Gomez},\ and\ \citenamefont
  {Helset}}]{Bjerrum-Bohr:2018jqe}%
  \BibitemOpen
  \bibfield  {author} {\bibinfo {author} {\bibfnamefont {N.~E.~J.}\
  \bibnamefont {Bjerrum-Bohr}}, \bibinfo {author} {\bibfnamefont
  {H.}~\bibnamefont {Gomez}}, \ and\ \bibinfo {author} {\bibfnamefont
  {A.}~\bibnamefont {Helset}},\ }\href {\doibase 10.1103/PhysRevD.99.045009}
  {\bibfield  {journal} {\bibinfo  {journal} {Phys. Rev. D}\ }\textbf {\bibinfo
  {volume} {99}},\ \bibinfo {pages} {045009} (\bibinfo {year} {2019})},\
  \Eprint {http://arxiv.org/abs/1811.06024} {arXiv:1811.06024 [hep-th]}
  \BibitemShut {NoStop}%
\bibitem [{\citenamefont {Gomez}\ and\ \citenamefont
  {Helset}(2019)}]{Gomez:2019cik}%
  \BibitemOpen
  \bibfield  {author} {\bibinfo {author} {\bibfnamefont {H.}~\bibnamefont
  {Gomez}}\ and\ \bibinfo {author} {\bibfnamefont {A.}~\bibnamefont {Helset}},\
  }\href {\doibase 10.1007/JHEP05(2019)129} {\bibfield  {journal} {\bibinfo
  {journal} {JHEP}\ }\textbf {\bibinfo {volume} {05}},\ \bibinfo {pages} {129}
  (\bibinfo {year} {2019})},\ \Eprint {http://arxiv.org/abs/1902.02633}
  {arXiv:1902.02633 [hep-th]} \BibitemShut {NoStop}%
\bibitem [{\citenamefont {Carrasco}\ and\ \citenamefont
  {Rodina}(2019)}]{Carrasco:2019qwr}%
  \BibitemOpen
  \bibfield  {author} {\bibinfo {author} {\bibfnamefont {J.~J.~M.}\
  \bibnamefont {Carrasco}}\ and\ \bibinfo {author} {\bibfnamefont
  {L.}~\bibnamefont {Rodina}},\ }\href {\doibase 10.1103/PhysRevD.100.125007}
  {\bibfield  {journal} {\bibinfo  {journal} {Phys. Rev. D}\ }\textbf {\bibinfo
  {volume} {100}},\ \bibinfo {pages} {125007} (\bibinfo {year} {2019})},\
  \Eprint {http://arxiv.org/abs/1908.08033} {arXiv:1908.08033 [hep-th]}
  \BibitemShut {NoStop}%
\bibitem [{\citenamefont {Carrillo~Gonz\'alez}\ \emph
  {et~al.}(2020)\citenamefont {Carrillo~Gonz\'alez}, \citenamefont {Penco},\
  and\ \citenamefont {Trodden}}]{Carrillo-Gonzalez:2019aao}%
  \BibitemOpen
  \bibfield  {author} {\bibinfo {author} {\bibfnamefont {M.}~\bibnamefont
  {Carrillo~Gonz\'alez}}, \bibinfo {author} {\bibfnamefont {R.}~\bibnamefont
  {Penco}}, \ and\ \bibinfo {author} {\bibfnamefont {M.}~\bibnamefont
  {Trodden}},\ }\href {\doibase 10.1103/PhysRevD.102.105011} {\bibfield
  {journal} {\bibinfo  {journal} {Phys. Rev. D}\ }\textbf {\bibinfo {volume}
  {102}},\ \bibinfo {pages} {105011} (\bibinfo {year} {2020})},\ \Eprint
  {http://arxiv.org/abs/1908.07531} {arXiv:1908.07531 [hep-th]} \BibitemShut
  {NoStop}%
\bibitem [{\citenamefont {Bijnens}\ \emph {et~al.}(2019)\citenamefont
  {Bijnens}, \citenamefont {Kampf},\ and\ \citenamefont
  {Sj\"o}}]{Bijnens:2019eze}%
  \BibitemOpen
  \bibfield  {author} {\bibinfo {author} {\bibfnamefont {J.}~\bibnamefont
  {Bijnens}}, \bibinfo {author} {\bibfnamefont {K.}~\bibnamefont {Kampf}}, \
  and\ \bibinfo {author} {\bibfnamefont {M.}~\bibnamefont {Sj\"o}},\ }\href
  {\doibase 10.1007/JHEP11(2019)074} {\bibfield  {journal} {\bibinfo  {journal}
  {JHEP}\ }\textbf {\bibinfo {volume} {11}},\ \bibinfo {pages} {074} (\bibinfo
  {year} {2019})},\ \Eprint {http://arxiv.org/abs/1909.13684} {arXiv:1909.13684
  [hep-th]} \BibitemShut {NoStop}%
\bibitem [{\citenamefont {Carrasco}\ \emph
  {et~al.}(2017{\natexlab{a}})\citenamefont {Carrasco}, \citenamefont {Mafra},\
  and\ \citenamefont {Schlotterer}}]{Carrasco:2016ldy}%
  \BibitemOpen
  \bibfield  {author} {\bibinfo {author} {\bibfnamefont {J.~J.~M.}\
  \bibnamefont {Carrasco}}, \bibinfo {author} {\bibfnamefont {C.~R.}\
  \bibnamefont {Mafra}}, \ and\ \bibinfo {author} {\bibfnamefont
  {O.}~\bibnamefont {Schlotterer}},\ }\href {\doibase 10.1007/JHEP06(2017)093}
  {\bibfield  {journal} {\bibinfo  {journal} {JHEP}\ }\textbf {\bibinfo
  {volume} {06}},\ \bibinfo {pages} {093} (\bibinfo {year}
  {2017}{\natexlab{a}})},\ \Eprint {http://arxiv.org/abs/1608.02569}
  {arXiv:1608.02569 [hep-th]} \BibitemShut {NoStop}%
\bibitem [{\citenamefont {Carrasco}\ \emph
  {et~al.}(2017{\natexlab{b}})\citenamefont {Carrasco}, \citenamefont {Mafra},\
  and\ \citenamefont {Schlotterer}}]{Carrasco:2016ygv}%
  \BibitemOpen
  \bibfield  {author} {\bibinfo {author} {\bibfnamefont {J.~J.~M.}\
  \bibnamefont {Carrasco}}, \bibinfo {author} {\bibfnamefont {C.~R.}\
  \bibnamefont {Mafra}}, \ and\ \bibinfo {author} {\bibfnamefont
  {O.}~\bibnamefont {Schlotterer}},\ }\href {\doibase 10.1007/JHEP08(2017)135}
  {\bibfield  {journal} {\bibinfo  {journal} {JHEP}\ }\textbf {\bibinfo
  {volume} {08}},\ \bibinfo {pages} {135} (\bibinfo {year}
  {2017}{\natexlab{b}})},\ \Eprint {http://arxiv.org/abs/1612.06446}
  {arXiv:1612.06446 [hep-th]} \BibitemShut {NoStop}%
\bibitem [{\citenamefont {Strominger}(2017)}]{Strominger:2017zoo}%
  \BibitemOpen
  \bibfield  {author} {\bibinfo {author} {\bibfnamefont {A.}~\bibnamefont
  {Strominger}},\ }\href@noop {} {\  (\bibinfo {year} {2017})},\ \Eprint
  {http://arxiv.org/abs/1703.05448} {arXiv:1703.05448 [hep-th]} \BibitemShut
  {NoStop}%
\bibitem [{\citenamefont {Kampf}\ \emph {et~al.}(2020)\citenamefont {Kampf},
  \citenamefont {Novotny}, \citenamefont {Shifman},\ and\ \citenamefont
  {Trnka}}]{Kampf:2019mcd}%
  \BibitemOpen
  \bibfield  {author} {\bibinfo {author} {\bibfnamefont {K.}~\bibnamefont
  {Kampf}}, \bibinfo {author} {\bibfnamefont {J.}~\bibnamefont {Novotny}},
  \bibinfo {author} {\bibfnamefont {M.}~\bibnamefont {Shifman}}, \ and\
  \bibinfo {author} {\bibfnamefont {J.}~\bibnamefont {Trnka}},\ }\href
  {\doibase 10.1103/PhysRevLett.124.111601} {\bibfield  {journal} {\bibinfo
  {journal} {Phys. Rev. Lett.}\ }\textbf {\bibinfo {volume} {124}},\ \bibinfo
  {pages} {111601} (\bibinfo {year} {2020})},\ \Eprint
  {http://arxiv.org/abs/1910.04766} {arXiv:1910.04766 [hep-th]} \BibitemShut
  {NoStop}%
\bibitem [{\citenamefont {Adler}(1965)}]{Adler:1964um}%
  \BibitemOpen
  \bibfield  {author} {\bibinfo {author} {\bibfnamefont {S.~L.}\ \bibnamefont
  {Adler}},\ }\href {\doibase 10.1103/PhysRev.137.B1022} {\bibfield  {journal}
  {\bibinfo  {journal} {Phys. Rev.}\ }\textbf {\bibinfo {volume} {137}},\
  \bibinfo {pages} {B1022} (\bibinfo {year} {1965})}\BibitemShut {NoStop}%
\bibitem [{\citenamefont {Cheung}\ \emph {et~al.}(2015)\citenamefont {Cheung},
  \citenamefont {Kampf}, \citenamefont {Novotny},\ and\ \citenamefont
  {Trnka}}]{Cheung:2014dqa}%
  \BibitemOpen
  \bibfield  {author} {\bibinfo {author} {\bibfnamefont {C.}~\bibnamefont
  {Cheung}}, \bibinfo {author} {\bibfnamefont {K.}~\bibnamefont {Kampf}},
  \bibinfo {author} {\bibfnamefont {J.}~\bibnamefont {Novotny}}, \ and\
  \bibinfo {author} {\bibfnamefont {J.}~\bibnamefont {Trnka}},\ }\href
  {\doibase 10.1103/PhysRevLett.114.221602} {\bibfield  {journal} {\bibinfo
  {journal} {Phys. Rev. Lett.}\ }\textbf {\bibinfo {volume} {114}},\ \bibinfo
  {pages} {221602} (\bibinfo {year} {2015})},\ \Eprint
  {http://arxiv.org/abs/1412.4095} {arXiv:1412.4095 [hep-th]} \BibitemShut
  {NoStop}%
\bibitem [{\citenamefont {Cheung}\ \emph {et~al.}(2016)\citenamefont {Cheung},
  \citenamefont {Kampf}, \citenamefont {Novotny}, \citenamefont {Shen},\ and\
  \citenamefont {Trnka}}]{Cheung:2015ota}%
  \BibitemOpen
  \bibfield  {author} {\bibinfo {author} {\bibfnamefont {C.}~\bibnamefont
  {Cheung}}, \bibinfo {author} {\bibfnamefont {K.}~\bibnamefont {Kampf}},
  \bibinfo {author} {\bibfnamefont {J.}~\bibnamefont {Novotny}}, \bibinfo
  {author} {\bibfnamefont {C.-H.}\ \bibnamefont {Shen}}, \ and\ \bibinfo
  {author} {\bibfnamefont {J.}~\bibnamefont {Trnka}},\ }\href {\doibase
  10.1103/PhysRevLett.116.041601} {\bibfield  {journal} {\bibinfo  {journal}
  {Phys. Rev. Lett.}\ }\textbf {\bibinfo {volume} {116}},\ \bibinfo {pages}
  {041601} (\bibinfo {year} {2016})},\ \Eprint
  {http://arxiv.org/abs/1509.03309} {arXiv:1509.03309 [hep-th]} \BibitemShut
  {NoStop}%
\bibitem [{\citenamefont {Cheung}\ \emph {et~al.}(2017)\citenamefont {Cheung},
  \citenamefont {Kampf}, \citenamefont {Novotny}, \citenamefont {Shen},\ and\
  \citenamefont {Trnka}}]{Cheung:2016drk}%
  \BibitemOpen
  \bibfield  {author} {\bibinfo {author} {\bibfnamefont {C.}~\bibnamefont
  {Cheung}}, \bibinfo {author} {\bibfnamefont {K.}~\bibnamefont {Kampf}},
  \bibinfo {author} {\bibfnamefont {J.}~\bibnamefont {Novotny}}, \bibinfo
  {author} {\bibfnamefont {C.-H.}\ \bibnamefont {Shen}}, \ and\ \bibinfo
  {author} {\bibfnamefont {J.}~\bibnamefont {Trnka}},\ }\href {\doibase
  10.1007/JHEP02(2017)020} {\bibfield  {journal} {\bibinfo  {journal} {JHEP}\
  }\textbf {\bibinfo {volume} {02}},\ \bibinfo {pages} {020} (\bibinfo {year}
  {2017})},\ \Eprint {http://arxiv.org/abs/1611.03137} {arXiv:1611.03137
  [hep-th]} \BibitemShut {NoStop}%
\bibitem [{\citenamefont {Elvang}\ \emph {et~al.}(2019)\citenamefont {Elvang},
  \citenamefont {Hadjiantonis}, \citenamefont {Jones},\ and\ \citenamefont
  {Paranjape}}]{Elvang:2018dco}%
  \BibitemOpen
  \bibfield  {author} {\bibinfo {author} {\bibfnamefont {H.}~\bibnamefont
  {Elvang}}, \bibinfo {author} {\bibfnamefont {M.}~\bibnamefont
  {Hadjiantonis}}, \bibinfo {author} {\bibfnamefont {C.~R.~T.}\ \bibnamefont
  {Jones}}, \ and\ \bibinfo {author} {\bibfnamefont {S.}~\bibnamefont
  {Paranjape}},\ }\href {\doibase 10.1007/JHEP01(2019)195} {\bibfield
  {journal} {\bibinfo  {journal} {JHEP}\ }\textbf {\bibinfo {volume} {01}},\
  \bibinfo {pages} {195} (\bibinfo {year} {2019})},\ \Eprint
  {http://arxiv.org/abs/1806.06079} {arXiv:1806.06079 [hep-th]} \BibitemShut
  {NoStop}%
\bibitem [{\citenamefont {Low}\ and\ \citenamefont {Yin}(2019)}]{Low:2019ynd}%
  \BibitemOpen
  \bibfield  {author} {\bibinfo {author} {\bibfnamefont {I.}~\bibnamefont
  {Low}}\ and\ \bibinfo {author} {\bibfnamefont {Z.}~\bibnamefont {Yin}},\
  }\href {\doibase 10.1007/JHEP11(2019)078} {\bibfield  {journal} {\bibinfo
  {journal} {JHEP}\ }\textbf {\bibinfo {volume} {11}},\ \bibinfo {pages} {078}
  (\bibinfo {year} {2019})},\ \Eprint {http://arxiv.org/abs/1904.12859}
  {arXiv:1904.12859 [hep-th]} \BibitemShut {NoStop}%
\bibitem [{\citenamefont {Dai}\ \emph {et~al.}(2020)\citenamefont {Dai},
  \citenamefont {Low}, \citenamefont {Mehen},\ and\ \citenamefont
  {Mohapatra}}]{Dai:2020cpk}%
  \BibitemOpen
  \bibfield  {author} {\bibinfo {author} {\bibfnamefont {L.}~\bibnamefont
  {Dai}}, \bibinfo {author} {\bibfnamefont {I.}~\bibnamefont {Low}}, \bibinfo
  {author} {\bibfnamefont {T.}~\bibnamefont {Mehen}}, \ and\ \bibinfo {author}
  {\bibfnamefont {A.}~\bibnamefont {Mohapatra}},\ }\href {\doibase
  10.1103/PhysRevD.102.116011} {\bibfield  {journal} {\bibinfo  {journal}
  {Phys. Rev. D}\ }\textbf {\bibinfo {volume} {102}},\ \bibinfo {pages}
  {116011} (\bibinfo {year} {2020})},\ \Eprint
  {http://arxiv.org/abs/2009.01819} {arXiv:2009.01819 [hep-ph]} \BibitemShut
  {NoStop}%
\bibitem [{\citenamefont {Arkani-Hamed}\ \emph {et~al.}(2018)\citenamefont
  {Arkani-Hamed}, \citenamefont {Rodina},\ and\ \citenamefont
  {Trnka}}]{Arkani-Hamed:2016rak}%
  \BibitemOpen
  \bibfield  {author} {\bibinfo {author} {\bibfnamefont {N.}~\bibnamefont
  {Arkani-Hamed}}, \bibinfo {author} {\bibfnamefont {L.}~\bibnamefont
  {Rodina}}, \ and\ \bibinfo {author} {\bibfnamefont {J.}~\bibnamefont
  {Trnka}},\ }\href {\doibase 10.1103/PhysRevLett.120.231602} {\bibfield
  {journal} {\bibinfo  {journal} {Phys. Rev. Lett.}\ }\textbf {\bibinfo
  {volume} {120}},\ \bibinfo {pages} {231602} (\bibinfo {year} {2018})},\
  \Eprint {http://arxiv.org/abs/1612.02797} {arXiv:1612.02797 [hep-th]}
  \BibitemShut {NoStop}%
\bibitem [{\citenamefont {Rodina}(2019{\natexlab{a}})}]{Rodina:2016jyz}%
  \BibitemOpen
  \bibfield  {author} {\bibinfo {author} {\bibfnamefont {L.}~\bibnamefont
  {Rodina}},\ }\href {\doibase 10.1007/JHEP09(2019)084} {\bibfield  {journal}
  {\bibinfo  {journal} {JHEP}\ }\textbf {\bibinfo {volume} {09}},\ \bibinfo
  {pages} {084} (\bibinfo {year} {2019}{\natexlab{a}})},\ \Eprint
  {http://arxiv.org/abs/1612.06342} {arXiv:1612.06342 [hep-th]} \BibitemShut
  {NoStop}%
\bibitem [{\citenamefont {Rodina}(2019{\natexlab{b}})}]{Rodina:2018pcb}%
  \BibitemOpen
  \bibfield  {author} {\bibinfo {author} {\bibfnamefont {L.}~\bibnamefont
  {Rodina}},\ }\href {\doibase 10.1103/PhysRevLett.122.071601} {\bibfield
  {journal} {\bibinfo  {journal} {Phys. Rev. Lett.}\ }\textbf {\bibinfo
  {volume} {122}},\ \bibinfo {pages} {071601} (\bibinfo {year}
  {2019}{\natexlab{b}})},\ \Eprint {http://arxiv.org/abs/1807.09738}
  {arXiv:1807.09738 [hep-th]} \BibitemShut {NoStop}%
\bibitem [{\citenamefont {Cachazo}\ \emph {et~al.}(2016)\citenamefont
  {Cachazo}, \citenamefont {Cha},\ and\ \citenamefont
  {Mizera}}]{Cachazo:2016njl}%
  \BibitemOpen
  \bibfield  {author} {\bibinfo {author} {\bibfnamefont {F.}~\bibnamefont
  {Cachazo}}, \bibinfo {author} {\bibfnamefont {P.}~\bibnamefont {Cha}}, \ and\
  \bibinfo {author} {\bibfnamefont {S.}~\bibnamefont {Mizera}},\ }\href
  {\doibase 10.1007/JHEP06(2016)170} {\bibfield  {journal} {\bibinfo  {journal}
  {JHEP}\ }\textbf {\bibinfo {volume} {06}},\ \bibinfo {pages} {170} (\bibinfo
  {year} {2016})},\ \Eprint {http://arxiv.org/abs/1604.03893} {arXiv:1604.03893
  [hep-th]} \BibitemShut {NoStop}%
\bibitem [{\citenamefont {Cachazo}\ \emph
  {et~al.}(2014{\natexlab{a}})\citenamefont {Cachazo}, \citenamefont {He},\
  and\ \citenamefont {Yuan}}]{Cachazo:2013gna}%
  \BibitemOpen
  \bibfield  {author} {\bibinfo {author} {\bibfnamefont {F.}~\bibnamefont
  {Cachazo}}, \bibinfo {author} {\bibfnamefont {S.}~\bibnamefont {He}}, \ and\
  \bibinfo {author} {\bibfnamefont {E.~Y.}\ \bibnamefont {Yuan}},\ }\href
  {\doibase 10.1103/PhysRevD.90.065001} {\bibfield  {journal} {\bibinfo
  {journal} {Phys. Rev. D}\ }\textbf {\bibinfo {volume} {90}},\ \bibinfo
  {pages} {065001} (\bibinfo {year} {2014}{\natexlab{a}})},\ \Eprint
  {http://arxiv.org/abs/1306.6575} {arXiv:1306.6575 [hep-th]} \BibitemShut
  {NoStop}%
\bibitem [{\citenamefont {Cachazo}\ \emph
  {et~al.}(2014{\natexlab{b}})\citenamefont {Cachazo}, \citenamefont {He},\
  and\ \citenamefont {Yuan}}]{Cachazo:2013hca}%
  \BibitemOpen
  \bibfield  {author} {\bibinfo {author} {\bibfnamefont {F.}~\bibnamefont
  {Cachazo}}, \bibinfo {author} {\bibfnamefont {S.}~\bibnamefont {He}}, \ and\
  \bibinfo {author} {\bibfnamefont {E.~Y.}\ \bibnamefont {Yuan}},\ }\href
  {\doibase 10.1103/PhysRevLett.113.171601} {\bibfield  {journal} {\bibinfo
  {journal} {Phys. Rev. Lett.}\ }\textbf {\bibinfo {volume} {113}},\ \bibinfo
  {pages} {171601} (\bibinfo {year} {2014}{\natexlab{b}})},\ \Eprint
  {http://arxiv.org/abs/1307.2199} {arXiv:1307.2199 [hep-th]} \BibitemShut
  {NoStop}%
\bibitem [{\citenamefont {Cachazo}\ \emph
  {et~al.}(2014{\natexlab{c}})\citenamefont {Cachazo}, \citenamefont {He},\
  and\ \citenamefont {Yuan}}]{Cachazo:2013iea}%
  \BibitemOpen
  \bibfield  {author} {\bibinfo {author} {\bibfnamefont {F.}~\bibnamefont
  {Cachazo}}, \bibinfo {author} {\bibfnamefont {S.}~\bibnamefont {He}}, \ and\
  \bibinfo {author} {\bibfnamefont {E.~Y.}\ \bibnamefont {Yuan}},\ }\href
  {\doibase 10.1007/JHEP07(2014)033} {\bibfield  {journal} {\bibinfo  {journal}
  {JHEP}\ }\textbf {\bibinfo {volume} {07}},\ \bibinfo {pages} {033} (\bibinfo
  {year} {2014}{\natexlab{c}})},\ \Eprint {http://arxiv.org/abs/1309.0885}
  {arXiv:1309.0885 [hep-th]} \BibitemShut {NoStop}%
\bibitem [{\citenamefont {Cachazo}\ \emph
  {et~al.}(2015{\natexlab{a}})\citenamefont {Cachazo}, \citenamefont {He},\
  and\ \citenamefont {Yuan}}]{Cachazo:2014xea}%
  \BibitemOpen
  \bibfield  {author} {\bibinfo {author} {\bibfnamefont {F.}~\bibnamefont
  {Cachazo}}, \bibinfo {author} {\bibfnamefont {S.}~\bibnamefont {He}}, \ and\
  \bibinfo {author} {\bibfnamefont {E.~Y.}\ \bibnamefont {Yuan}},\ }\href
  {\doibase 10.1007/JHEP07(2015)149} {\bibfield  {journal} {\bibinfo  {journal}
  {JHEP}\ }\textbf {\bibinfo {volume} {07}},\ \bibinfo {pages} {149} (\bibinfo
  {year} {2015}{\natexlab{a}})},\ \Eprint {http://arxiv.org/abs/1412.3479}
  {arXiv:1412.3479 [hep-th]} \BibitemShut {NoStop}%
\bibitem [{\citenamefont {Low}\ and\ \citenamefont
  {Yin}(2018{\natexlab{a}})}]{Low:2017mlh}%
  \BibitemOpen
  \bibfield  {author} {\bibinfo {author} {\bibfnamefont {I.}~\bibnamefont
  {Low}}\ and\ \bibinfo {author} {\bibfnamefont {Z.}~\bibnamefont {Yin}},\
  }\href {\doibase 10.1103/PhysRevLett.120.061601} {\bibfield  {journal}
  {\bibinfo  {journal} {Phys. Rev. Lett.}\ }\textbf {\bibinfo {volume} {120}},\
  \bibinfo {pages} {061601} (\bibinfo {year} {2018}{\natexlab{a}})},\ \Eprint
  {http://arxiv.org/abs/1709.08639} {arXiv:1709.08639 [hep-th]} \BibitemShut
  {NoStop}%
\bibitem [{\citenamefont {Low}\ and\ \citenamefont
  {Yin}(2018{\natexlab{b}})}]{Low:2018acv}%
  \BibitemOpen
  \bibfield  {author} {\bibinfo {author} {\bibfnamefont {I.}~\bibnamefont
  {Low}}\ and\ \bibinfo {author} {\bibfnamefont {Z.}~\bibnamefont {Yin}},\
  }\href {\doibase 10.1007/JHEP10(2018)078} {\bibfield  {journal} {\bibinfo
  {journal} {JHEP}\ }\textbf {\bibinfo {volume} {10}},\ \bibinfo {pages} {078}
  (\bibinfo {year} {2018}{\natexlab{b}})},\ \Eprint
  {http://arxiv.org/abs/1804.08629} {arXiv:1804.08629 [hep-th]} \BibitemShut
  {NoStop}%
\bibitem [{\citenamefont {Yin}(2019)}]{Yin:2018hht}%
  \BibitemOpen
  \bibfield  {author} {\bibinfo {author} {\bibfnamefont {Z.}~\bibnamefont
  {Yin}},\ }\href {\doibase 10.1007/JHEP03(2019)158} {\bibfield  {journal}
  {\bibinfo  {journal} {JHEP}\ }\textbf {\bibinfo {volume} {03}},\ \bibinfo
  {pages} {158} (\bibinfo {year} {2019})},\ \Eprint
  {http://arxiv.org/abs/1810.07186} {arXiv:1810.07186 [hep-th]} \BibitemShut
  {NoStop}%
\bibitem [{\citenamefont {Low}\ and\ \citenamefont {Yin}(2020)}]{Low:2019wuv}%
  \BibitemOpen
  \bibfield  {author} {\bibinfo {author} {\bibfnamefont {I.}~\bibnamefont
  {Low}}\ and\ \bibinfo {author} {\bibfnamefont {Z.}~\bibnamefont {Yin}},\
  }\href {\doibase 10.1016/j.physletb.2020.135544} {\bibfield  {journal}
  {\bibinfo  {journal} {Phys. Lett. B}\ }\textbf {\bibinfo {volume} {807}},\
  \bibinfo {pages} {135544} (\bibinfo {year} {2020})},\ \Eprint
  {http://arxiv.org/abs/1911.08490} {arXiv:1911.08490 [hep-th]} \BibitemShut
  {NoStop}%
\bibitem [{\citenamefont {Low}\ \emph {et~al.}(2021)\citenamefont {Low},
  \citenamefont {Rodina},\ and\ \citenamefont {Yin}}]{Low:2020ubn}%
  \BibitemOpen
  \bibfield  {author} {\bibinfo {author} {\bibfnamefont {I.}~\bibnamefont
  {Low}}, \bibinfo {author} {\bibfnamefont {L.}~\bibnamefont {Rodina}}, \ and\
  \bibinfo {author} {\bibfnamefont {Z.}~\bibnamefont {Yin}},\ }\href {\doibase
  10.1103/PhysRevD.103.025004} {\bibfield  {journal} {\bibinfo  {journal}
  {Phys. Rev. D}\ }\textbf {\bibinfo {volume} {103}},\ \bibinfo {pages}
  {025004} (\bibinfo {year} {2021})},\ \Eprint
  {http://arxiv.org/abs/2009.00008} {arXiv:2009.00008 [hep-th]} \BibitemShut
  {NoStop}%
\bibitem [{\citenamefont {Bern}\ \emph {et~al.}(2019)\citenamefont {Bern},
  \citenamefont {Carrasco}, \citenamefont {Chiodaroli}, \citenamefont
  {Johansson},\ and\ \citenamefont {Roiban}}]{Bern:2019prr}%
  \BibitemOpen
  \bibfield  {author} {\bibinfo {author} {\bibfnamefont {Z.}~\bibnamefont
  {Bern}}, \bibinfo {author} {\bibfnamefont {J.~J.}\ \bibnamefont {Carrasco}},
  \bibinfo {author} {\bibfnamefont {M.}~\bibnamefont {Chiodaroli}}, \bibinfo
  {author} {\bibfnamefont {H.}~\bibnamefont {Johansson}}, \ and\ \bibinfo
  {author} {\bibfnamefont {R.}~\bibnamefont {Roiban}},\ }\href@noop {} {\
  (\bibinfo {year} {2019})},\ \Eprint {http://arxiv.org/abs/1909.01358}
  {arXiv:1909.01358 [hep-th]} \BibitemShut {NoStop}%
\bibitem [{\citenamefont {Carrasco}\ \emph {et~al.}(2020)\citenamefont
  {Carrasco}, \citenamefont {Rodina}, \citenamefont {Yin},\ and\ \citenamefont
  {Zekioglu}}]{Carrasco:2019yyn}%
  \BibitemOpen
  \bibfield  {author} {\bibinfo {author} {\bibfnamefont {J.~J.~M.}\
  \bibnamefont {Carrasco}}, \bibinfo {author} {\bibfnamefont {L.}~\bibnamefont
  {Rodina}}, \bibinfo {author} {\bibfnamefont {Z.}~\bibnamefont {Yin}}, \ and\
  \bibinfo {author} {\bibfnamefont {S.}~\bibnamefont {Zekioglu}},\ }\href
  {\doibase 10.1103/PhysRevLett.125.251602} {\bibfield  {journal} {\bibinfo
  {journal} {Phys. Rev. Lett.}\ }\textbf {\bibinfo {volume} {125}},\ \bibinfo
  {pages} {251602} (\bibinfo {year} {2020})},\ \Eprint
  {http://arxiv.org/abs/1910.12850} {arXiv:1910.12850 [hep-th]} \BibitemShut
  {NoStop}%
\bibitem [{\citenamefont {Bern}\ \emph {et~al.}(2008)\citenamefont {Bern},
  \citenamefont {Carrasco},\ and\ \citenamefont {Johansson}}]{Bern:2008qj}%
  \BibitemOpen
  \bibfield  {author} {\bibinfo {author} {\bibfnamefont {Z.}~\bibnamefont
  {Bern}}, \bibinfo {author} {\bibfnamefont {J.~J.~M.}\ \bibnamefont
  {Carrasco}}, \ and\ \bibinfo {author} {\bibfnamefont {H.}~\bibnamefont
  {Johansson}},\ }\href {\doibase 10.1103/PhysRevD.78.085011} {\bibfield
  {journal} {\bibinfo  {journal} {Phys. Rev. D}\ }\textbf {\bibinfo {volume}
  {78}},\ \bibinfo {pages} {085011} (\bibinfo {year} {2008})},\ \Eprint
  {http://arxiv.org/abs/0805.3993} {arXiv:0805.3993 [hep-ph]} \BibitemShut
  {NoStop}%
\bibitem [{\citenamefont {Cachazo}\ \emph
  {et~al.}(2015{\natexlab{b}})\citenamefont {Cachazo}, \citenamefont {He},\
  and\ \citenamefont {Yuan}}]{Cachazo:2015ksa}%
  \BibitemOpen
  \bibfield  {author} {\bibinfo {author} {\bibfnamefont {F.}~\bibnamefont
  {Cachazo}}, \bibinfo {author} {\bibfnamefont {S.}~\bibnamefont {He}}, \ and\
  \bibinfo {author} {\bibfnamefont {E.~Y.}\ \bibnamefont {Yuan}},\ }\href
  {\doibase 10.1103/PhysRevD.92.065030} {\bibfield  {journal} {\bibinfo
  {journal} {Phys. Rev. D}\ }\textbf {\bibinfo {volume} {92}},\ \bibinfo
  {pages} {065030} (\bibinfo {year} {2015}{\natexlab{b}})},\ \Eprint
  {http://arxiv.org/abs/1503.04816} {arXiv:1503.04816 [hep-th]} \BibitemShut
  {NoStop}%
\bibitem [{\citenamefont {Du}\ and\ \citenamefont {Luo}(2015)}]{Du:2015esa}%
  \BibitemOpen
  \bibfield  {author} {\bibinfo {author} {\bibfnamefont {Y.-J.}\ \bibnamefont
  {Du}}\ and\ \bibinfo {author} {\bibfnamefont {H.}~\bibnamefont {Luo}},\
  }\href {\doibase 10.1007/JHEP08(2015)058} {\bibfield  {journal} {\bibinfo
  {journal} {JHEP}\ }\textbf {\bibinfo {volume} {08}},\ \bibinfo {pages} {058}
  (\bibinfo {year} {2015})},\ \Eprint {http://arxiv.org/abs/1505.04411}
  {arXiv:1505.04411 [hep-th]} \BibitemShut {NoStop}%
\bibitem [{\citenamefont {Low}(2016)}]{Low:2015ogb}%
  \BibitemOpen
  \bibfield  {author} {\bibinfo {author} {\bibfnamefont {I.}~\bibnamefont
  {Low}},\ }\href {\doibase 10.1103/PhysRevD.93.045032} {\bibfield  {journal}
  {\bibinfo  {journal} {Phys. Rev. D}\ }\textbf {\bibinfo {volume} {93}},\
  \bibinfo {pages} {045032} (\bibinfo {year} {2016})},\ \Eprint
  {http://arxiv.org/abs/1512.01232} {arXiv:1512.01232 [hep-th]} \BibitemShut
  {NoStop}%
\bibitem [{\citenamefont {Wess}\ and\ \citenamefont
  {Zumino}(1971)}]{Wess:1971yu}%
  \BibitemOpen
  \bibfield  {author} {\bibinfo {author} {\bibfnamefont {J.}~\bibnamefont
  {Wess}}\ and\ \bibinfo {author} {\bibfnamefont {B.}~\bibnamefont {Zumino}},\
  }\href {\doibase 10.1016/0370-2693(71)90582-X} {\bibfield  {journal}
  {\bibinfo  {journal} {Phys. Lett. B}\ }\textbf {\bibinfo {volume} {37}},\
  \bibinfo {pages} {95} (\bibinfo {year} {1971})}\BibitemShut {NoStop}%
\bibitem [{\citenamefont {Witten}(1983)}]{Witten:1983tw}%
  \BibitemOpen
  \bibfield  {author} {\bibinfo {author} {\bibfnamefont {E.}~\bibnamefont
  {Witten}},\ }\href {\doibase 10.1016/0550-3213(83)90063-9} {\bibfield
  {journal} {\bibinfo  {journal} {Nucl. Phys. B}\ }\textbf {\bibinfo {volume}
  {223}},\ \bibinfo {pages} {422} (\bibinfo {year} {1983})}\BibitemShut
  {NoStop}%
\bibitem [{\citenamefont {Dixon}(1996)}]{Dixon:1996wi}%
  \BibitemOpen
  \bibfield  {author} {\bibinfo {author} {\bibfnamefont {L.~J.}\ \bibnamefont
  {Dixon}},\ }in\ \href@noop {} {\emph {\bibinfo {booktitle} {{Theoretical
  Advanced Study Institute in Elementary Particle Physics (TASI 95): QCD and
  Beyond}}}}\ (\bibinfo {year} {1996})\ pp.\ \bibinfo {pages} {539--584},\
  \Eprint {http://arxiv.org/abs/hep-ph/9601359} {arXiv:hep-ph/9601359}
  \BibitemShut {NoStop}%
\bibitem [{\citenamefont {Del~Duca}\ \emph {et~al.}(2000)\citenamefont
  {Del~Duca}, \citenamefont {Dixon},\ and\ \citenamefont
  {Maltoni}}]{DelDuca:1999rs}%
  \BibitemOpen
  \bibfield  {author} {\bibinfo {author} {\bibfnamefont {V.}~\bibnamefont
  {Del~Duca}}, \bibinfo {author} {\bibfnamefont {L.~J.}\ \bibnamefont {Dixon}},
  \ and\ \bibinfo {author} {\bibfnamefont {F.}~\bibnamefont {Maltoni}},\ }\href
  {\doibase 10.1016/S0550-3213(99)00809-3} {\bibfield  {journal} {\bibinfo
  {journal} {Nucl. Phys. B}\ }\textbf {\bibinfo {volume} {571}},\ \bibinfo
  {pages} {51} (\bibinfo {year} {2000})},\ \Eprint
  {http://arxiv.org/abs/hep-ph/9910563} {arXiv:hep-ph/9910563} \BibitemShut
  {NoStop}%
\bibitem [{\citenamefont {Kleiss}\ and\ \citenamefont
  {Kuijf}(1989)}]{Kleiss:1988ne}%
  \BibitemOpen
  \bibfield  {author} {\bibinfo {author} {\bibfnamefont {R.}~\bibnamefont
  {Kleiss}}\ and\ \bibinfo {author} {\bibfnamefont {H.}~\bibnamefont {Kuijf}},\
  }\href {\doibase 10.1016/0550-3213(89)90574-9} {\bibfield  {journal}
  {\bibinfo  {journal} {Nucl. Phys. B}\ }\textbf {\bibinfo {volume} {312}},\
  \bibinfo {pages} {616} (\bibinfo {year} {1989})}\BibitemShut {NoStop}%
\bibitem [{\citenamefont {Berends}\ and\ \citenamefont
  {Giele}(1988)}]{Berends:1987me}%
  \BibitemOpen
  \bibfield  {author} {\bibinfo {author} {\bibfnamefont {F.~A.}\ \bibnamefont
  {Berends}}\ and\ \bibinfo {author} {\bibfnamefont {W.~T.}\ \bibnamefont
  {Giele}},\ }\href {\doibase 10.1016/0550-3213(88)90442-7} {\bibfield
  {journal} {\bibinfo  {journal} {Nucl. Phys. B}\ }\textbf {\bibinfo {volume}
  {306}},\ \bibinfo {pages} {759} (\bibinfo {year} {1988})}\BibitemShut
  {NoStop}%
\bibitem [{\citenamefont {Chiodaroli}\ \emph {et~al.}(2017)\citenamefont
  {Chiodaroli}, \citenamefont {Gunaydin}, \citenamefont {Johansson},\ and\
  \citenamefont {Roiban}}]{Chiodaroli:2017ngp}%
  \BibitemOpen
  \bibfield  {author} {\bibinfo {author} {\bibfnamefont {M.}~\bibnamefont
  {Chiodaroli}}, \bibinfo {author} {\bibfnamefont {M.}~\bibnamefont
  {Gunaydin}}, \bibinfo {author} {\bibfnamefont {H.}~\bibnamefont {Johansson}},
  \ and\ \bibinfo {author} {\bibfnamefont {R.}~\bibnamefont {Roiban}},\ }\href
  {\doibase 10.1007/JHEP07(2017)002} {\bibfield  {journal} {\bibinfo  {journal}
  {JHEP}\ }\textbf {\bibinfo {volume} {07}},\ \bibinfo {pages} {002} (\bibinfo
  {year} {2017})},\ \Eprint {http://arxiv.org/abs/1703.00421} {arXiv:1703.00421
  [hep-th]} \BibitemShut {NoStop}%
\bibitem [{\citenamefont {Chiodaroli}\ \emph {et~al.}(2015)\citenamefont
  {Chiodaroli}, \citenamefont {G\"unaydin}, \citenamefont {Johansson},\ and\
  \citenamefont {Roiban}}]{Chiodaroli:2014xia}%
  \BibitemOpen
  \bibfield  {author} {\bibinfo {author} {\bibfnamefont {M.}~\bibnamefont
  {Chiodaroli}}, \bibinfo {author} {\bibfnamefont {M.}~\bibnamefont
  {G\"unaydin}}, \bibinfo {author} {\bibfnamefont {H.}~\bibnamefont
  {Johansson}}, \ and\ \bibinfo {author} {\bibfnamefont {R.}~\bibnamefont
  {Roiban}},\ }\href {\doibase 10.1007/JHEP01(2015)081} {\bibfield  {journal}
  {\bibinfo  {journal} {JHEP}\ }\textbf {\bibinfo {volume} {01}},\ \bibinfo
  {pages} {081} (\bibinfo {year} {2015})},\ \Eprint
  {http://arxiv.org/abs/1408.0764} {arXiv:1408.0764 [hep-th]} \BibitemShut
  {NoStop}%
\bibitem [{\citenamefont {Kawai}\ \emph {et~al.}(1986)\citenamefont {Kawai},
  \citenamefont {Lewellen},\ and\ \citenamefont {Tye}}]{Kawai:1985xq}%
  \BibitemOpen
  \bibfield  {author} {\bibinfo {author} {\bibfnamefont {H.}~\bibnamefont
  {Kawai}}, \bibinfo {author} {\bibfnamefont {D.~C.}\ \bibnamefont {Lewellen}},
  \ and\ \bibinfo {author} {\bibfnamefont {S.~H.~H.}\ \bibnamefont {Tye}},\
  }\href {\doibase 10.1016/0550-3213(86)90362-7} {\bibfield  {journal}
  {\bibinfo  {journal} {Nucl. Phys. B}\ }\textbf {\bibinfo {volume} {269}},\
  \bibinfo {pages} {1} (\bibinfo {year} {1986})}\BibitemShut {NoStop}%
\bibitem [{\citenamefont {Elvang}\ \emph {et~al.}(2017)\citenamefont {Elvang},
  \citenamefont {Jones},\ and\ \citenamefont {Naculich}}]{Elvang:2016qvq}%
  \BibitemOpen
  \bibfield  {author} {\bibinfo {author} {\bibfnamefont {H.}~\bibnamefont
  {Elvang}}, \bibinfo {author} {\bibfnamefont {C.~R.~T.}\ \bibnamefont
  {Jones}}, \ and\ \bibinfo {author} {\bibfnamefont {S.~G.}\ \bibnamefont
  {Naculich}},\ }\href {\doibase 10.1103/PhysRevLett.118.231601} {\bibfield
  {journal} {\bibinfo  {journal} {Phys. Rev. Lett.}\ }\textbf {\bibinfo
  {volume} {118}},\ \bibinfo {pages} {231601} (\bibinfo {year} {2017})},\
  \Eprint {http://arxiv.org/abs/1611.07534} {arXiv:1611.07534 [hep-th]}
  \BibitemShut {NoStop}%
\bibitem [{\citenamefont {Du}\ and\ \citenamefont {Luo}(2017)}]{Du:2016njc}%
  \BibitemOpen
  \bibfield  {author} {\bibinfo {author} {\bibfnamefont {Y.-J.}\ \bibnamefont
  {Du}}\ and\ \bibinfo {author} {\bibfnamefont {H.}~\bibnamefont {Luo}},\
  }\href {\doibase 10.1007/JHEP03(2017)062} {\bibfield  {journal} {\bibinfo
  {journal} {JHEP}\ }\textbf {\bibinfo {volume} {03}},\ \bibinfo {pages} {062}
  (\bibinfo {year} {2017})},\ \Eprint {http://arxiv.org/abs/1611.07479}
  {arXiv:1611.07479 [hep-th]} \BibitemShut {NoStop}%
\bibitem [{\citenamefont {Pham}\ and\ \citenamefont
  {Truong}(1985)}]{Pham:1985cr}%
  \BibitemOpen
  \bibfield  {author} {\bibinfo {author} {\bibfnamefont {T.~N.}\ \bibnamefont
  {Pham}}\ and\ \bibinfo {author} {\bibfnamefont {T.~N.}\ \bibnamefont
  {Truong}},\ }\href {\doibase 10.1103/PhysRevD.31.3027} {\bibfield  {journal}
  {\bibinfo  {journal} {Phys. Rev. D}\ }\textbf {\bibinfo {volume} {31}},\
  \bibinfo {pages} {3027} (\bibinfo {year} {1985})}\BibitemShut {NoStop}%
\bibitem [{\citenamefont {Adams}\ \emph {et~al.}(2006)\citenamefont {Adams},
  \citenamefont {Arkani-Hamed}, \citenamefont {Dubovsky}, \citenamefont
  {Nicolis},\ and\ \citenamefont {Rattazzi}}]{Adams:2006sv}%
  \BibitemOpen
  \bibfield  {author} {\bibinfo {author} {\bibfnamefont {A.}~\bibnamefont
  {Adams}}, \bibinfo {author} {\bibfnamefont {N.}~\bibnamefont {Arkani-Hamed}},
  \bibinfo {author} {\bibfnamefont {S.}~\bibnamefont {Dubovsky}}, \bibinfo
  {author} {\bibfnamefont {A.}~\bibnamefont {Nicolis}}, \ and\ \bibinfo
  {author} {\bibfnamefont {R.}~\bibnamefont {Rattazzi}},\ }\href {\doibase
  10.1088/1126-6708/2006/10/014} {\bibfield  {journal} {\bibinfo  {journal}
  {JHEP}\ }\textbf {\bibinfo {volume} {10}},\ \bibinfo {pages} {014} (\bibinfo
  {year} {2006})},\ \Eprint {http://arxiv.org/abs/hep-th/0602178}
  {arXiv:hep-th/0602178} \BibitemShut {NoStop}%
\bibitem [{\citenamefont {Arkani-Hamed}\ \emph {et~al.}(2020)\citenamefont
  {Arkani-Hamed}, \citenamefont {Huang},\ and\ \citenamefont
  {Huang}}]{Arkani-Hamed:2020blm}%
  \BibitemOpen
  \bibfield  {author} {\bibinfo {author} {\bibfnamefont {N.}~\bibnamefont
  {Arkani-Hamed}}, \bibinfo {author} {\bibfnamefont {T.-C.}\ \bibnamefont
  {Huang}}, \ and\ \bibinfo {author} {\bibfnamefont {Y.-T.}\ \bibnamefont
  {Huang}},\ }\href@noop {} {\  (\bibinfo {year} {2020})},\ \Eprint
  {http://arxiv.org/abs/2012.15849} {arXiv:2012.15849 [hep-th]} \BibitemShut
  {NoStop}%
\bibitem [{\citenamefont {Huang}\ \emph {et~al.}(2020)\citenamefont {Huang},
  \citenamefont {Liu}, \citenamefont {Rodina},\ and\ \citenamefont
  {Wang}}]{Huang:2020nqy}%
  \BibitemOpen
  \bibfield  {author} {\bibinfo {author} {\bibfnamefont {Y.-t.}\ \bibnamefont
  {Huang}}, \bibinfo {author} {\bibfnamefont {J.-Y.}\ \bibnamefont {Liu}},
  \bibinfo {author} {\bibfnamefont {L.}~\bibnamefont {Rodina}}, \ and\ \bibinfo
  {author} {\bibfnamefont {Y.}~\bibnamefont {Wang}},\ }\href@noop {} {\
  (\bibinfo {year} {2020})},\ \Eprint {http://arxiv.org/abs/2008.02293}
  {arXiv:2008.02293 [hep-th]} \BibitemShut {NoStop}%
\bibitem [{\citenamefont {Bern}\ \emph {et~al.}(2014)\citenamefont {Bern},
  \citenamefont {Davies},\ and\ \citenamefont {Nohle}}]{Bern:2014oka}%
  \BibitemOpen
  \bibfield  {author} {\bibinfo {author} {\bibfnamefont {Z.}~\bibnamefont
  {Bern}}, \bibinfo {author} {\bibfnamefont {S.}~\bibnamefont {Davies}}, \ and\
  \bibinfo {author} {\bibfnamefont {J.}~\bibnamefont {Nohle}},\ }\href
  {\doibase 10.1103/PhysRevD.90.085015} {\bibfield  {journal} {\bibinfo
  {journal} {Phys. Rev. D}\ }\textbf {\bibinfo {volume} {90}},\ \bibinfo
  {pages} {085015} (\bibinfo {year} {2014})},\ \Eprint
  {http://arxiv.org/abs/1405.1015} {arXiv:1405.1015 [hep-th]} \BibitemShut
  {NoStop}%
\bibitem [{\citenamefont {He}\ \emph {et~al.}(2014)\citenamefont {He},
  \citenamefont {Huang},\ and\ \citenamefont {Wen}}]{He:2014bga}%
  \BibitemOpen
  \bibfield  {author} {\bibinfo {author} {\bibfnamefont {S.}~\bibnamefont
  {He}}, \bibinfo {author} {\bibfnamefont {Y.-t.}\ \bibnamefont {Huang}}, \
  and\ \bibinfo {author} {\bibfnamefont {C.}~\bibnamefont {Wen}},\ }\href
  {\doibase 10.1007/JHEP12(2014)115} {\bibfield  {journal} {\bibinfo  {journal}
  {JHEP}\ }\textbf {\bibinfo {volume} {12}},\ \bibinfo {pages} {115} (\bibinfo
  {year} {2014})},\ \Eprint {http://arxiv.org/abs/1405.1410} {arXiv:1405.1410
  [hep-th]} \BibitemShut {NoStop}%
\bibitem [{\citenamefont {Bianchi}\ \emph {et~al.}(2015)\citenamefont
  {Bianchi}, \citenamefont {He}, \citenamefont {Huang},\ and\ \citenamefont
  {Wen}}]{Bianchi:2014gla}%
  \BibitemOpen
  \bibfield  {author} {\bibinfo {author} {\bibfnamefont {M.}~\bibnamefont
  {Bianchi}}, \bibinfo {author} {\bibfnamefont {S.}~\bibnamefont {He}},
  \bibinfo {author} {\bibfnamefont {Y.-t.}\ \bibnamefont {Huang}}, \ and\
  \bibinfo {author} {\bibfnamefont {C.}~\bibnamefont {Wen}},\ }\href {\doibase
  10.1103/PhysRevD.92.065022} {\bibfield  {journal} {\bibinfo  {journal} {Phys.
  Rev. D}\ }\textbf {\bibinfo {volume} {92}},\ \bibinfo {pages} {065022}
  (\bibinfo {year} {2015})},\ \Eprint {http://arxiv.org/abs/1406.5155}
  {arXiv:1406.5155 [hep-th]} \BibitemShut {NoStop}%
\end{thebibliography}%

\end{document}